\documentclass[longauth,useAMS,usenatbib]{aa}  

\usepackage{graphicx}
\usepackage{txfonts}
\usepackage{hyperref}
\usepackage{longtable}
\usepackage{appendix}
\begin{document}

   \title{TESS and ESPRESSO discover a super-Earth and a mini-Neptune orbiting the K-dwarf TOI--238 \thanks{Based [in part] on Guaranteed Time Observations collected at the European Southern Observatory under ESO programme 1102.C-0744. by the ESPRESSO Consortium.} }

   \author{A.~Su{\'a}rez Mascare{\~n}o\inst{\ref{iac},\ref{uiac}},
           V.\,M.~Passegger\inst{\ref{iac},\ref{uiac},\ref{hs}}, 
           J.\,I.~Gonz{\'a}lez Hern{\'a}ndez\inst{\ref{iac},\ref{uiac}},  
           D.\,J.~Armstrong\inst{\ref{Warwick},\ref{Warwick2}},
           L.\,D.~Nielsen\inst{\ref{ESO}},
           C.~Lovis\inst{\ref{unige1}}, 
           B.~Lavie\inst{\ref{unige1}},
           S.\,G.~Sousa\inst{\ref{porto1}},    
           A.\,M.~Silva\inst{\ref{porto1},\ref{porto2}},
           R.~Allart\inst{\ref{montreal},\ref{unige1}},           
           R.~Rebolo\inst{\ref{iac},\ref{uiac},\ref{csic}},
           F.~Pepe\inst{\ref{unige1}},  
           N.\,C.~Santos\inst{\ref{porto1},\ref{porto2}},
           S.~Cristiani\inst{\ref{INAF-Trieste}},                        
           A.~Sozzetti\inst{\ref{INAF-Torino}}, 
           M.~R.~Zapatero~Osorio\inst{\ref{cab}},
           H.\,M.~Tabernero\inst{\ref{ucm}},
           X.~Dumusque\inst{\ref{unige1}},
           S.~Udry\inst{\ref{unige1}},
           %Alphabetical order
           V.~Adibekyan\inst{\ref{porto1},\ref{porto2}},
           C.~Allende~Prieto\inst{\ref{iac},\ref{uiac}},
           Y.~Alibert\inst{\ref{bern}},
           S.\,C.\,C.~Barros\inst{\ref{porto1},\ref{porto2}},
           F.~Bouchy\inst{\ref{unige1}},
           A.~Castro-Gonz\'{a}lez\inst{\ref{cab}},
           K.\,A.~Collins\inst{\ref{Hardvard_Smith}},
           M.~Damasso\inst{\ref{INAF-Torino}},
           V.~D'Odorico \inst{\ref{INAF-Trieste},\ref{Pisa}},
           O.\,D.\,S. ~Demangeon\inst{\ref{porto1},\ref{porto2}},
           P. ~Di ~Marcantonio\inst{\ref{INAF-Trieste}},
           D.~Ehrenreich\inst{\ref{unige1}},
           A.~Hadjigeorghiou\inst{\ref{Warwick},\ref{Warwick2}},
           N.~Hara\inst{\ref{unige1}},
           F.~Hawthorn\inst{\ref{Warwick},\ref{Warwick2}},
           J.\,M.~Jenkins\inst{\ref{NASA-Ames}}, 
           J.~Lillo-Box\inst{\ref{cab}},
           G.~Lo Curto\inst{\ref{ESO}},
           C.\,J.\,A.\,P.~Martins\inst{\ref{porto1},\ref{porto3}},
           A.~Mehner\inst{\ref{ESO}},
           G.~Micela\inst{\ref{INAF-Palermo}},
           P.~Molaro\inst{\ref{INAF-Trieste}},
           N.~Nunes\inst{\ref{Lisbon}},
           N.~Nari\inst{\ref{uiac},\ref{lb},\ref{iac}},
           A.~Osborn\inst{\ref{Warwick},\ref{Warwick2}},
           E.~Pall\'e\inst{\ref{iac},\ref{uiac}},
           G.\,R.~Ricker\inst{\ref{MIT-Physics}},
           J.~Rodrigues\inst{\ref{porto1},\ref{porto2}},
           P.~Rowden\inst{\ref{RAS}},
           S.~Seager\inst{\ref{MIT-Physics},\ref{MIT-Earth},\ref{MIT-Aero}},
           A.\,K.~Stefanov\inst{\ref{iac},\ref{uiac}},
           P.\,A.~Str{\o}m\inst{\ref{Warwick},\ref{Warwick2}},
           J.\,N.\,S.~Villase{\~n}or\inst{\ref{MIT-Physics}},
           C.\,N.~Watkins\inst{\ref{Hardvard_Smith}},
           J.~Winn\inst{\ref{Princeton}},
           B.~Wohler\inst{\ref{SETI},\ref{NASA-Ames}} and 
           R.~Zambelli\inst{\ref{Lunae}}
          }

    \authorrunning{A.~Su{\'a}rez Mascare{\~n}o et al.}
    
   \institute{Instituto de Astrof\'{\i}sica de Canarias, c/ V\'ia L\'actea s/n, 38205 La Laguna, Tenerife, Spain\label{iac}\\
    \email{asm@iac.es} 
    \and Departamento de Astrof\'{\i}sica, Universidad de La Laguna, 38206 La Laguna, Tenerife, Spain \label{uiac}
    \and Hamburger Sternwarte, Gojenbergsweg 112, 21029 Hamburg, Germany \label{hs}  
    \and Department of Physics, University of Warwick, Gibbet Hill Road, Coventry CV4 7AL, UK \label{Warwick} 
    \and Centre for Exoplanets and Habitability, University of Warwick, Gibbet Hill Road, Coventry CV4 7AL, UK \label{Warwick2}
    \and European Southern Observatory, Karl-Schwarzschild-Straße 2, 85748 Garching bei München, Germany \label{ESO}
    \and Observatoire astronomique de l’Universit\'e de Gen\`{e}ve, Chemin Pegasi 51b, 1290 Versoix, Switzerland \label{unige1}
    \and Instituto de Astrof\'isica e Ci\^{e}ncias do Espa\c{c}o, CAUP, Universidade do Porto, Rua das Estrelas, 4150-762 Porto, Portugal \label{porto1}
    \and Departamento de F\'isica e Astronomia, Faculdade de Ci\^{e}ncias, Universidade do Porto, Rua do Campo Alegre, 4169-007 Porto, Portugal \label{porto2}
    \and  D\'epartement de Physique, Institut Trottier de Recherche sur les Exoplan\`etes, Universit\'e de Montr\'eal, Montr\'eal, Qu\'ebec, H3T 1J4, Canada
      \label{montreal}
    \and    Consejo Superior de Investigaciones Cient{\'\i}ficas (CSIC), E-28006 Madrid, Spain \label{csic}    
    \and    INAF – Osservatorio Astronomico di Trieste, Via Tiepolo 11, I-34143 Trieste, Italy \label{INAF-Trieste}
    \and  INAF – Osservatorio Astrofisico di Torino, Strada Osservatorio, 20 10025 Pino Torinese (TO), Italy \label{INAF-Torino} 
    \and Centro de Astrobiolog\'{i}a, CSIC-INTA, Camino Bajo del Castillo s/n, E-28692 Villanueva de la Ca{\~n}ada, Madrid, Spain \label{cab}
    \and Departamento de F{\'i}sica de la Tierra y Astrof{\'i}sica \& IPARCOS-UCM (Instituto de F\'{i}sica de Part\'{i}culas y del Cosmos de la UCM), Facultad de Ciencias F{\'i}sicas, Universidad Complutense de Madrid, E-28040 Madrid, Spain \label{ucm}
    \and Physics Institute of University of Bern, Gesellschafts strasse 6, 3012, Bern, Switzerland\label{bern}
    \and Scuola Normale Superiore, Piazza dei Cavalieri 7, I-56126, Pisa,  Italy \label{Pisa}
    \and NASA Ames Research Center, Moffett Field, CA 94035, USA \label{NASA-Ames}
    \and Center for Astrophysics - Harvard \& Smithsonian, 60 Garden Street, Cambridge, MA 02138, USA \label{Hardvard_Smith}
    \and Centro de Astrof\'{i}sica da Universidade do Porto, Rua das Estrelas, 4150-762 Porto, Portugal \label{porto3}
    \and INAF - Osservatorio Astronomico di Palermo, Piazza del Parlamento 1, 90134 Palermo, Italy \label{INAF-Palermo}
    \and Instituto de Astrof\'{i}sica e Ci\^{e}ncias do Espa\c{C}o, Faculdade de Ci\^{e}ncias da Universidade de Lisboa, Campo Grande, 1749-016, Lisboa, Portugal \label{Lisbon}
    \and Light Bridges S. L., Avda. Alcalde Ram{\'i}rez Bethencourt, 17, E-35004, Las Palmas de Gran Canaria, Canarias, Spain \label{lb}
    \and Royal Astronomical Society, Burlington House, Piccadilly, London W1J 0BQ, United Kingdom \label{RAS}
    \and Department of Physics and Kavli Institute for Astrophysics and Space Research, Massachusetts Institute of Technology, Cambridge, MA 02139, USA \label{MIT-Physics}
    \and Department of Earth, Atmospheric and Planetary Sciences, Massachusetts Institute of Technology, Cambridge, MA 02139, USA \label{MIT-Earth}
    \and Department of Aeronautics and Astronautics, MIT, 77 Massachusetts Avenue, Cambridge, MA 02139, USA \label{MIT-Aero}
    \and Department of Astrophysical Sciences, Princeton University, Princeton, NJ 08544, USA \label{Princeton}
    \and SETI Institute, Mountain View, CA 94043 USA\label{SETI}
    \and Societ\`{a} Astronomica Lunae, Castelnuovo Magra, Italy \label{Lunae}
        }

   \date{Written October-December 2023}

% \abstract{}{}{}{}{} 
% 5 {} token are mandatory
 
  \abstract
   {The number of super-Earth and mini-Neptune planet discoveries has increased significantly in the last two decades thanks to transit and radial velocity surveys. When it is possible to apply both techniques, we can characterise the internal composition of exoplanets, which in turn provides unique insights on their architecture, formation and evolution.
   
   We performed a combined photometric and radial velocity analysis of TOI--238 (TYC 6398--132--1), which has one short-orbit super-Earth planet candidate announced by NASA's TESS team. We aim to confirm its planetary nature using radial velocities taken with the ESPRESSO and HARPS spectrographs, to measure its mass and to detect the presence of other possible planetary companions. We carried out a joint analysis by including Gaussian processes and Keplerian orbits to account for the stellar activity and planetary signals simultaneously.
   
   We detected the signal induced by TOI--238 b in the radial velocity time-series, and the presence of a second transiting planet, TOI--238 c, whose signal appears in RV and TESS data. TOI--238 b is a planet with a radius of 1.402$^{+0.084}_{-0.086}$ R$_{\oplus}$ and a mass of 3.40$^{+0.46}_{-0.45}$ M$_{\oplus}$. It orbits at a separation of 0.02118 $\pm$ 0.00038 AU of its host star, with an orbital period of 1.2730988 $\pm$ 0.0000029 days, and has an equilibrium temperature of 1311 $\pm$ 28 K. TOI--238 c has a radius of 2.18$\pm$ 0.18 R$_{\oplus}$ and a mass of 6.7 $\pm$ 1.1 M$_{\oplus}$. It orbits at a separation of 0.0749 $\pm$ 0.0013 AU of its host star, with an orbital period of 8.465652 $\pm$ 0.000031 days, and has an equilibrium temperature of 696 $\pm$ 15 K. The mass and radius of planet b are fully consistent with an Earth-like composition, making it likely a rocky super-Earth. Planet c could be a water-rich planet or a rocky planet with a small H-He atmosphere.}


   \maketitle
%-------------------------------------------------------------------

\section{Introduction}

The discovery and characterisation of exoplanets is one of the most exciting fields in astronomy today. The diversity and complexity of these worlds challenge our understanding of planetary formation and evolution, as well as offer potential clues for the origin and distribution of life in the universe. To fully explore these aspects, it is essential to obtain accurate measurements of the physical properties of exoplanets, such as their masses, radii, densities, atmospheres, orbits, and interactions with their host stars. 

The total number of known exoplanets keeps increasing. With main contributions from transit satellites such as the \textit{Kepler} Space Telescope \citep{Borucki2010, Howell2014} and the Transiting Exoplanet Survey Satellite (TESS, \citealt{Ricker2015}), as well as radial velocity surveys, the number of confirmed detections has surpassed 5500\footnote{NASA Exoplanet Archive \citep{Akeson2013}} \citep{Christiansen2022}. Despite this promising evolution in terms of detection, the characterisation of these objects is a matter that has not been addressed with the same level of success. From the large sample of confirmed exoplanets, only $\sim$20$\%$ have their true dynamical mass constrained, and only $\sim$10$\%$ of the complete sample have a measurement of their density with uncertainties lower than 33$\%$, with one of the main limitations being the uncertainty in the planetary mass. This limits studies of planetary formation or atmospheric characterisation, for which a precise mass measurement is required \citep{Batalha2019}.

There is an observed bimodal distribution of the size of small exoplanets \citep{Fulton2017}. Atmospheric mass loss mechanisms such as photo evaporation \citep{OwenWu2017} or core-powered mass loss \citep{Ginzburg2018} are both able to adequately reproduce it, assuming the planets have rocky cores. This has led to the interpretation that super-Earths and sub-Neptunes have formed accreting only dry condensates within the water ice line, and their difference in sizes is attributed to the absence or presence of extended volatile-rich atmospheres. However, given their bulk densities, another possibility is that sub-Neptune planets are water-rich worlds \citep{LegerSelsis2004}. The most promising way of discriminating between both scenarios is through atmospheric characterisation of exoplanets around the valley of the distribution, which first requires producing a significant sample of transiting low-mass exoplanets with precise masses. 

The most widely used method to measure exoplanet masses is radial velocity (RV), which relies on detecting the Doppler shift induced by the gravitational tug of an orbiting planet on its host star. The amplitude of this shift depends on several factors, such as the orbital period, eccentricity, inclination, and the planet-to-star mass ratio. In the past years, RV instruments such as HARPS~\citep{Mayor2003}, \mbox{HARPS-N}~\citep{Cosentino2012}, and more recently, CARMENES \citep{Quirrenbach2014} and ESPRESSO \citep{Pepe2021}, have demonstrated that it is possible to detect planets with masses similar to the mass of the Earth using RVs \citep[e.g.,][]{AngladaEscude2016, AstudilloDefru2017,Masca2020,SuarezMascareno2023}. However, this is only possible in the case of low-mass stars. For planets orbiting solar-type and K-type stars, we continue to be restricted to the detection and mass measurement of super-Earths (rocky planets with masses between 1 and 10 M$_{\oplus}$ \citealt{Mayor2011}). 

TESS is a NASA mission that was launched in April 2018 with the primary goal of finding transiting planets around bright, nearby stars \citet{Ricker2015}. TESS observes a large fraction of the sky using four cameras that cover 24x96 degree fields of view. Each field is observed for about 27 days, with a cadence of 2 minutes for selected targets and 200 seconds for full-frame images. TESS image data are reduced and analyzed by the Science Processing Operations Center (SPOC) at NASA Ames Research Center. The SPOC pipeline performs photometry on the images to produce light curves (LC), which then removes instrumental and systematic effects and performs transit searches on them \citep{Jenkins2016}. This pipeline also performs various validation tests to assess the reliability of transit signals and to rule out false positives caused by instrumental or astrophysical effects. The most promising candidates are then labelled as TESS Objects of Interest (TOIs) and are made available to the community for further follow-up observations. 

TOI--238  (TYC 6398--132--1, TIC 9006668) is a bright  \mbox{(V = 10.75 mag)} K-dwarf. It was observed by the TESS satellite in sectors 2, 29 and 69 (August-September of 2018, 2020 and 2023, respectively). It is suspected to host a super-Earth (Radius $\sim$ 1.6 R$_{\oplus}$) in a very short orbit (orbital period $\sim$ 1.27 days). This result was recently validated by \citet{Mistry2023}. 

In this paper we present a combined analysis of the TESS data with RV measurements from the ultra-stable spectrographs ESPRESSO and HARPS that allow us to obtain a high-precision mass determination for the transiting candidate TOI--238.01. Our combined analysis reveals the presence of a transiting sub-Neptune with a period of $\sim$8.46 days.

\section{Observations and data} \label{obs_data}

\begin{figure*}[ht]
    \begin{center}
        \begin{minipage}[b]{0.3\textwidth}
        \includegraphics[width=6cm]{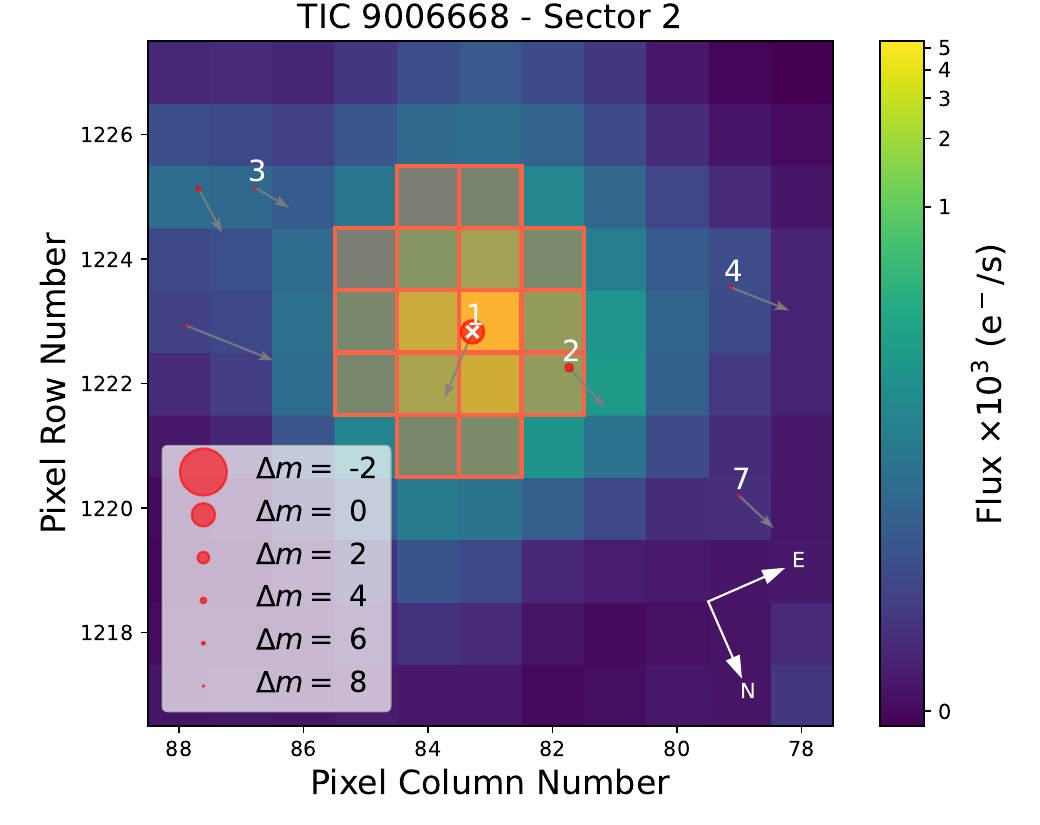}
        \end{minipage}    
        \begin{minipage}[b]{0.3\textwidth}
        \includegraphics[width=6cm]{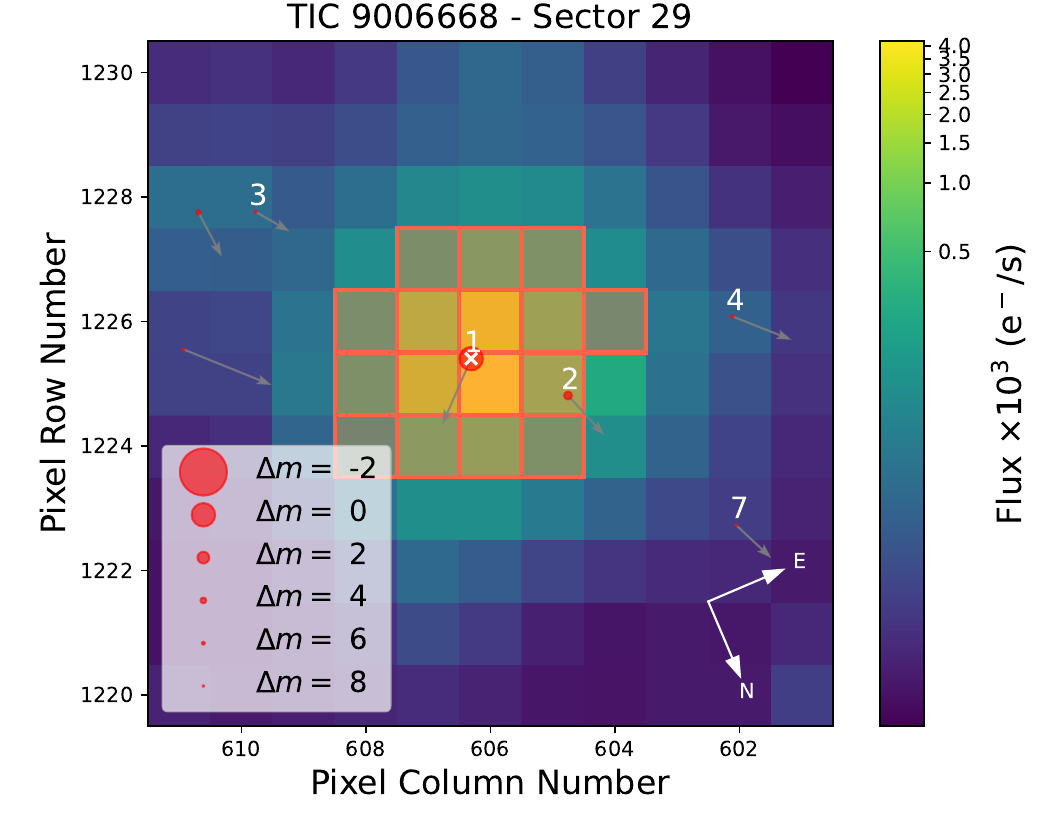}
        \end{minipage}    
        \begin{minipage}[b]{0.3\textwidth}
        \includegraphics[width=6cm]{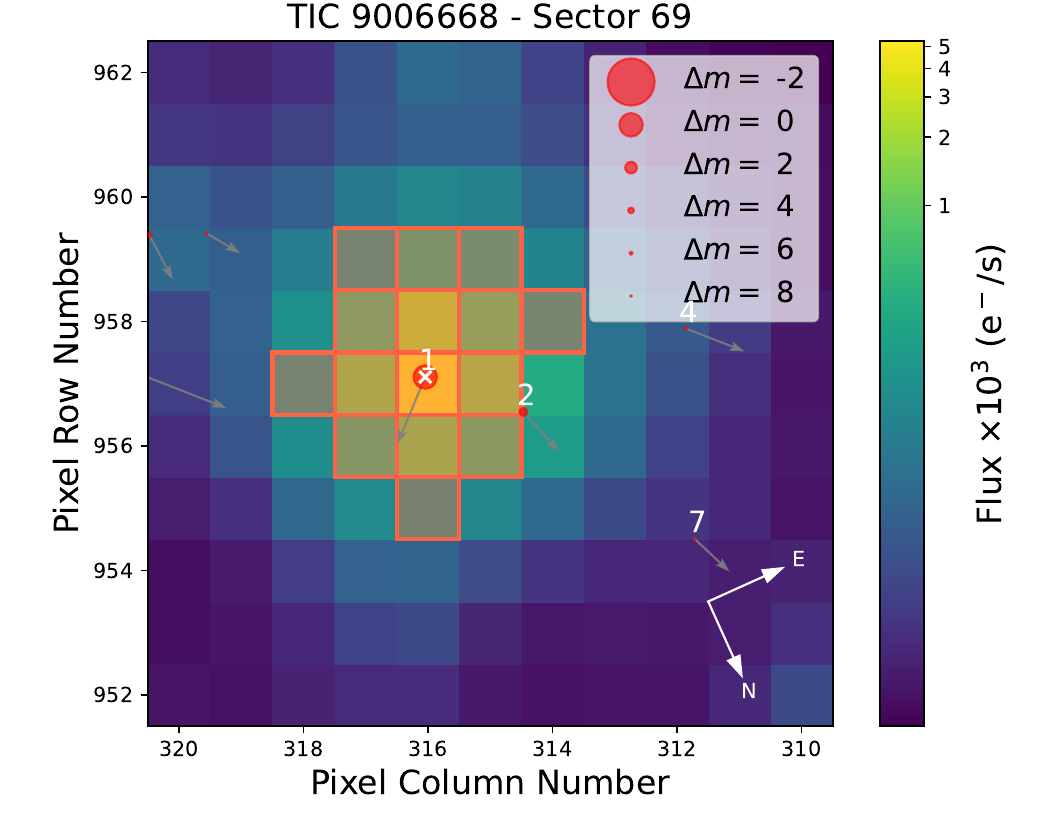}
        \end{minipage}

        \caption{\textbf{TPF files of TOI--238 for sectors 2, 29 and 69 }. TOI--238 is marked with a cross-symbol, and the number 1. The rest of the numbered red symbols show the positions of the closest stars in the field.}
        \label{tess_img}
    \end{center} 
    \end{figure*}

\begin{figure*}[ht]
    \centering
	\includegraphics[width=18cm]{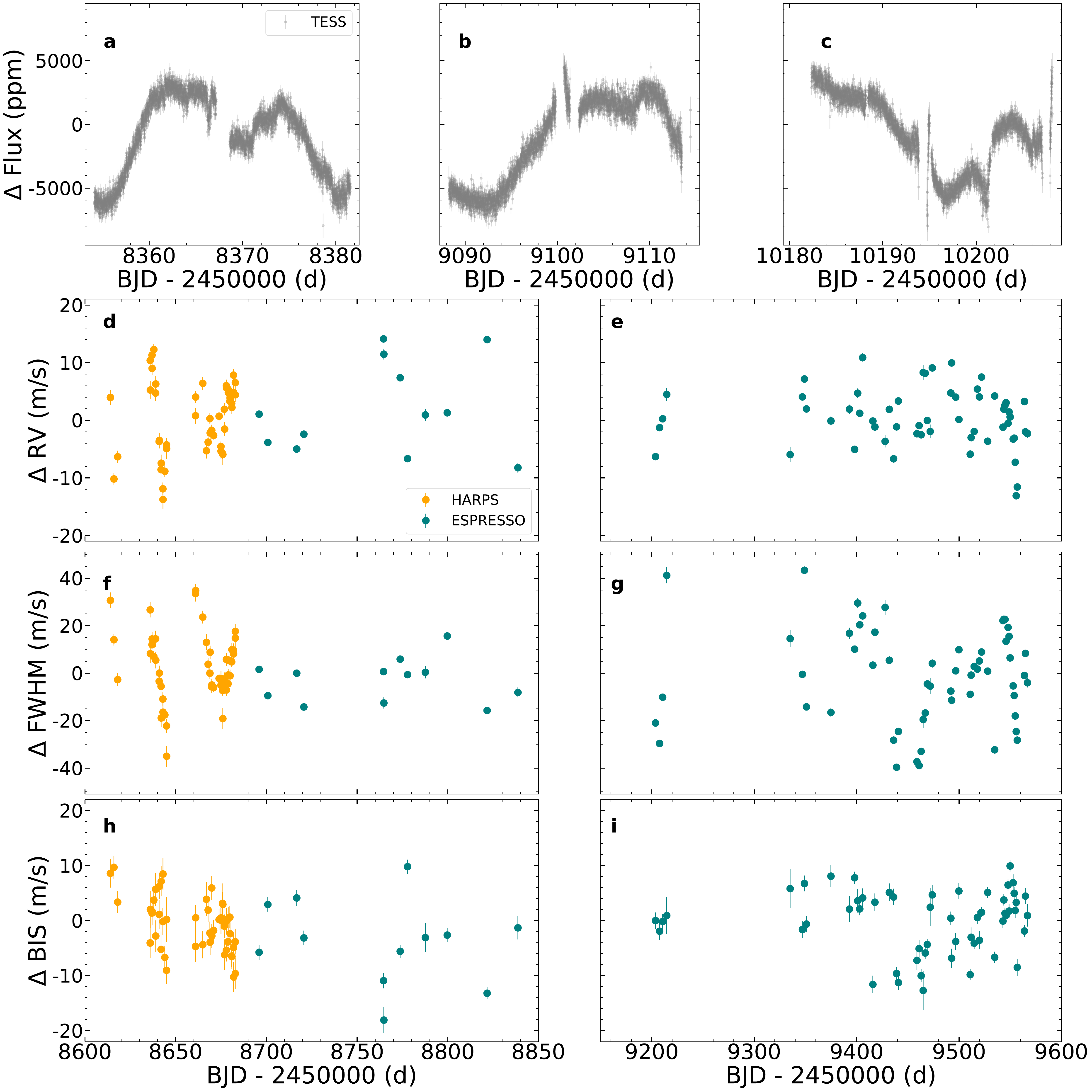}
	\caption{\textbf{Data used in the global analysis.} Panels \textbf{a}, \textbf{b} and \textbf{c} show TESS data of sectors 2, 29, and 69, respectively. Panels \textbf{d} and \textbf{e} show the RV data obtained during the 2019 and 2020-2021 campaigns, respectively. Panels \textbf{f} and \textbf{g} show the FWHM time-series, and panels \textbf{h} and \textbf{i} show the bisector span time-series.}
	\label{data_1}
\end{figure*}

\subsection{TESS observations}

TESS is an all-sky photometric transit survey with the main objective of finding planets smaller than Neptune orbiting bright stars. Such stars are amenable to follow-up observations that aim to determine planetary masses and atmospheric compositions. In its primary mission, TESS has conducted high-precision photometry of more than 200,000 stars over two years of observations until 4 July 2020. TESS is currently in its second Extended Mission. All targets are made available to the community as target pixel files (TPFs) and calibrated LCs. TESS light curve files include the timestamps, simple aperture photometry (SAP) fluxes \citep{Twicken2010, Morris2020}, and pre-search data conditioned simple aperture photometry (PDCSAP) fluxes \citep{Smith2012, Stumpe2012, Stumpe2014}. The SAP flux comes from the sum of the calibrated pixels within the TESS optimal photometric aperture, while the PDCSAP flux corresponds to the SAP flux values corrected for instrumental variations and for crowding. In the process, it usually also removes activity variations with periods similar to the length of a sector. The optimal photometric aperture is defined such that the stellar signal has a high signal-to-noise ratio, with minimal contamination from the background. The TESS detector bandpass spans from $\sim$ 530 to 1060 nm. Figure~\ref{tess_img} shows the TPF files of the field of TOI--238, obtained using the publicly available \texttt{tpfplotter} code \citep{Aller2020}. This code overplots all sources from the \textit{Gaia} Data Release 3 (DR3) catalogue \citep{GaiaEDR3} with a magnitude contrast up to $\Delta m$=8\,mag on top of the TESS TPFs. There is one faint \textit{Gaia} sources at the edge of the photometric aperture around TOI--238 automatically selected by the pipeline ($\Delta m$=4\,mag). Considering the flux-ratio between both, this will have a negligible effect on the planetary radius, assuming the planet transits the brighter star \citep{Ciardi2015}. \citet{Mistry2023}, and our own efforts (see Sect.\ref{subsubsec:lcogt}), showed TOI--238.01 transits the central brightest star within aperture, which leads us to consider the extracted TESS light curve to be free of significant contamination from nearby stars. 

TOI--238 was observed by TESS at the 2-min cadence integrations in sectors 2, 29 and 69. These sectors were processed by the SPOC pipeline \citep{Jenkins2016} and searched for transiting planet signatures with an adaptive, wavelet-based transit detection algorithm \citep{Jenkins2010}. The SPOC pipeline identified a planet candidate at an orbital period of 1.27241 days\footnote{\url{https://exofop.ipac.caltech.edu/tess/target.php?toi=238.01}}, and TOI--238 was announced as a TESS object of interest (TOI) via the dedicated MIT TESS data alerts public website\footnote{\url{https://tess.mit.edu/toi-releases/}}. The SPOC conducted a transit search of Sector 2 on 04 October 2018 with an adaptive, noise-compensating matched filter \citep{Jenkins2002,Jenkins2010,Jenkins2020}, producing a threshold crossing event for which an initial limb-darkened transit model was fitted \citep{Li2019} and a suite of diagnostic tests were conducted to help make or break the planetary nature of the signal \citep{Twicken2018}. The TESS Science Office reviewed the vetting information and issued an alert on 29 November 2018 \citep{Guerrero2021}.The signal was repeatedly recovered as additional observations were made in sector 29. The host star is located within 18.07 $\pm$ 2.69 arcsec of the source of the transit signal.

We used SAP flux data in our study. The TESS light curve of TOI--238 consists of three sectors of 27.4, 26.2 and 25.8 days, respectively, separated by 706 and 1067 days, respectively. The data has a median precision of 430 parts per million (ppm) with a root mean square (RMS) of 3007 ppm, 3295, and 2881 ppm  for sectors 2, 29 and 69, respectively. Figure~\ref{data_1} shows the three TESS light curves. All LCs and TPF files were downloaded from the Mikulski Archive for Space Telescopes \footnote{\url{https://archive.stsci.edu}}, which is a NASA funded project.

\subsection{LCOGT\label{subsubsec:lcogt}}

The TESS pixel scale is $\sim 21\arcsec$ pixel$^{-1}$ and photometric apertures typically extend out to roughly $1\arcmin$, generally causing multiple stars to blend in the TESS photometric aperture. To determine the true source of the TESS detection, we acquired ground-based time-series follow-up photometry of the field around TOI--238 as part of the TESS Follow-up Observing Program \citep[TFOP;][]{collins:2019}\footnote{\url{https://tess.mit.edu/followup}}.

We observed a full transit window of TOI--238.01 continuously for 128 minutes in Pan-STARRS $z$-short band on UTC 2022 November 26 at the Las Cumbres Observatory Global Telescope (LCOGT) \citep{Brown:2013} 1\,m network node of the Teide Observatory, Spain. The $4096\times4096$ LCOGT SINISTRO cameras have an image scale of $0\farcs389$ per pixel, resulting in a $26\arcmin\times26\arcmin$ field of view. The images were calibrated by the standard LCOGT {\tt BANZAI} pipeline \citep{McCully:2018} and differential photometric data were extracted using {\tt AstroImageJ} \citep{Collins:2017}. The TOI--238.01 SPOC pipeline transit depth of 379\,ppm is generally too shallow to reliably detect with ground-based observations, so we instead checked for possible nearby eclipsing binaries (NEBs) that could be contaminating the TESS photometric aperture and causing the TESS detection. To account for possible contamination from the wings of neighboring star point spread functions (PSFs), we searched for NEBs out to $2\farcm5$ from TOI--238. If fully blended in the SPOC aperture, a neighboring star that is fainter than the target star by 8.6 magnitudes in TESS-band could produce the SPOC-reported flux deficit at mid-transit (assuming a 100\% eclipse). To account for possible TESS magnitude uncertainties and possible delta-magnitude differences between TESS-band and Pan-STARRS $z$-short band, we included an extra 0.5 magnitudes fainter (down to TESS-band magnitude 19.0). We calculated the RMS of each of the 13 nearby star light curves (binned in 10 minute bins) that meet our search criteria and found that the values are smaller by at least a factor of 3 compared to the required NEB depth in each respective star. Our analysis ruled out an NEB blend as the cause of the detection of the planet candidate TOI--238.01 by the SPOC pipeline. All light curve data are available on the {\tt EXOFOP-TESS} website\footnote{\href{https://exofop.ipac.caltech.edu/tess/target.php?id=9006668}{\url{https://exofop.ipac.caltech.edu/tess/target.php?id=9006668}}}

\subsection{ESPRESSO observations}

The Échelle SPectrograph for Rocky Exoplanets and Stable Spectroscopic Observations (ESPRESSO) is a fibre-fed high resolution echelle spectrograph installed at the Very Large Telescope (VLT) telescope array in the ESO Paranal Observatory, Chile \citep{Pepe2013, GonzalezHernandez2018, Pepe2021}. ESPRESSO has a resolving power of approximately $R\sim 140\,000$ over a spectral range from $\sim$380 to $\sim$788 nm and has been designed to attain a long-term radial velocity precision of 10 cm$\cdot$s$^{-1}$. It is contained in a vacuum vessel to avoid spectral drifts due to temperature and air pressure variations, thus  ensuring its stability. Observations were carried out using a Fabry-Perot (FP) etalon as simultaneous calibration. The FP offers the possibility of monitoring the instrumental drift with a precision better than 10 cm$\cdot$s$^{-1}$ without the risk of contamination of the stellar spectra by the ThAr saturated lines \citep{Wildi2010}. ESPRESSO can be used on any VLT unit telescope (UT), allowing for an efficient observation and a high-cadence observations. More information can be found on the ESPRESSO user manual.\footnote{\url{https://www.eso.org/sci/facilities/paranal/instruments/espresso/doc.html}}

ESPRESSO is equipped with its own Data Reduction Software (DRS) and provides extracted and wavelength-calibrated spectra, as well as RV measurements. The DRS automatically measures radial-velocity through a Gaussian fit of the cross correlation function (CCF) of the spectrum with a binary mask, computed from a stellar template \citep{Fellgett1955, Baranne1996, PepeMayor2000}. The DRS, version 3.0.0, is available to download from the ESO pipeline website \footnote{\url{http://eso.org/sci/software/pipelines/}}. 

We obtained 77 individual spectra  as part of the ESPRESSO Guaranteed Time Observations (GTO), within programme ID 1102.C-744 (PI: F.Pepe), between May 2019 and December 2021. Measurements were taken in ESPRESSO's 1UT high resolution  mode, with fast readout and 1$\times$1 detector binning (HR11). We used 15 minutes of integration time, for a signal to noise ratio (S/N) of 68 in each slice at 550 nm. More information on the different observing modes can be found on the ESO instrument page\footnote{\url{https://www.eso.org/sci/facilities/paranal/instruments/espresso.html}}. 

In June 2019, ESPRESSO underwent an intervention to update the fibre link, improving the instrument's efficiency by up to 50\% \citep{Pepe2021}. This intervention introduced an RV offset, leading us to consider separate ESPRESSO18 and ESPRESSO19 datasets. We obtained four ESPRESSO spectra before the fibre link update of 2019 and 73 spectra after. Taking into account the expected complexity of the model (including planetary and stellar activity signals), we decided to exclude these four points as they would increase the complexity without adding meaningful new information. Previously, a potential second offset was identified due to a change in calibration lamps in late December 2020. Data reduced with DRS version 2.2.8 showed a $\sim$ 2 m$\cdot$s$^{-1}$ jump in the time-series of standard stars between 19 December and 26 December 2020 due to an imperfect characterisation of the new lamps. This led \citet{Faria2021} and \citet{SuarezMascareno2023} to consider an offset before and after the lamp change. DRS version 3.0.0 removed this offset by updating its characterisation of the lamps and corrected the chromatic drift of the Fabry-Perot etalon identified by \citet{Schmidt2022}. We used data reduced with DRS 3.0.0 and decided not to split the time-series before and after the lamp change. Therefore, we consider a single dataset of 73 ESPRESSO spectra.

\subsection{HARPS observations}

In combination with the ESPRESSO data, we include 50 spectra obtained between May 2019 and July 2019 with the High Accuracy Radial velocity Planet Searcher (HARPS) \citep{Mayor2003} under the programme ID 1102.C-0249 (PI: D. Armstrong). HARPS is a fibre-fed high resolution echelle spectrograph installed at the 3.6 m ESO telescope in La Silla Observatory, Chile. It has a resolving power of $R\sim 115\,000$ over a spectral range from $\sim$380 to $\sim$690 nm and has been designed to attain very high long-term radial-velocity precision. It is contained in temperature- and pressure-controlled vacuum vessels to avoid spectral drifts due to temperature and air pressure variations, thus  ensuring its stability. HARPS is equipped with its own pipeline providing extracted and wavelength-calibrated spectra, as well as RV measurements and other data products such as cross-correlation functions and their bisector profiles. All observations have been carried out with simultaneous calibration, using the FP. TOI--238 was typically observed once per night using an exposure time of 1800 s, achieving a median S/N of 63 at 550 nm. As in the case of ESPRESSO, the DRS provides RV measurements determined by a Gaussian fit of the CCF of the spectrum with a binary mask computed from a stellar template.

\subsection{Radial velocities}

Using the ESPRESSO and HARPS spectra, we extracted RVs using the \texttt{S-BART} algorithm \citep{Silva2022}. \texttt{S-BART} builds a high signal-to-noise template by co-adding all the existing observations. Then, unlike usual template-matching algorithms, \texttt{S-BART} uses a single RV shift to describe simultaneously the RV differences between all orders of a given spectrum and the template. The algorithm estimates the posterior distribution of RV shifts after marginalising with respect to a linear model for the continuum levels of the spectra and template, using a Laplace approximation of the posterior probability distribution, an analytical expression of the posterior probability derived from a Gaussian fitting with a mean equal to the maximum a posteriori probability. From the Laplace approximation to the posterior distribution, the mean is used as the estimated RV and the standard deviation as the estimated RV uncertainty for each epoch. With \texttt{S-BART}, and after a sigma-clip based on the measured uncertainty, we measured an RV RMS of 6.2 m$\cdot$s$^{-1}$ and 5.55 m$\cdot$s$^{-1}$ for HARPS and ESPRESSO data, respectively. We measure median RV internal uncertainties of 1.1 m$\cdot$s$^{-1}$ and 0.58 m$\cdot$s$^{-1}$, respectively. The data is spread along a baseline of 952 days. The velocities obtained with \texttt{S-BART} are consistent with those extracted using other template matching software, such as SERVAL~\citep{Zechmeister2018} or using the CCF technique. For more details on the motivation behind the choice of RV extraction, see Appendix ~\ref{append_rv}. Figure~\ref{data_1} shows the RV time-series. 

We investigated the effect of telluric contamination and the different corrections by comparing RVs coming from uncorrected spectra, RVs from spectra using the automatic correction by \texttt{S-BART}, based on \texttt{TELFIT} \citep{Gullikson2014}, and RVs from spectra corrected following \citet{Allart22}. We opted to use \texttt{S-BART}'s automatic correction, which performed very similar to the correction of \citet{Allart22}, and could be applied to all our data in a homogenous way. Appendix ~\ref{append_tellurics} provides with more details.

\subsection{CCF derived products}

The presence of active regions on the stellar disc affects the flux emitted by the star and its velocity field, distorting the shape of the lines and the PSF measured by the instrument. This effect manifests itself as changes in the width, depth, and symmetry of the CCF. These changes are usually related to the changes in flux, or their gradient, caused by the active regions; and their scale is related to the coverage of active regions, their contrast, and $v \sin i$ of the star. 

To measure these variations, we use the full width at half maximum (FWHM), the bisector span \citep{Queloz2001} and the contrast of the CCF. All those quantities are automatically provided by the ESPRESSO and HARPS DRS. Over our complete time-series, the CCF FWHM shows a dispersion of 17.0 m$\cdot$s$^{-1}$, with a median uncertainty of 1.9 m$\cdot$s$^{-1}$. The dispersion of the bisector span time-series is 6.2 m$\cdot$s$^{-1}$ with a median uncertainty of 1.9 m$\cdot$s$^{-1}$. The contrast of the CCF has a dispersion of 0.249 with a median uncertainty of 0.011. Figure~\ref{data_1} shows the FWHM and bisector span time-series. Figure~\ref{cont_model} shows the CCF contrast time-series.

\subsection{Telemetry data}

Modern RV spectrographs are designed to minimise instrumental effects caused by the changes in their environments. However, small effects either linked to the stability of both instruments or to the extraction of the velocities can be present in the data. This potentially biases some of the results \citep{SuarezMascareno2023}. We collected the time-series of the temperature of the \'Echelle gratings of ESPRESSO and HARPS, and the Barycentric Earth Radial Velocity (BERV) as proxies for variations potentially induced by instrumental changes or imperfect data extraction. We measure an RMS of the variations of the  temperature of the \'Echelle gratings of ESPRESSO and HARPS of 11 mK and 5 mK, respectively. See Appendix~\ref{append_tel} for more details. 

\subsection{Spectral activity indicators}

In addition to the CCF indicators, we studied the time-series of several standard chromospheric indicators. 

\subsubsection{Ca II H\&K}

The intensity of the emission of the cores of the Ca II H\&K lines is linked to the strength of the magnetic field of the star, which in turn is well correlated with the stellar rotation period of FGKM stars. The measured emission intensity of the line cores also changes when active regions move across the stellar disc, helping us trace the rotation of the star. We calculate the Mount Wilson $S$-index for the ESPRESSO data following \citet{Vaughan1978}. We define two triangular-shaped passbands with a FWHM of 1.09~{\AA}  centred at 3968.470~{\AA} and at 3933.664~{\AA} for the Ca II H\&K line cores. For the continuum we use two bands 20~{\AA} in width centred at 3901.070~{\AA} (V) and 4001.070~{\AA} (R).
Then the S-index is defined as: 

\begin{equation}
   S=\alpha {{\tilde{N}_{H}+\tilde{N}_{K}}\over{\tilde{N}_{R}+\tilde{N}_{V}}} + \beta,
\end{equation}
\noindent where $\tilde{N}_{H},\tilde{N}_{K},\tilde{N}_{R}$, and $\tilde{N}_{V}$ are the mean fluxes per wavelength unit in each passband,  while $\alpha$ and $\beta$ are calibration constants fixed at $\alpha = 1.111$ and $\beta = 0.0153$ . The S index (S$_{MW}$) serves as a measurement of the Ca II H\&K core flux normalised to the neighbour continuum. Figure~\ref{smw_model} shows the S-index time-series.

We measure a median S-index of 0.320 $\pm$ 0.064, which using the calibration of \citet{Masca2015} translates to a log$_{10}$ (R$_{\rm HK}^{'})$ of -- 4.74 $\pm$ 0.27. Following the activity-rotation calibration for K-dwarfs of \citet{Masca2016}, we estimate a rotation period of 28 $\pm$ 16 days, with the uncertainties being a result of the large uncertainties in the photometric magnitudes shown in Table~\ref{tab:parameters}.

\subsubsection{H$\alpha$}

Similar to Ca II H\&K, the emission in the core of the H$\alpha$ line (or filling, in the case of low activity stars) is related to the strength of the magnetic field, and the presence of active regions on the stellar disc. We can also use it to track the motion of said regions across the stellar disc, and thus the stellar rotation. We compute the H$\alpha$  values following the definition of~\citet{GomesdaSilva2011}. Figure~\ref{ha_model} shows the H$\alpha$ index time-series. We measure a median H$\alpha$ index value of 0.02845 $\pm$ 0.00012.

\subsubsection{Na I}

The sodium D lines have been shown to be good chromospheric indicators for low-mass stars. We compute the time-series of the Na I D fluxes following the definition of ~\citet{Diaz2007}. Figure~\ref{nai_model} shows the Na I index time-series. We measure a median Na I index value of 0.00534 $\pm$ 0.00045.

\section{Stellar parameters of TOI--238}

We obtained stellar atmospheric parameters from the high-resolution ESPRESSO spectra via the ARES+MOOG method \citep{Sousa2014}. The ARES+MOOG method is a curve-of-growth technique based on neutral and ionised iron lines. Equivalent widths of these spectral lines were measured automatically from the stacked ESPRESSO spectrum using ARESv2 \citep{Sousa2015}. We determine effective temperature ($T_\text{eff}$), surface gravity (log $g$), iron abundance ([Fe/H]), and microturbulent velocity by imposing excitation and ionisation equilibria. For this purpose, we used the radiative transfer code MOOG \citep{Sneden1973}, that assumes local thermodynamic equilibrium (LTE) employing a grid of ATLAS plane-parallel model atmospheres \citep{Kurucz1993}. Subsequently, we corrected the surface gravity for accuracy \citep{Mortier2014} and added systematic errors in quadrature to our precision errors for the effective temperature, surface gravity, and iron abundance \citep{Sousa2011}. To compute the stellar mass and radius, we used PARAM \citep{daSilva2006, Rodrigues2017}. This code matches the stellar parameters $T_\text{eff}$, log $g$, and [Fe/H] obtained in the previous analysis, along with the Gaia DR3 parallax and the V magnitude published in the literature, to a grid of stellar evolutionary tracks and isochrones from PARSEC \citep{Bressan2012}. The iron-to-silicate mass fraction ($f^{star}_{iron}$) of the planetbuilding blocks, as estimated from their host star abundances, can be used as a proxy for the composition of small planets in the planetary system \citep{Adibekyan-21}. We derived the abundances of Mg and Si using the same tools and models as for stellar parameter determination as well as using the classical curve-of-growth analysis method assuming local thermodynamic equilibrium. Although the EWs of the spectral lines were automatically measured with ARES, for Mg which has only three lines available we performed careful visual inspection of the EWs measurements. For the derivation of abundances we closely followed the methods described in \citet{Adibekyan-12,Adibekyan-15}. Table~\ref{tab:parameters} shows the full list of parameters. 

We crosschecked our results using the STEPAR code \citep{Tabernero2019}, and by computing the parameters from the spectral energy distribution (SED) using the available photometry. STEPAR provided us with consistent stellar parameters: T$_\text{eff}$ = 5054 $\pm$ 78 K, log g = 4.52 $\pm$ 0.20, $[Fe/H]$ = -- 0.06 $\pm$ 0.04, Age = 4.0 $\pm$ 3.8 Gyr, M$_{*}$ = 0.800 $\pm$ 0.020 M$_{\odot}$ and R$_{*}$ = 0.741 $\pm$ 0.017 R$_{\odot}$. Using the SED, combined with the parallax from Gaia DR3 \citep{GaiaEDR3}, we computed a luminosity L$_{*}$/L$_{\odot}$ = 0.3297 $\pm$ 0.0058. Then, through the Stefan-Boltzmann law, we estimate a R$_{*}$ = 0.747 $\pm$ 0.034 R$_{\odot}$ and M$_{*}$ = 0.785 $\pm$ 0.048 M$_{\odot}$. Both sets of parameters are consistent with the stellar parameters presented in Table~\ref{tab:parameters}. It is important to note that the uncertainties of derived stellar masses and radii are typically smaller than the uncertainties of direct measurements. As these parameters can affect the computation of planetary parameters, we enlarged the error bars of the stellar parameters in all our calculations following the reccomendations of \citet{Tayar2022}.

Using Gaia DR3 positions, proper movements, RV and parallax, we computed the galactic velocities of TOI--238 following \citet{JohnsonSoderblom1987}. We obtain galactic velocities U = 14.13 $\pm$ 0.09 km$\cdot$s$^{-1}$, V = 2.83 $\pm$ 0.10 km$\cdot$s$^{-1}$, W = 19.50 $\pm$ 0.26 km$\cdot$s$^{-1}$. These values are fully compatible with the thin disk of the galaxy and do not match any known young moving group \citep{Gagne2018}. This result indicates the star to be older than 0.8 Gyr, which is consistent with the ages derived from the spectroscopic modelling.

\begin{table}
\begin{center}
\caption{Stellar properties of TOI--238 \label{tab:parameters}}
\begin{tabular}[center]{l l l}
\hline
Parameter & TOI--238 & Ref. \\ \hline
RA [J2000] & 23:16:55.5175 & 1 \\
DEC [J2000] & --18:36:23.9218 & 1\\
$\mu \alpha \cdot cos(\delta)$ [$mas$ yr$^{-1}$]& --53.147 & 1 \\
$\mu \delta$ [$mas$ yr$^{-1}$]& --0.315 & 1 \\
Parallax [$mas$] &  	12.475 $\pm$ 0.015 & 1\\
Distance [pc] & 80.162 $\pm$ 0.096 & 1\\
U [km$\cdot$s$^{-1}$] & 14.13 $\pm$ 0.09 & 0\\
V [km$\cdot$s$^{-1}$] & 2.83 $\pm$ 0.10 & 0\\
W [km$\cdot$s$^{-1}$] & 19.50 $\pm$ 0.26 & 0\\
$m_{B}$	 [mag] & 11.67 $\pm$ 0.13  & 2 \\
$m_{V}$	 [mag] & 10.75 $\pm$ 0.09 & 2 \\
Spectral Type  & K2V & 0\\
L$_{*}$/L$_{\odot}$ & 0.3382 $\pm$ 0.0097 & 3 \\
T$_{eff}$ [K] & 5059 $\pm$ 89 & 0 \\
$[\rm Fe/H]$ [dex] & --0.114 $\pm$ 0.051 & 0 \\
$[\rm Mg/H]$ [dex] & --0.12  $\pm$ 0.06  & 0\\
$[\rm Si/H]$ [dex] & --0.09  $\pm$ 0.06  & 0\\
$f^{star}_{iron}$ (\%) & 32.7 $\pm$ 1.8 & 0\\
M$_{*}$ [M$_{\odot}$] & 0.790  $\pm$ 0.022 & 0 \\
R$_{*}$ [R$_{\odot}$] & 0.733  $\pm$ 0.015 & 0 \\
log g [cgs] & 4.577 $\pm$ 0.021 & 0\\
Age [Gyr] & 4.3 $\pm$ 3.8 & 0\\
v sin i [km$\cdot$s$^{-1}$] & 1.84 $\pm$ 0.45 & 0 \\
log$_{10}$ (R$_{\rm HK}^{'})$ & -- 4.74 $\pm$ 0.27 & 0 \\
P$_{\rm rot ~GP}$ [days] & 26.0 $\pm$ 1.3 & 0 \\
\hline
\end{tabular}
\end{center}
\textbf{References:} 0 - This work, 1 -  \citet{GaiaEDR3}, 2 - \citet{Hog2000}, 3 - TESS Input Catalog Stellar Parameters
\end{table}

\section{Analysis}

\subsection{Telemetry and stellar activity} \label{sect_tel_stel}

We started by analysing the possible effect of the changes in environmental conditions and imperfect data extraction in the science data. We studied the correlations between the BERV and temperature of the \'Echelle gratings of ESPRESSO and HARPS, and the different science data. Similar to \citet{SuarezMascareno2023}, we did not detect any significant correlation between any of them and the radial velocities. In some of the CCF and spectroscopic indicators, we found correlation between the temperature of the optical elements and the BERV. We interpret this as an effect of a subtle change in focus (correlation with \'Ech. Temp.) and leftover telluric contamination (correlation with BERV). Appendices~\ref{append_tellurics} and ~\ref{append_tel} provide a more detailed analysis. 

We then analysed all stellar indicators described in Sect.~\ref{obs_data} to search for periodicities related to stellar activity and that could create false-positive detections in RV or bias the amplitude measurements in RV. We modelled each time-series independently, using Gaussian Processes regression (GP; see \citealt{Rasmussen2006} and \citealt{Roberts2012}). These results are later used to decide the characteristics of the global model. 

The GP framework has become one of the most successful methods in the analysis of stellar activity in RV time-series (e.g. \citealt{Haywood2014}). The stellar noise is described by a covariance with a prescribed functional form and the parameters attempt to describe the physical phenomena to be modelled. The GP framework can be used to characterise the activity signal without requiring a detailed knowledge of the distribution of active regions on the stellar surface, their lifetime, or their temperature contrast. One of the biggest advantage of GPs is that they are flexible enough to effortlessly model quasi-periodic signals, and account for changes in the amplitude, phase, or even small changes in the period of the signal. This flexibility is also one of their biggest drawbacks, as they can easily overfit the data, suppressing potential planetary signals. 

We used the newly presented \texttt{S+LEAF} code \citep{Delisle2022} \footnote{\url{https://gitlab.unige.ch/delisle/spleaf}}, which extends the formalism of semi-separable matrices introduced with \texttt{celerite} \citep{Foreman-Mackey2017} to allow for fast evaluation of GP models even in the case of large datasets. The \texttt{S+LEAF} code supports a wide variety of GP kernels with fairly different properties. We opted for the \textit{ESP} Kernel, which is a very close approximation of the widely used \textit{Quasi-Periodic} Kernel:

\begin{equation} \label{eq_qp}
\fontsize{8}{11}\selectfont
 k(\tau) = {A^{2}} \cdot exp \Bigg[ - {\tau^2 \over{2L^{2}}} - {{sin^2(\pi \tau / P_{rot})} \over{2 \omega^2} }\Bigg] 
 + (\sigma^2 (t) + \sigma^2_{j}) \cdot \delta_{\tau},
\end{equation}

\noindent where $A$ is the variance of the data, $P_{rot}$ is the periodicity of the data, $L$ is the evolutionary timescale and $\omega$ is the length scale of the periodic component. The covariance matrix also includes a term of uncorrelated noise ($\sigma$), independent for every instrument. This term is added in quadrature to the diagonal of the covariance matrix to account for all unmodelled noise components, such as uncorrected activity or instrumental instabilities. The \textit{ESP} Kernel in the \texttt{S+LEAF} package allows for a explicit configuration of the number of harmonics of the main variability to include. We used the default configuration, which includes the first two harmonics.

Considering the correlations with the telemetry mentioned before, we included low-order polynomials with respect to the temperature of the \'Echelle grating and with respect to the BERV, when we deemed appropriate. We found that none of them improved on the GP-only models. All the polynomial parameters where consistent with zero within 1$\sigma$. Contrary to what was seen in \citet{SuarezMascareno2023}, it seems that any small variation induced by instrumental changes is too subtle to be confidently separated from the activity model. TOI--238 is more active than GJ 1002, which means the ratio between activity and instrumental variations will be much larger. The analysis of some indicators also included a low-order polynomial or a sinusoidal to account for long-term variations. None of these models improved on the GP-only model. 

We opted to restrict the analysis to a GP-only model, with floating offsets and jitter terms.  We modelled the data using Bayesian inference through the nested sampling \citep{Skilling2004, Skilling2006} code \texttt{Dynesty}~\citep{Speagle2020, kosopov2023}. All models had seven free parameters ($N_{free}$). We went for a number of live points of $10 \times N_{free}$. We used the default bounding option, multi-ellipsoidal decomposition~\citep{Feroz2009}, and chose to sample our parameter space using random walk~\citep{Skilling2006}, which is very efficient at exploring low-to-mid dimension parameter spaces. 

Additionally, for every indicator we study the relationship between the indicator and the derivative of its best fit GP model, with the RV measurements. We measured the slope between both quantities by fitting a linear function, using the measurement uncertainties as weights. We then used the diagonal of the covariance matrix to evaluate the significance of the slope. We consider that there might be a relationship when the slope is $\ge$2$\sigma$ different from zero. This information can later be leveraged when building a multi-dimensional GP model.

For all models, we use log-normal priors for the amplitude of the GP, with a mean corresponding to ln(RMS) of the data, and a standard deviation of ln(RMS) of the data. We use the same priors for the jitter values. We explore rotation periods between 5--50 days, and evolutionary timescales of 10--200 days, both in log-uniform scale. Main-sequence K-dwarfs are not expected to have rotation periods longer than 50 days, and the timescale of coherence of the rotation signal is typically of 2-3 rotations \citep{Giles2017}. For the zero-points of the time-series we use a normal prior with a mean of zero and a standard deviation of the RMS of the data.

We obtained good fits for the FWHM and BIS of the CCF, with significant correlations between the activity proxies and their derivatives, and the RV data. We find that correlations between FWHM and BIS, and the RV, show opposite slopes. We also obtained a good fit for the H$\alpha$ time-series, very similar to that of the FWHM and showing a very similar correlation with the RV as the FWHM. We opted for using the FWHM and the BIS of the CCF in our global model, as both provided different information about the stellar activity variations. 

\begin{figure*}[ht]
    \centering
	\includegraphics[width=18cm]{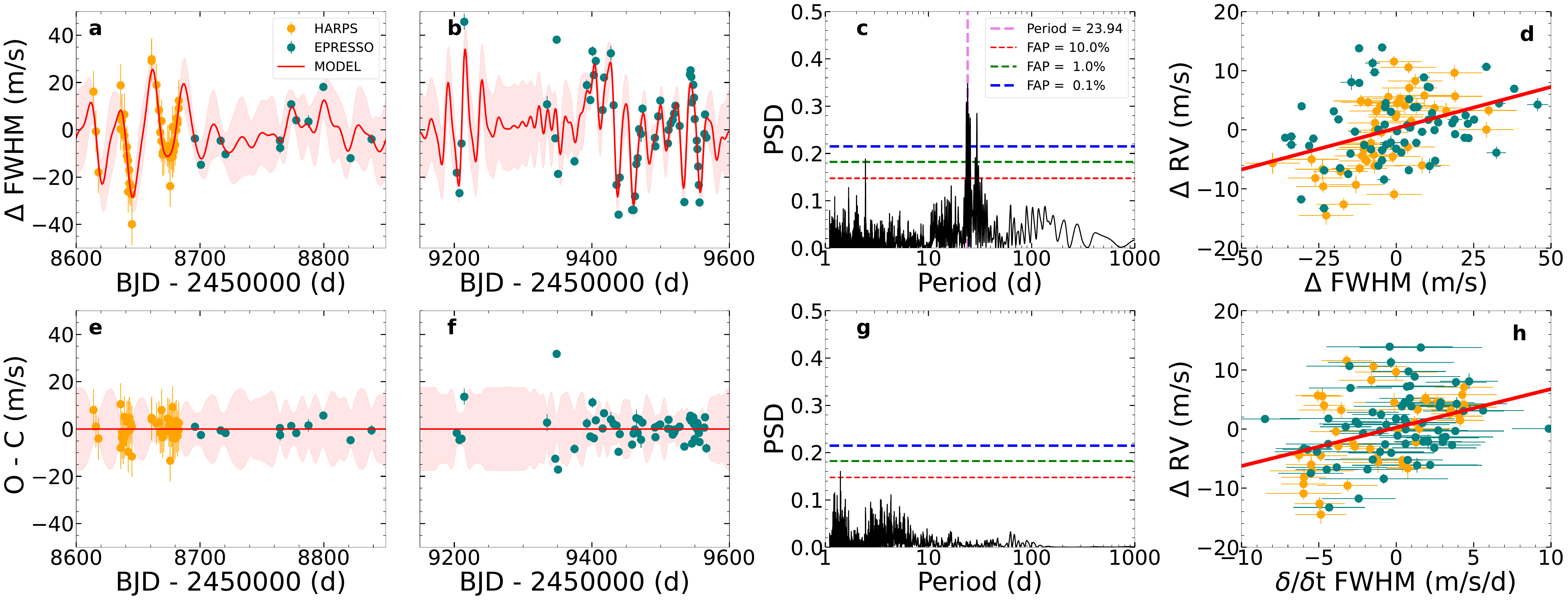}
	\caption{\textbf{Analysis of the FWHM of the CCF.} \textbf{a, b}: FWHM time-series with the best-model fit. The data is split into two panels because of a large period with no observations between the two campaigns. The shaded area shows the variance of the GP model. \textbf{c}: GLS periodogram of the FWHM data. The grey vertical line highlights the most significant period. \textbf{d}: Relationship between the RV and FWHM data. The best fit is shown when the slope is $\ge$2$\sigma$ different from zero. \textbf{e, f}: Residuals of the FWHM after subtracting the best model fit. \textbf{g}: GLS periodogram of the residuals. \textbf{h}: Comparison of the RV and gradient of the FWHM model.}
	\label{fwhm_model}
\end{figure*}

Figure~\ref{fwhm_model} shows the data, model, periodogram and correlation plots of the FWHM of the CCF. The time-series of the FWHM of the CCF show an RMS of 17.0 m$\cdot$s$^{-1}$, with the data having a median uncertainty of 2.1 m$\cdot$s$^{-1}$. For this time-series, we obtain a period of 24.3$^{+1.4}_{-1.1}$ days, with a damping timescale of 22.7$^{+5.7}_{-4.7}$ days. The period measured with the GP is very close to the period detected by the GLS periodogram~\citep{Zechmeister2009}. We obtain a good fit to the data, with reasonably clean variations during all observing campaigns. After subtracting the model, we measure an RMS of the residuals of 6.7 m$\cdot$s$^{-1}$, and find no significant signals in their periodogram. We report moderate positive relationships between the FWHM and the RV measurements, as well as the gradient of the best-fit model and the RV measurements. 

\begin{figure*}[h!]
    \centering
	\includegraphics[width=18cm]{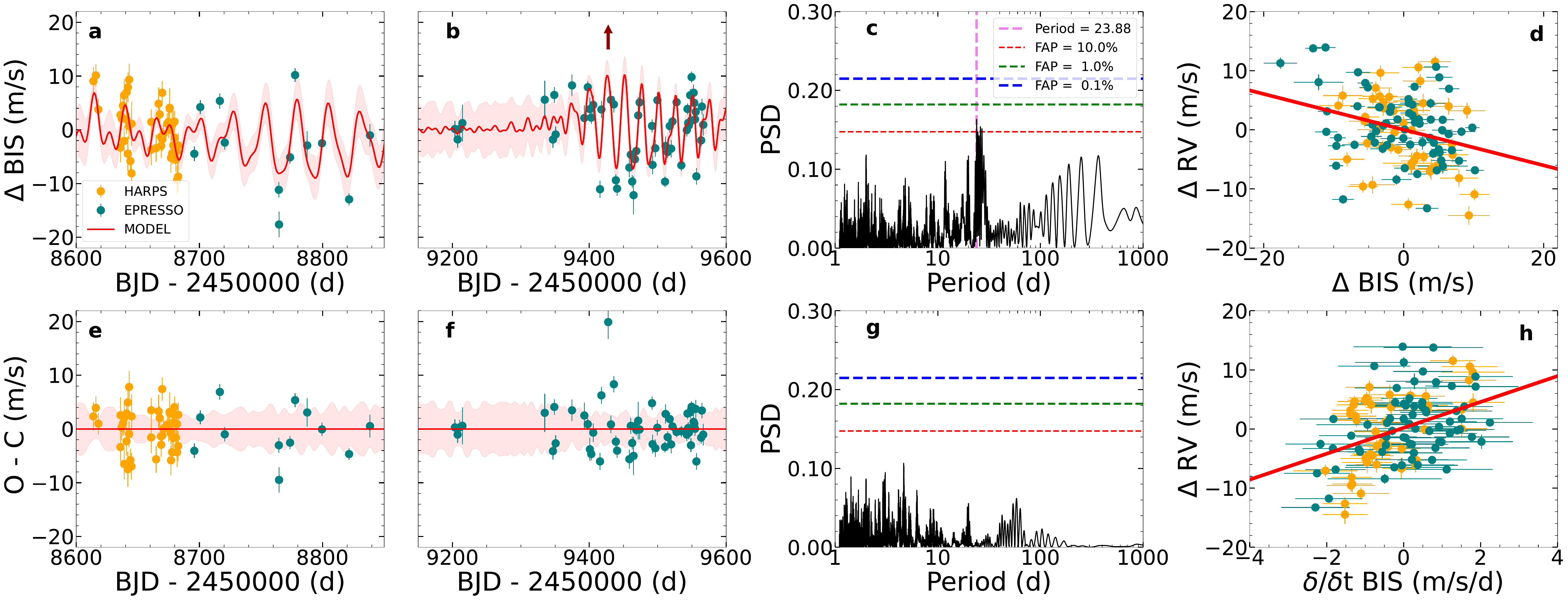}
	\caption{\textbf{Analysis of the bisector span of the CCF.} Same as Figure~\ref{fwhm_model}, only with the bisector span of the CCF instead.}
	\label{bis_model}
\end{figure*}

Figure~\ref{bis_model} shows the data, model, periodogram and correlation plots of the bisector span of the CCF \citep{Queloz2001}. We measure an RMS of the data of  6.4 m$\cdot$s$^{-1}$, and a median uncertainty of 1.9 m$\cdot$s$^{-1}$. We obtain a period of 25.28$^{+0.82}_{-0.65}$ days, with a damping timescale of 52$^{+24}_{-18}$ days. The period measured with the GP is close to the period detected by the GLS periodogram and to the period measured in the FWHM time-series. The GLS of the data shows a structure with a periodicity of around 1 year. However, the inclusion of additional parameters to account for it is disfavoured in the fitting.  After subtracting the model, we measure an RMS of the residuals of 3.7 m$\cdot$s$^{-1}$, and find no significant signals in their periodogram. We find moderate anti-correlations between the bisector span and the RV measurements, and a moderate positive correlation between the gradient of the best-fit model and the RV measurements. This indicates that the RV data is significantly affected by stellar activity variations. 

The models for the additional activity proxies can be found in Appendix~\ref{append_act}. 

\subsection{Global model - Activity only} \label{sect_gp}

We built the global model combining TESS photometry and RV, and FWHM and BIS as activity proxies. We modelled the stellar variations using a GP with \texttt{S+LEAF}, again using the ESP Kernel. The \texttt{S+LEAF} code allows to fit simultaneously a GP to several time-series, based on a linear combination of the GP and its derivative, with different amplitudes for each time-series.

The usual assumption is that there exists an underlying function governing the behaviour of the stellar activity, which we denote $G(t)$. $G(t)$ manifests in each time series as a linear combination of itself and its gradient, $G'(t)$, with a set of amplitudes for each time series, following the idea of the $FF'$ formalism \citep{Aigrain2012, Rajpaul2015}, as described in equation~\ref{eq_gp_grad}.

\begin{equation} \label{eq_gp_grad}
\begin{split}
&\Delta ~TS_{1} = A_{1} \cdot G(\tau) + B_{1} \cdot G'(\tau)~, \\
&\Delta ~TS_{2} = A_{2} \cdot G(\tau) + B_{2} \cdot G'(\tau)~, \\
&...
\end{split}
\end{equation}

\noindent with $\Delta~TS_{i}$ representing the variations of each time-series, and $A_{i}$ and $B_{i}$ the scaling coefficients of the underlying funciton $G(t)$ and its gradient, $G'(t)$, respectively. 

We modelled the data using Bayesian inference via nested sampling with \texttt{Dynesty}, with the same configuration described before, but with one exception. Now, we sampled the parameter space using random slice sampling, which is well suited for the high-dimensional spaces \citep{Handley2015a,Handley2015b} resultant of modelling several time-series at once. We used a number of slices equal to  2 $\times$ $N_{free}$. 

We fitted the RV combined with the FWHM and/or the BIS following this idea, with moderate success. Depending on the specific model, we derived different timescales, and the residuals after the fit contained periodic signals that we could still attribute to stellar activity. For example, the combined fit of the RV and FWHM left some periodic signals present in the BIS data and vice versa. The combined model of the three time-series resulted in moderate under-fit that left a few periodic signals in the RV which also appeared in the FWHM or BIS data. 

Following the conclusions of \citet{Tran2023}, we attempted a model with two latent GPs. \citet{Tran2023} demonstrated that in some circumstances a single multi-dimensional GP is not capable of following all activity timescales simultaneously. As the variations in FWHM and BIS appear to show different timescales (see Sect~\ref{sect_tel_stel}), and their effect in the RV seems to be different, we built a model that linked the FWHM to the RV, the BIS to the RV, but not the BIS to the FWHM. 

We included the TESS data in the global model with its own independent activity model, that is not linked to the RV or to any of the activity proxies. TESS SAP photometry is often contaminated by instrumental trends, and the rotation period of this star is too long to be preserved in the PDCSAP photometry. The variations seen in the SAP photometry are likely a combination of stellar and instrumental variability. We modelled TESS variability with a GP with a simple harmonic oscillator (SHO) to take advantage of its scalability to datasets with thousands of points. We resampled the TESS data into 10 minutes bins to avoid an excessive increase of computing time. The cost of the evaluation of the GP model scales linearly with the number of points. Modelling the data in 2-minute bins takes five times longer than in 10-minute bins, while it does not offer any advantage given the expected characteristics of the transit. 

Following \citet{Foreman-Mackey2017}, the $k_{SHO}$ kernel is defined as

\begin{equation} \label{eq_sho}
\fontsize{8}{11}\selectfont
 k_{i}(\tau) = {C_{i}^{2}} e^{-\tau/L}  \left\{\begin{array}{cc}\cosh(\eta {2 \pi \tau}/P_{i})+{{P_{i}}\over{2 \pi \eta L }}\sinh(\eta {2 \pi \tau}/P_{i}) ; ~\rm if ~P_{i} > 2 \pi L\\2 (1 + {{2 \pi \tau}\over{P_{i}}}) ; ~\rm if ~P_{i} = 2\pi L\\ \cos(\eta {2 \pi \tau}/P_{i}) + {{P_{i}}\over{2 \pi \eta L}} \sin(\eta {2 \pi \tau}/P_{i}); ~\rm if ~P_{i} < 2 \pi L\end{array}\right\}~,
\end{equation}

\noindent where $\eta = (1 - (2L/P_{i})^{-2})^{1/2}$ controls the damping of the oscillator.

This Kernel has a power spectrum density:  
\begin{equation} \label{psd_kernel}
S(\omega) = \sqrt {{2} \over {\pi}} {{S_{i} ~\omega_{i}^{4}} \over {(\omega^{2} - \omega_{i}^{2})^2 + \omega_{i}^{2}~\omega^{2} / Q^{2}}}~,
\end{equation}

\noindent where $\omega$ is the angular frequency, $\omega_{i}$ is the undamped angular frequency for each component ($\omega_{i}$ = 2 $\pi$ / $P_{i}$), $S_{i}$ is the power at $\omega$ = $\omega_{i}$ for each component, and $Q_{i}$ is the quality factor. $S_{i}$, $P_{i}$ and $Q_{i}$ are the parameters sampled in the covariance matrix, which are related to the amplitude ($C_{i}$), rotation period ($P_{rot}$) and timescale of evolution ($L$) in the following way: 

\begin{equation} \label{eq_params}
\begin{split}
&P_{1} = P_{rot} ~,~ S_{1} = {{C_{1}} \over {2 \cdot L}} \left( {{P_{1}} \over {\pi}} \right)^{2} ~,~  Q_{1} = {\pi {L}\over{P_{1}}}~
\end{split}
\end{equation}

Our global activity model, including four different time-series and three GP kernels, is then defined as follows: 

\begin{equation} \label{full_model}
\begin{split}
\Delta ~\rm Flux & =  G_{\rm ~TESS}~(A_{\rm ~TESS},P_{\rm ~TESS},L_{\rm ~TESS},\tau),  \\
\Delta ~\rm FWHM & = A_{\rm ~FWHM} \cdot G_{\rm ~FWHM}~(1, P_{\rm ~FWHM}, L_{\rm ~FWHM}, \omega_{\rm ~FWHM}, \tau),  \\
\Delta ~\rm BIS  & = A_{\rm ~BIS}\cdot G_{\rm ~BIS}~ (1, P_{\rm ~BIS},  L_{\rm ~BIS},  \omega_{\rm ~BIS},  \tau),  \\
\Delta ~\rm RV   & = A11_{\rm~RV} \cdot G_{\rm ~FWHM}(\tau) + A12_{\rm~RV} \cdot G'_{\rm ~FWHM}(\tau)  \\
& + A21_{\rm~RV} \cdot G_{\rm ~BIS}(t) + A22_{\rm~RV} \cdot G'_{\rm ~BIS}(\tau),
\end{split}
\end{equation}

\noindent with G$_{\rm ~TESS}$ being a SHO Kernel (following eq.~\ref{eq_sho}), and G$_{\rm ~FWHM}$ and G$_{\rm ~BIS}$ being ESP Kernels (following eq.~\ref{eq_qp}), with variance equal to unity. The model also includes a floating zero-point for every time-series and instrument, and jitter terms for the RV, FWHM and BIS data. In the case of TESS data, we fix the jitter by estimating the RMS of the data in the flat region between the start of the observations and BJD 2458356 (see panel a of Fig.~\ref{data_1}).

For the model of the TESS data we use the following parameters and priors: $A_{\rm ~TESS}$ is the amplitude of the TESS variability and uses a log-uniform ($\mathcal{LU}$) prior with a range of [0 , 20]. $P_{\rm ~TESS}$ is the period of the variability of the TESS data and uses a normal ($\mathcal{N}$) prior around 28 days (highest peak in the periodogram) with a standard deviation of 10\%. $L_{\rm ~TESS}$ is the evolutionary timescale of the TESS variability. Based on \citet{Giles2017} we use a $\mathcal{N}$ prior 4.0 $\pm$ 0.3 (54$^{+20}_{-14}$ days). The zero-point of the TESS data uses a $\mathcal{N}$ prior centred at 0, with a standard deviation equal to the RMS of the data.

The combined FWHM+BIS+RV model uses the following parameters and priors: $A_{\rm ~FWHM}$ and $A_{\rm ~BIS}$ are the amplitudes of the variability in the FWHM and BIS time-series and use uniform ($\mathcal{U}$) priors between 0 and 5$\times$ the RMS of their respective time-series. $P_{\rm ~FWHM}$ and $P_{\rm ~BIS}$ are the periods of variability of the FWHM and BIS, and use $\mathcal{N}$ priors of 25 $\pm$ 2.5 days ($\sim$ the weighted mean of all periods measured in the activity indicators). $L_{\rm ~FWHM}$ and $L_{\rm ~BIS}$ are the evolutionary timescales of the FWHM and BIS variability and, once again based on \citet{Giles2017}, use $\mathcal{N}$ priors 4.0 $\pm$ 0.3 (54$^{+20}_{-14}$ days, consistent although slightly longer than the values measured in the activity indicators). These priors for the $L$ parameters are 1$\sigma$ compatibles with the values obtained for the $L$ parameters in the independent FWHM and BIS models. $\omega_{\rm ~FWHM}$ and $\omega_{\rm ~FWHM}$ are the length scale of the periodic components of the variability, and use $\mathcal{LU}$ priors with a range of [-5 , 5]. $A11_{\rm~RV}$ and $A21_{\rm~RV}$ are the amplitudes of the variability in RV related to the FWHM and BIS variations, respectively, and use $\mathcal{U}$ priors [-5$\times$ RMS , 5$\times$ RMS].  $A12_{\rm~RV}$ and $A22_{\rm~RV}$ are the amplitudes of the variability in RV related to the gradient of the FWHM and BIS variations, respectively, and use $\mathcal{U}$ priors [-10$\times$ RMS , 10$\times$ RMS]. The priors in the RV amplitude allow for negative values to take into account possible negative correlations between RV and FWHM. The zero points of all datasets use $\mathcal{N}$ priors centred at 0, with sigmas equal to the RMS of their respective dataset. The white noise component of all datasets uses $\mathcal{LN}$ priors centred around the $log(\rm RMS)$ of their respective datasets, with a sigma of $log(\rm RMS)$.

\begin{figure*}[h!]
    \centering
	\includegraphics[width=18cm]{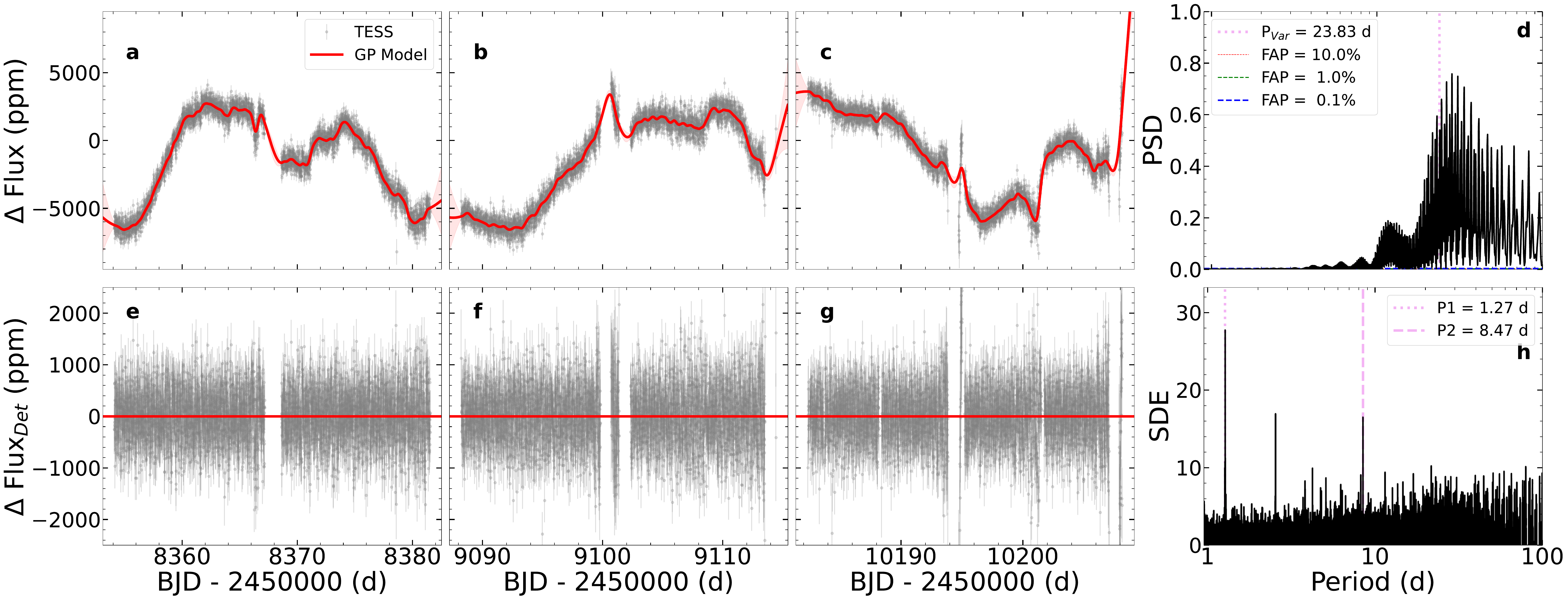}
	\caption{\textbf{TESS data with the best model fit (GP-only).} Panels \textbf{a}, \textbf{b}, and \textbf{c}, show the TESS light curve with the best GP model. Panel \textbf{d} shows the GLS periodogram of the TESS data. Panels \textbf{e}, \textbf{f}, and \textbf{g}, show the detrended TESS data. Panel \textbf{h} shows the TLS periodogram of the detrended light curve.}
	\label{data_tess_gp}
\end{figure*}

\begin{figure*}[h!]
    \centering
    \includegraphics[width=18cm]{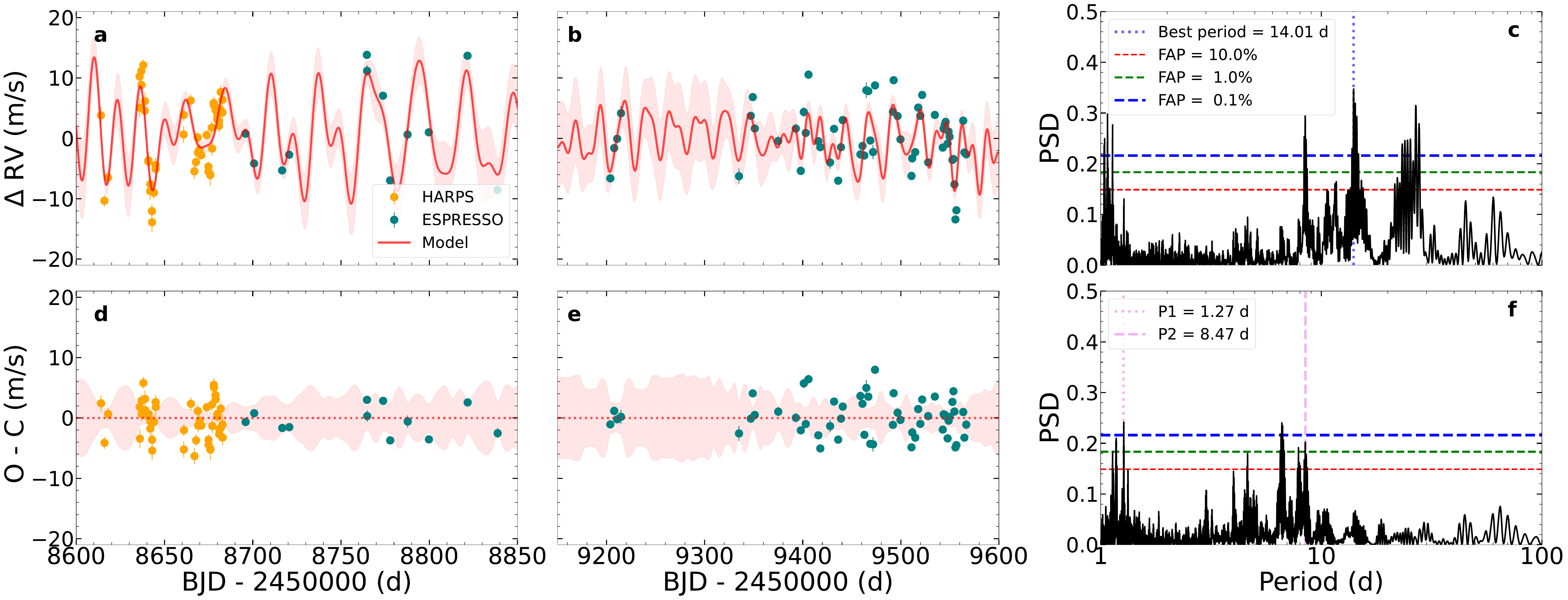}
	\caption{\textbf{RV time-series with the best model fit (GP-only).} Panels \textbf{a} and \textbf{b} show the RV time-series with the best GP model. Panel \textbf{c} shows the GLS periodogram of the raw RV data. Panels \textbf{d} and \textbf{e} show the detrended RV data. Panel \textbf{f} shows the GLS periodogram of the detrended RV time-series.}
	\label{data_rv_gp}
\end{figure*}

Figures~\ref{data_tess_gp} and ~\ref{data_rv_gp} show the best-fit model, using the parameters obtained from the posterior distribution of the nested sampling. Table~\ref{table_measured} shows the priors and measured values for all parameters. 

In the case of the TESS data, we do not gain a lot of information about the stellar activity of the star. However, the GP-model manages to perform a smooth fit that we use to detrend the data. We apply a Transit Least Squares (TLS, \citealt{Hippke2019}) periodogram to the residuals under the fit. The periodogram shows its highest peak at 1.27 days, the period of the candidate TOI 238.01. The periodogram also shows peaks at $\sim$ 2.54 days (2$\times$ 1.27 days) and 8.47 days. 

The model on the FWHM+BIS+RV provides significantly different timescales of variability for the FWHM+RV and BIS+RV components. The FWHM+RV component shows a periodicity of 24.49 $\pm$ 0.86 days, with a timescale of evolution of 35.5$^{+6.7}_{-5.8}$ days, while the BIS+RV shows a periodicity of \mbox{26.53 $\pm$ 0.46} days, with a timescale of evolution of 47.8$^{+10.1}_{-9.6}$ days. The omega parameters are 0.76$^{+0.36}_{-0.25}$, and 0.74 (3$\sigma$), respectively. The variability shows a significant RV amplitude related to the FWHM variations ($A11_{\rm~RV}$ = 4.7 $^{+2.1}_{-1.5}$ m$\cdot$s$^{-1}$), a non significant RV amplitude correlated with the gradient of the FWHM variations ($A12_{\rm~RV}$ = -1.6 $^{+4.6}_{-5.0}$ m$\cdot$s$^{-1}$), a significant RV amplitude anti-correlated with the BIS variations ($A21_{\rm~RV}$ = -5.1 $^{+1.2}_{-1.4}$ m$\cdot$s$^{-1}$) and an almost-significant RV amplitude correlated with the gradient of the BIS variations ($A22_{\rm~RV}$ = 8.6 $^{+3.9}_{-4.0}$ m$\cdot$s$^{-1}$). These results show once again that both indicators are affected by different types of variability, and they are both present in the RV data. We apply the best model to the data and subtract it to study the residuals. 

The RMS of the residuals after the fit of the RV data goes down to 3.0 m$\cdot$s$^{-1}$ from the original 5.86 m$\cdot$s$^{-1}$ (49\% reduction), with the ESPRESSO and HARPS data retaining a very similar RMS. The RMS of residuals after the fit to the FWHM, and BIS, data is of 6.9 m$\cdot$s$^{-1}$ and 4.2 m$\cdot$s$^{-1}$, respectively. These values correspond to a reduction of the variability of 59 \% for the FWHM data and 32\% for the BIS data.

The GLS periodogram of the residuals has its most dominant peak at 1.27 days (see panel f of Fig.~\ref{data_rv_gp}) again the period of the candidate TOI 238.01. Some other peaks appear in the periodogram. Most of them are related to the aliases of the 1.27 days periodicity and the gaps inbetween the observations. It is worth noting that the periodogram of the raw velocities shows several significant peaks. A forest of peaks around 25 days (stellar rotation), another at 14 days, and a third one at 8.47 days (see panels c and f of Fig.~\ref{data_rv_gp}). The latter has a period that matches one of the peaks in the \texttt{TLS} periodogram of the TESS detrended data \citep{Hippke2019}. It is possible that all of them are artefacts of the rotation signal. Yet, it may be that the GP is incorrectly suppressing some of them due to their closeness to the rotation and its harmonics. 

\subsection{Global model - One planet} \label{sect_gp_1p}

Following the results obtained after detrending the data with an activity-only model, we included the possibility of one planet in the TESS and RV data. 

To model the transit, we used the \texttt{pytransit} package~\citep{Parviainen2015} with quadratic limb darkening~\citep{MandelAgol2002}. For the orbital period, we used a $\mathcal{N}$ prior of 1.27 $\pm$ 0.1 days. To avoid a multi-modal posterior for the time of transit (T0), we parameterised it as a phase centred around the expected last transit of the TESS light curve. The time of transit is defined as:

\begin{equation}
T0_{b} = x[-1]_{TESS} + P_{b} \cdot (Ph_{b} - 1),
\end{equation}

\noindent where x[-1]$_{TESS}$ is the last TESS observation, $P_{b}$ is the orbital period of the planet, and $Ph_{b}$ is the phase, using a $\mathcal{U}$ prior [0,1]. We sampled the planet radius directly, with $\mathcal{LU}$ prior [-5,5] in R$_{\oplus}$ (corresponding to $\sim$ 0 to 148 R$_{\oplus}$). The $\mathcal{LU}$ prior allows for better sensitivity when sampling very small values (e.g. in the case of having no transit signal). For the impact parameter, we used a $\mathcal{U}$ [0,1] prior. We parameterised the eccentricity as $e = (\sqrt{e} ~\cos(\omega))^{2} + (\sqrt{e} ~\sin(\omega))^{2}$ and $\omega = \arctan2(\sqrt{e} ~sin(\omega),\sqrt{e} ~\cos(\omega))$. We then sample $\sqrt{e} ~cos(\omega)$ and $\sqrt{e} ~\sin(\omega)$ with normal priors of 0 $\pm$ 0.3. Eccentricity is typically overestimated in noisy data and datasets with unmodelled sources of variability \citep{Hara2019}. The parameterisation described above favours low eccentricities, which are expected for close-in planets, while still allowing for eccentricities up to 1. To estimate the limb darkening priors we used the \texttt{LDTK} package \citep{Parviainen2015b}, which provides limb darkening coefficients, and uncertainties, for a given set of the stellar parameters and a given observing passband. We used $\mathcal{N}$ priors centred in the results given by \texttt{LDTK}, with sigmas of 3$\times$ the provided uncertainty. Even with simple 2 parameter models, such as the quadratic limb darkening, there can be degeneracies between them. Several combinations can yield indistinguishable results, in particular at low signal to noise. \texttt{LDTK} provides a likelihood term that estimates the log-likelihood of the combination of both parameters. We included it as an additional term in our likelihood evaluation. 

For the planetary component in the RV data, we used a Keplerian model:

\begin{equation} \label{eq_kepler}
  y(t)=K \left(\cos(\eta+\omega) + e \ \cos(\omega)\right)
\end{equation} 

\noindent where the true anomaly $\eta$ is related to the solution of the Kepler equation that depends on the orbital period of the planet $P_{\rm orb}$ and the orbital phase $\phi$. This phase corresponds to the periastron time, which depends on the mid-point transit time $T_{0}$, the argument of periastron $\omega$, and the eccentricity of the orbit $e$.

The $P_{\rm orb}$, mid-point transit time $T_{0}$ argument of periastron $\omega$, and eccentricity, are shared with the transit model and their parameterisation is described above. Instead of the amplitude ($K$), we sampled the planet mass directly, with $\mathcal{LU}$ prior [-5,5] in M$_{\oplus}$ (corresponding to $\sim$ 0 to 148 M$_{\oplus}$).

 We included the stellar radius and mass in the model as they are needed for the computation of several transit parameters, and the transformations between planet radius and mass, and their respective amplitudes. We sampled them using $\mathcal{N}$ priors, centred around the values shown in Table~\ref{tab:parameters}. The standard deviation of the distributions is the uncertainty derived in the stellar parameters, and enlarged following the results of \citet{Tayar2022}, to account for the systematic uncertainties of the model, which result in R$_{s}$ = 0.733 $\pm$ 0.034 R$_{\odot}$ and M$_{s}$ = 0.790 $\pm$ 0.043 M$_{\odot}$. This way, we ensure proper propagation of uncertainties to the rest of the parameters sampled in the model.

\begin{figure*}[ht]
    \centering
	\includegraphics[width=18cm]{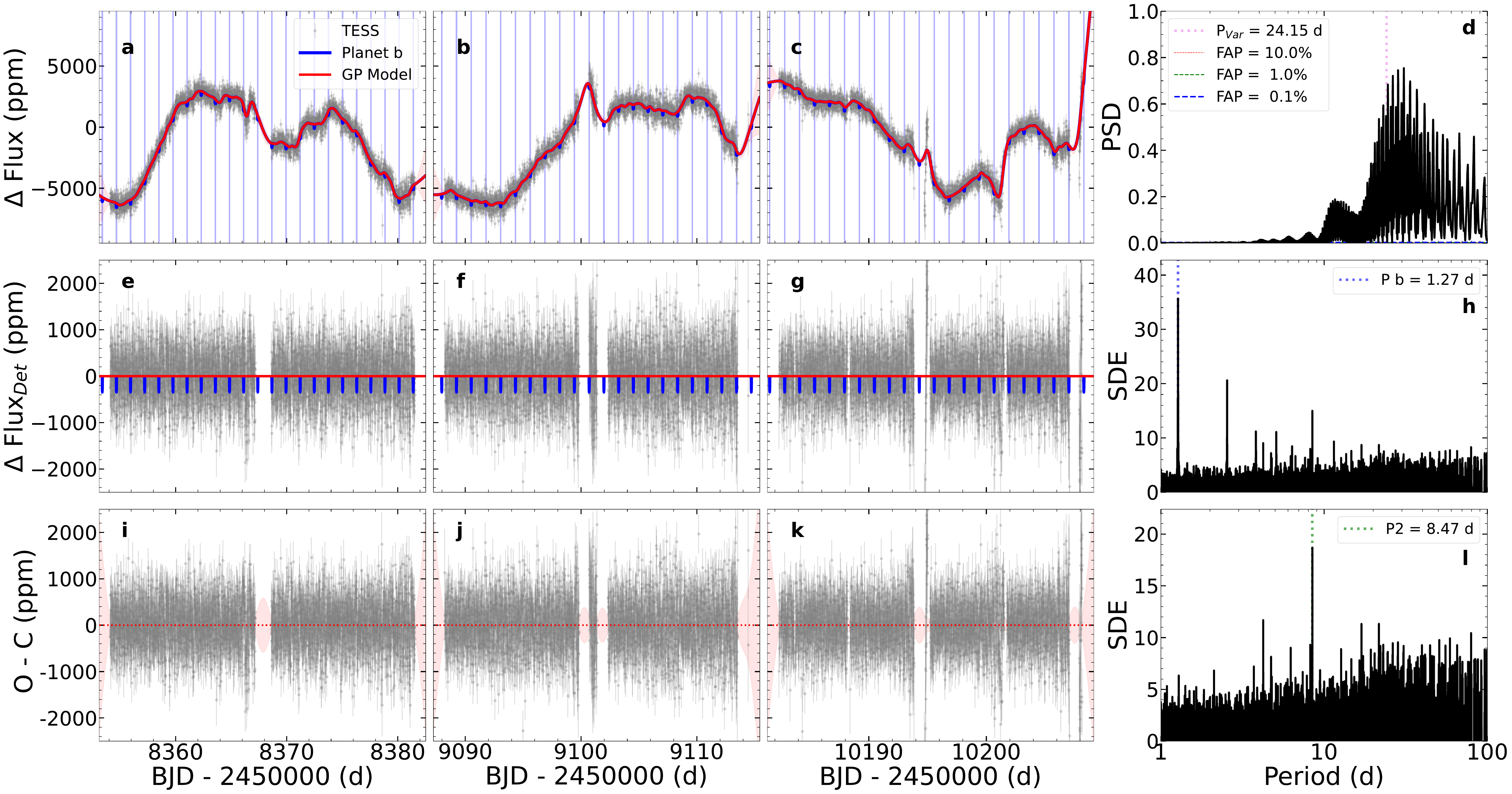}
	\caption{\textbf{TESS data with the best model fit (GP+1p).} Panels \textbf{a}, \textbf{b}, and \textbf{c}, show the TESS light curve with the best GP+1p model. The blue lines show the transits of TOI 238.01. Panel \textbf{d} shows the GLS periodogram of the TESS data. Panels \textbf{e}, \textbf{f}, and \textbf{g}, show the detrended TESS data along with the best fit model of TOI 238.01. Panel \textbf{h} shows the TLS periodogram of the detrended light curve. Panels \textbf{i}, \textbf{j}, and \textbf{k} show the residuals after the fit of the best GP+1p model.  Panel \textbf{i} shows the TLS periodogram of the residuals.}
	\label{data_tess_gp_1p}
\end{figure*}

\begin{figure*}[ht]
    \centering
    \includegraphics[width=18cm]{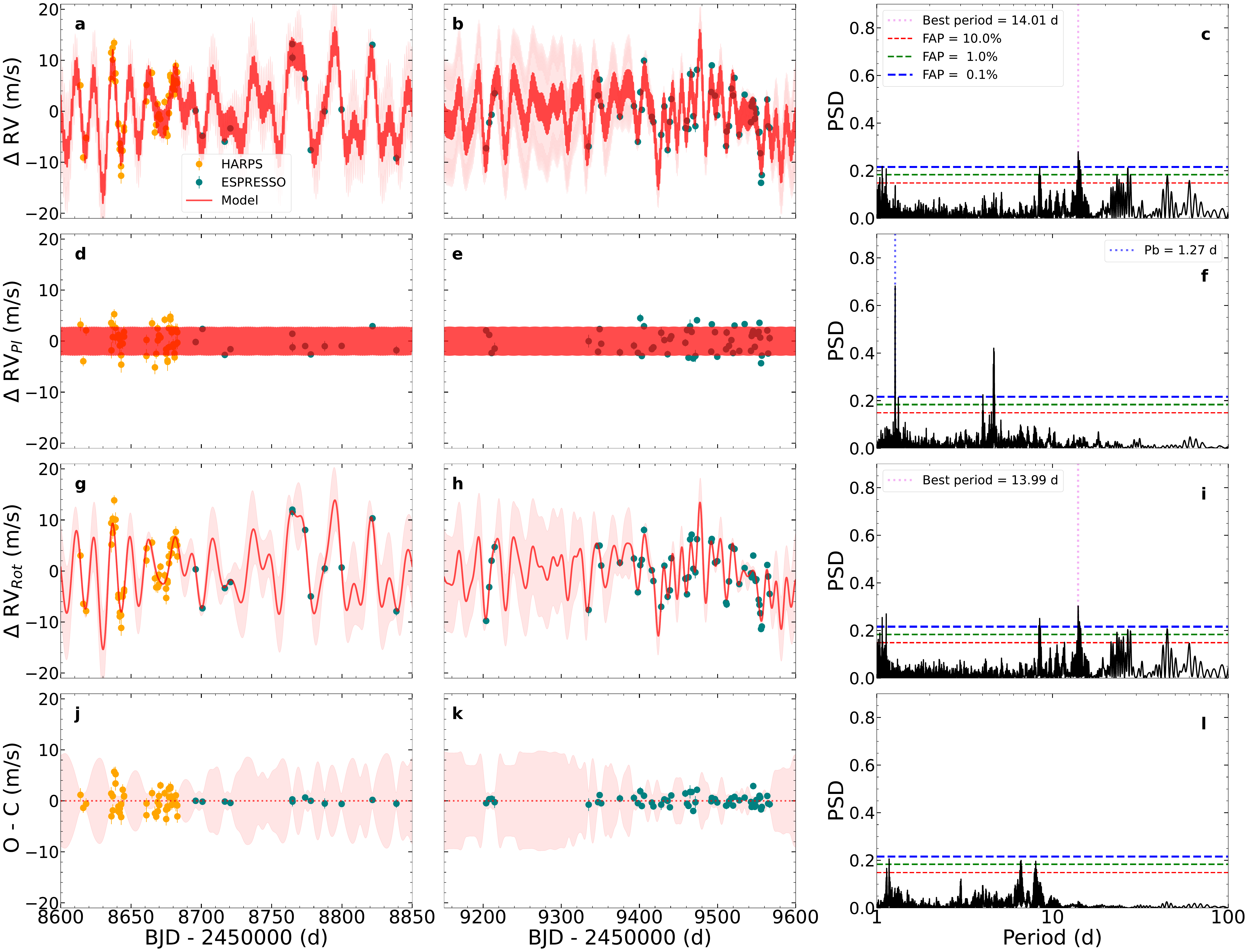}
	\caption{\textbf{RV time-series with the best model fit (GP+1p).} Panels \textbf{a} and \textbf{b} show the RV time-series with the best GP+1p model. Panel \textbf{c} shows the GLS periodogram of the raw RV data. Panels \textbf{d} and \textbf{e} show the RV data detrended from stellar activity (i.e planetary component). Panel \textbf{f} shows the GLS periodogram of the detrended RV time-series. Panels \textbf{g} and \textbf{h} show the RV data detrended from the planetary component (i.e stellar activity). Panel \textbf{i} shows the GLS periodogram of the activity-induced RVs. Panels \textbf{j} and \textbf{k} show the residuals after the fit. Panel \textbf{i} shows the GLS periodogram of the residuals.}
	\label{data_rv_gp_1p}
\end{figure*}

We obtain a significant detection of the candidate TOI 238.01, both in the TESS light curve and the RV data. We measure an orbital period $P_{b}$ of 1.2730991 $\pm$ 0.0000029 days and a time of mid-transit at BJD 2460207.0240 $\pm$ 0.0033 days. We measure a radius $R_{b}$ of 1.387 $\pm$ 0.086 R$_{\oplus}$, and a mass $m_{b}$ of 3.90 $\pm$ 0.51 M$_{\oplus}$. These values correspond to a planet-star contrast (R$_{\rm planet}$/R$_{\rm star}$) of 0.01704 $\pm$ 0.00084 and a RV semi-amplitude of 2.72 $\pm$ 0.32 m$\cdot$s$^{-1}$. We measure an impact parameter $\textless$ 0.61 ($\sigma$). We measure a $\Delta$ $ln$Z of +495.7 with respect to the activity-only model, which extremely favours the model with a planet with respect to the activity-only model. The RMS of the residuals after the fit of the RV data goes down to 1.53 m$\cdot$s$^{-1}$, from the original 5.86 m$\cdot$s$^{-1}$ (74\% reduction), with the ESPRESSO data showing an RMS of the residuals of 0.93 m$\cdot$s$^{-1}$ and the HARPS data showing 2.11 m$\cdot$s$^{-1}$. 

Figures~\ref{data_tess_gp_1p} and ~\ref{data_rv_gp_1p} show the best-fit model, using the parameters obtained from the posterior distribution of the nested sampling. Table~\ref{table_measured} shows the priors and measured values for all parameters. The \texttt{TLS} periodogram of the residuals of the TESS light curve shows a dominant peak at 8.47 days. The same peak appears in the GLS periodogram of the activity-induced signal in RV. This suggests that the GP is potentially absorbing an additional planetary signal in the RV data.

\subsection{Global model - Two planets} \label{sect_gp_2p}

We investigated the 8.47 days peak in the \texttt{TLS} periodogram by including a second Keplerian component in the model. We sample the period using a $\mathcal{N}$ prior of 8.5 $\pm$ 0.5 days. The rest of the components of the planet model are the same described in section~\ref{sect_gp_1p}.

\begin{figure*}[ht]
    \centering
	\includegraphics[width=18cm]{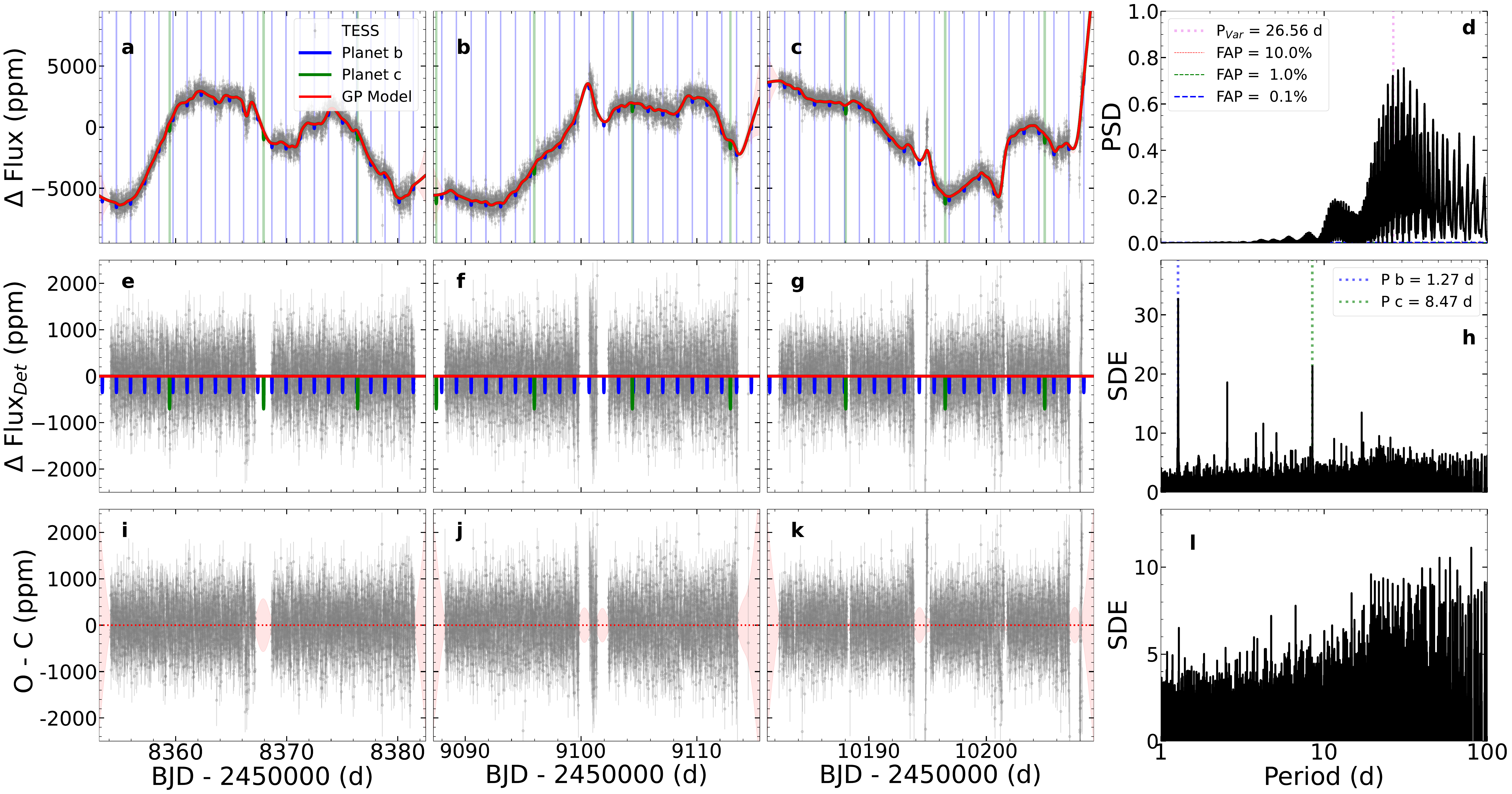}
		\caption{\textbf{TESS data with the best model fit (GP+2p).} Panels \textbf{a}, \textbf{b}, and \textbf{c}, show the TESS light curve with the best GP+1p model. The blue and green lines show the transits at the periods of 1.27 days, and 8.47 days, respectively. Panel \textbf{d} shows the GLS periodogram of the TESS data. Panels \textbf{e}, \textbf{f}, and \textbf{g}, show the detrended TESS data along with the best fit model of both planets. Panel \textbf{h} shows the TLS periodogram of the detrended light curve. Panels \textbf{i}, \textbf{j}, and \textbf{k} show the residuals after the fit of the best GP+1p model.  Panel \textbf{i} shows the TLS periodogram of the residuals.}
	\label{data_tess_gp_2p}
\end{figure*}

\begin{figure*}[ht]
    \centering
    \includegraphics[width=18cm]{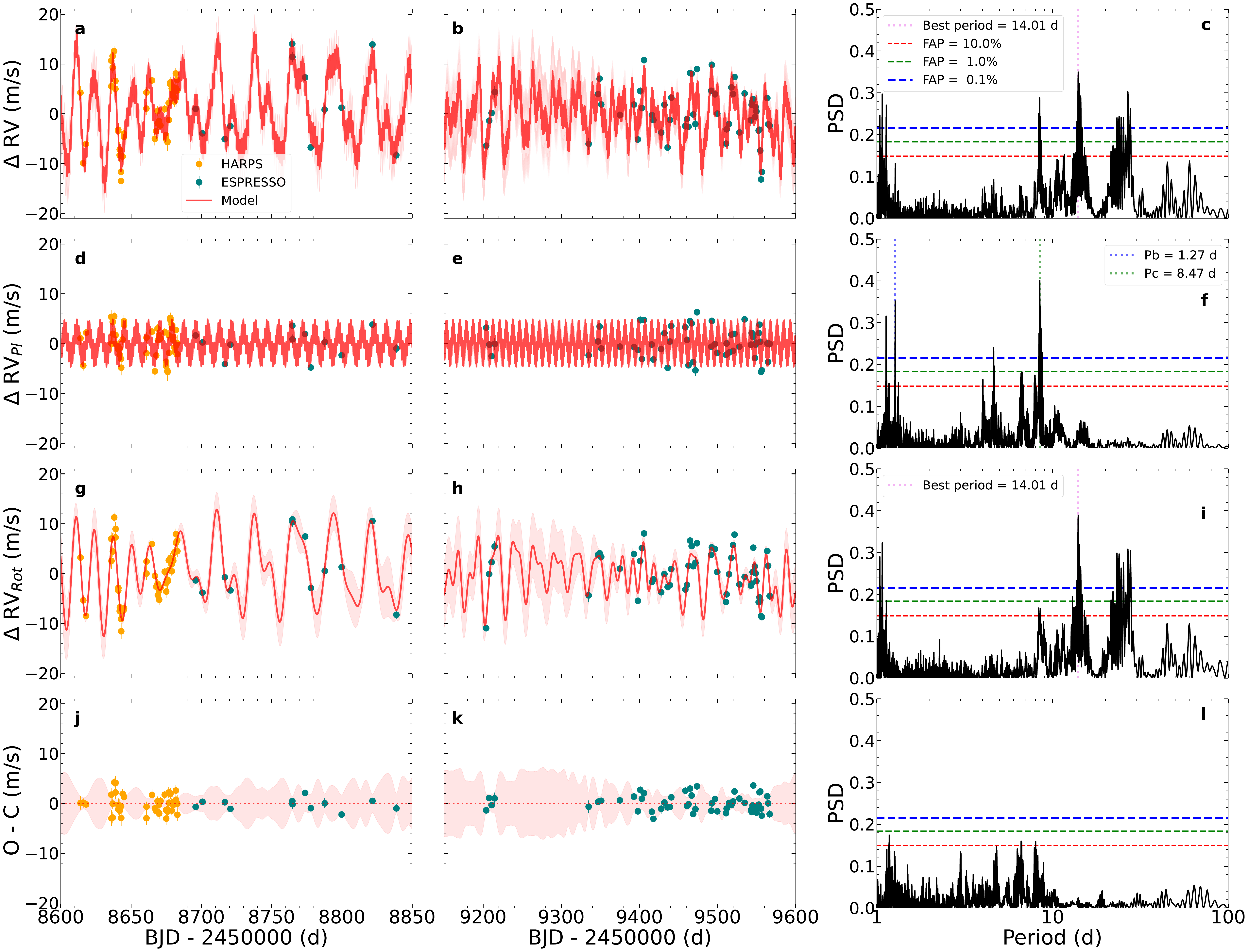}
	\caption{\textbf{RV time-series with the best model fit (GP+2p).} Panels \textbf{a} and \textbf{b} show the RV time-series with the best GP+1p model. Panel \textbf{c} shows the GLS periodogram of the raw RV data. Panels \textbf{d} and \textbf{e} show the RV data detrended from stellar activity (i.e planetary component). Panel \textbf{f} shows the GLS periodogram of the detrended RV time-series. Panels \textbf{g} and \textbf{h} show the RV data detrended from the planetary component (i.e stellar activity). Panel \textbf{i} shows the GLS periodogram of the activity-induced RVs. Panels \textbf{j} and \textbf{k} show the residuals after the fit. Panel \textbf{l} shows the GLS periodogram of the residuals.}
	\label{data_rv_gp_2p}
\end{figure*}

We obtain a significant detection of a planetary signal in both the TESS light curve and the RV data. We measure an orbital period $P_{c}$ of 8.465651 $\pm$ 0.000035 days and a time of mid-transit at BJD 2460204.9675 $\pm$ 0.0050 days, a radius $R_{c}$ of 2.18 $\pm$ 0.18 R$_{\oplus}$, and a mass $M_{c}$ of 6.7 $\pm$ 1.1 M$_{\oplus}$. These values correspond to a R$_{\rm planet}$/R$_{\rm star}$ of 0.0267 $\pm$ 0.0021 and a RV semi-amplitude of 2.46 $\pm$ 0.41 m$\cdot$s$^{-1}$. We measure an impact parameter of 0.774 $\pm$ 0.057. This model yields a $\Delta$ $ln$Z of +30.4 with respect to the 1-planet model, and +528.1 with respect to the activity-only model. These values paint the 2-planet model as much more likely than either the 1-planet model or the activity-only model. The RMS of the residuals after the fit of the RV data goes down to 1.61 m$\cdot$s$^{-1}$, from the original 5.86 m$\cdot$s$^{-1}$ (72\% reduction), with the ESPRESSO data showing an RMS of the residuals of 1.50 m$\cdot$s$^{-1}$ and the HARPS data showing 1.75 m$\cdot$s$^{-1}$. The RMS of the residuals is slightly higher than with the 1-planet model, which might be explained by the GP overfitting the data in the 1-planet model. 

Figures~\ref{data_tess_gp_2p} and ~\ref{data_rv_gp_2p} show the best-fit model, using the parameters obtained from the posterior distribution of the nested sampling. Table~\ref{table_measured} and figure~\ref{posterior_2pk} show the priors and measured values for all parameters. 

\begin{figure}[ht]
    \centering
    \includegraphics[width=9cm]{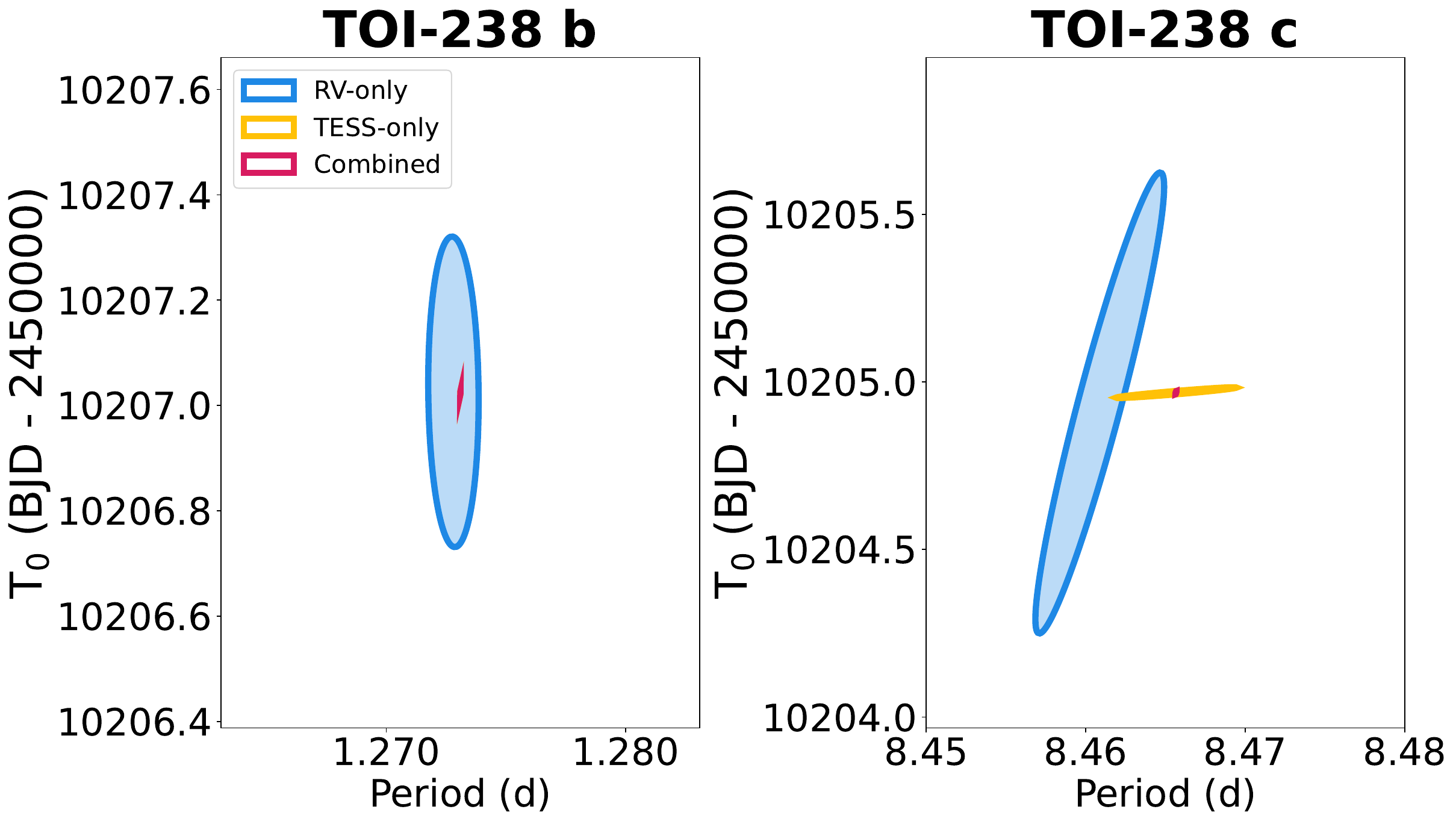}
	\caption{\textbf{Time of mid-transit against orbital period for different models.} The blue ellipse shows the RV-only solution, the orange ellipse shows the TESS-only solution and the red ellipse shows the combined solution. The orange ellipse in the left panel is completely covered by the red ellipse.}
	\label{tess_rv_comp}
\end{figure}

Additionally, we repeated the modelling process in the RV and TESS data independently. We found consistent results for the planetary parameters in our combined analysis, a RV-only analysis and a TESS-only analysis. Figure~\ref{tess_rv_comp} shows the comparison between the periods and times of mid transits obtained in these three models. We tested the effect of different GP Kernels (such as \texttt{S+LEAF}'s \textit{MEP} Kernel or different combinations of SHO Kernels). We found no significant differences in the planetary parameters, however the activity model would be sometimes very different. We found the \textit{ESP} Kernel to be the most reliable at producing a smooth model that did not show obvious signs of overfitting. 

After fitting a model with a GP and two planetary signals, the periodograms of the residuals of the TESS light curve and the RV don't show any clear peak that could be attributed to a planetary signal. There is, however, a peak at 14 days in the GLS of the activity-induced component of the RV that does not fully fit as one of the harmonics of the rotation period.

\subsection{Additional planets}

Although we find no more prominent periodicities in the \texttt{TLS} periodogram of the residuals of the TESS light curve, or the GLS periodogram of the residuals of the RV, we test for the presence of additional planets that might have been missed. After fitting the model that includes a GP and two planetary signals, the RMS of the residuals of the RV data is 1.61 m$\cdot$s$^{-1}$, with the ESPRESSO data showing an RMS of 1.50 m$\cdot$s$^{-1}$ and the HARPS data 1.75 m$\cdot$s$^{-1}$. These values are significantly larger than the internal errors of the RV of 0.58 m$\cdot$s$^{-1}$ and 1.10 m$\cdot$s$^{-1}$ for ESPRESSO and HARPS, respectively. This large difference can be due to a combination of systematic issues, unmodelled activity variations and/or additional planets in the system. Considering the impact parameter of planet c, we do not expect any additional detectable transiting planet in the data. However, it is possible that the signal of a non-transiting planet has been absorbed by the GP model. The GLS periodogram of the suspected activity-induced RV data showed a peak at 14 days that does not fully fit as one of the harmonics of the measured rotation period (see figure~\ref{data_rv_gp_2p}). To investigate this signal, and others that might have gone unnoticed, we add an additional sinusoidal component in the RV data. We write the RV variation as:

\begin{equation} \label{eq_sine}
  y(t)=-K  \sin(2\pi \cdot (t - T_{0})/P) 
\end{equation} 

\noindent which ensures that T$_{0}$ coincides with the time of mid transit. 

We test two different models: a blind search, in which we define the prior for the orbital period as $\mathcal{LU}$ between 10 and 100 days, and a guided fit in which we define the prior for the orbital period as $\mathcal{N}$ around 14 $\pm$ 2 days. The $T0$ and mass are defined as described in section~\ref{sect_gp_1p}. 

In the case of the blind search, we obtain a power excess at $\sim$ 30 days orbital period, with a non-significant mass measurement and a wide range of possible values for the phase. While this periodicity is not consistent with the variability measured by the combined GPs, it is consistent with a secondary peak in the periodogram of the FWHM (see Fig.~\ref{fwhm_model}). Althought this might not be enough information to fully establish the origin of the signal, it is sufficient to not consider a planetary origin. 

In the guided search, the model converges to a period of 13.992 $\pm$ 0.033 days and a possible mass of 8.4$^{+2.2}_{-3.0}$ M$_{\oplus}$ (2.8 $\sigma$). We measure a $\Delta$ $Log$Z of $\sim$ 3. These results hint at the presence of an additional signal, independent from the activity model and the other two planets, but are not decisive enough to claim a detection.  

\begin{figure}[ht]
    \centering
    \includegraphics[width=9cm]{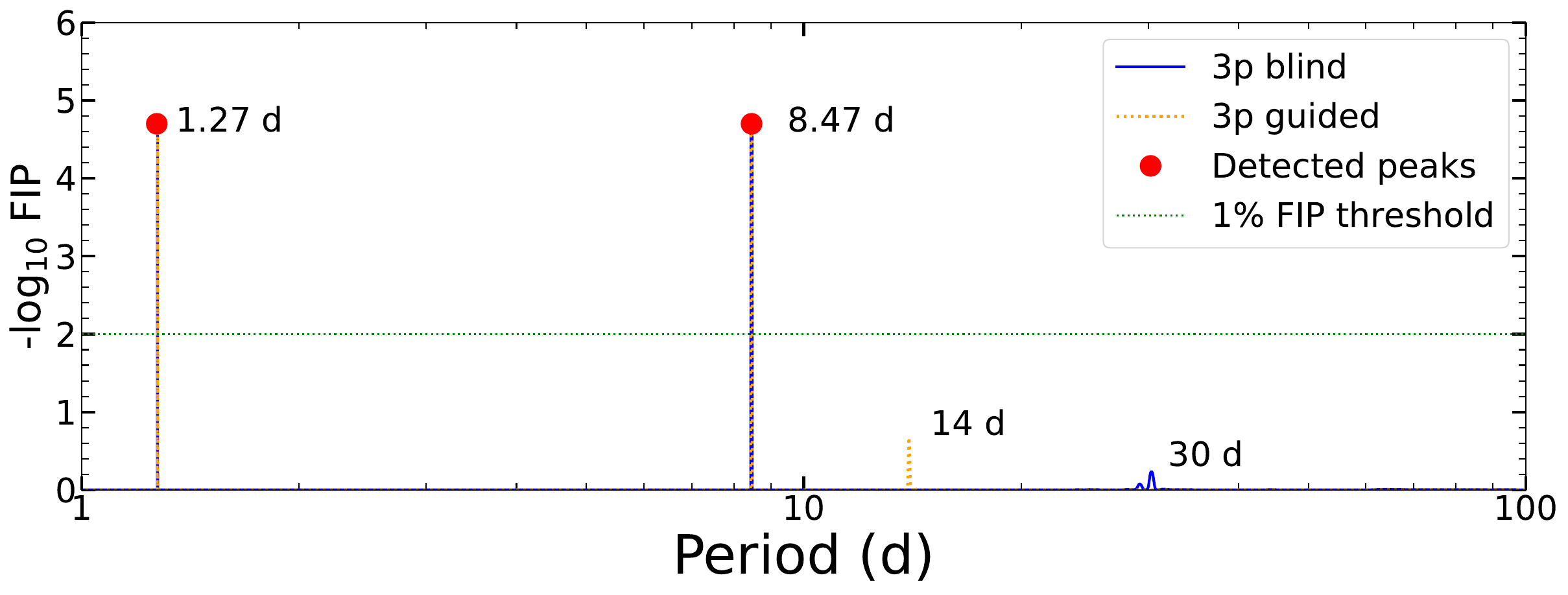}
	\caption{\textbf{FIP periodogram of the different 3-planet models of TOI--238.} The blue solid line shows the result of the blind search for a potential third planet. The dotted orange line shows the result of the guided search. The red solid circles show the detected periodicities. The dotted green line shows the 1\% FIP threshold.}
	\label{fip_3p}
\end{figure}

We assesed the significance of the detection of those two signals using the False inclusion probability (FIP) framework \mbox{\citep{hara2022a}}. This framework uses the posterior distribution of the nested sampling run, and compute the probability of having a planet within a certain orbital frequency interval based on the fraction of samples within that frequency interval. Figure~\ref{fip_3p} shows the FIP periodogram of the two 3-planet models tested. We identify clear significant peaks for the signals at 1.27 days and 8.47 days. However, the peaks at 14 days and 30 days appear well below the 1\% threshold. 

With the data at hand, and within the framework of our analysis, there is no clear evidence of the presence of additional planets in the system. There might be an additional non-transiting planet with a period of 14 days, and a minimum mass of $\sim$ 8.4 M$_{\oplus}$, but with the current dataset and analysis technique, we cannot confidently disentangle the signal from the activity-induced signal. 

\section{Discussion}

\subsection{The planetary system of TOI--238}

We detect the presence of two planetary signals both in the TESS and RV data. TOI--238 b is a planet with an orbital period of 1.2730991 $\pm$ 0.0000029 days and a time of mid-transit at BJD 2460207.0240 $\pm$ 0.0033 days. It has a radius of 1.402 $\pm$ 0.086 R$_{\oplus}$ (R$_{\rm planet}$/R$_{\rm star}$ = 0.01721 $\pm$ 0.00083), and a mass of 3.40 $\pm$ 0.46 M$_{\oplus}$ (K = 2.36 $\pm$ 0.32 m$\cdot$s$^{-1}$). It orbits at a distance of 0.02118 $\pm$ 0.00038 au from its parent star. The planet's orbit is consistent with circular, with an upper limit to the eccentricity of $e = 0.18$ to within $3\sigma$. Assuming a bond albedo of 0.3, we estimate an equilibrium temperature of equilibrium ($T_{\rm eq}$) of 1311 $\pm$ 28 K. TOI--238 c is a planet with an orbital period of 8.465651 $\pm$ 0.000035 days and a time of mid-transit at BJD 2460204.9675 $\pm$ 0.0050 days. It has a radius of 2.18 $\pm$ 0.18 R$_{\oplus}$ (R$_{\rm planet}$/R$_{\rm star}$ = 0.0267 $\pm$ 0.0021), and a mass of 6.7 $\pm$ 1.1 M$_{\oplus}$ (K = 2.46 $\pm$ 0.41 m$\cdot$s$^{-1}$). The planet's orbit is consistent with circular, with an upper limit to the eccentricity of $e = 0.34$ to within $3\sigma$. It orbits at a distance of 0.0749 $\pm$ 0.0013 au. Assuming bond albedos of 0.3 we estimate an equilibrium temperature of 696 $\pm$ 15 K. Table~\ref{param_planets} shows all the parameters of the planets in the system. 

\begin{figure}[ht]
    \includegraphics[width=9cm]{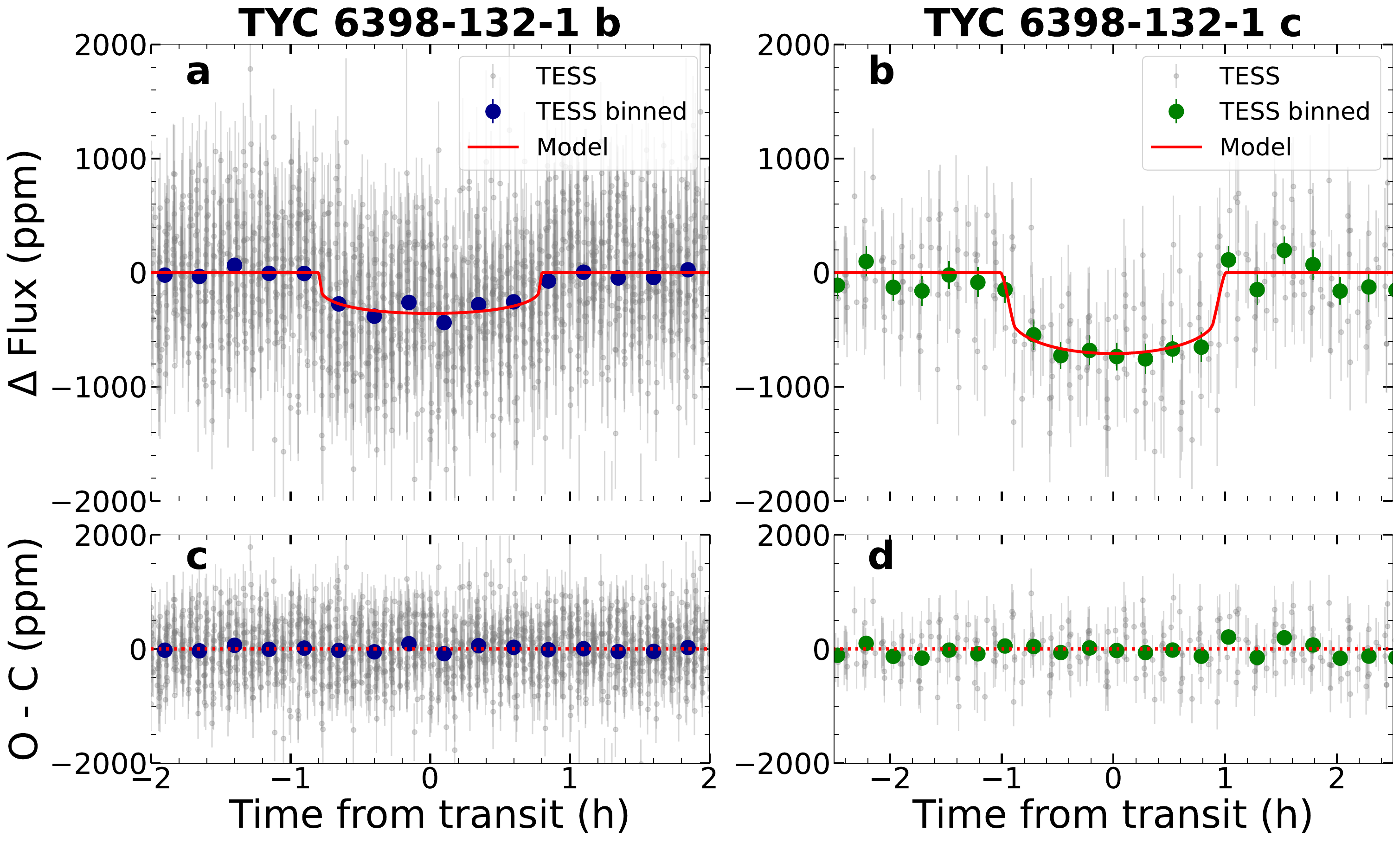}
	\caption{\textbf{Phase folded light curves.} Panels \textbf{a} and \textbf{b} show the phase folded plot of the TESS light curves for planets b and c, respectively, after subtracting the best fit GP model and, in each case, the other planetary signal. Panels \textbf{c} and \textbf{d} show the residuals after the fit.}
	\label{data_tess_phase}
\end{figure}

\begin{figure}[ht]
    \includegraphics[width=9cm]{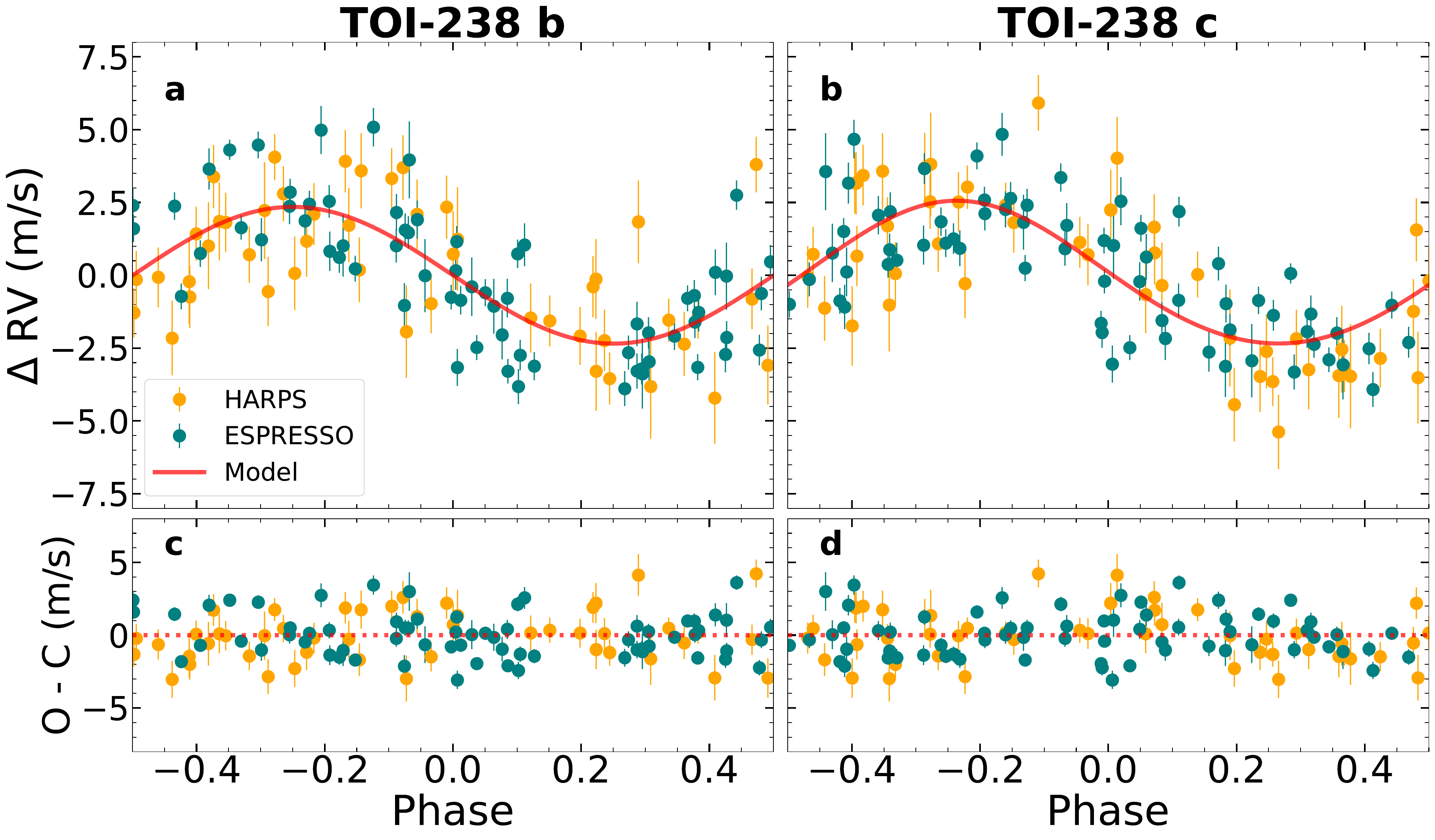}
	\caption{\textbf{Phase folded RV time-series.} Panels \textbf{a} and \textbf{b} show the phase folded plot of the RV time-series for planets b and c, respectively, after subtracting the best fit GP model and, in each case, the other planetary signal. Panels \textbf{c} and \textbf{d} show the residuals after the fit.}
	\label{data_rv_phase}
\end{figure}

\begin{table} 
\begin{center}
\caption{Parameters of the two planets detected orbiting TOI--238 \label{param_planets}}
\begin{tabular}[centre]{l c}
\hline
Parameter  & Value \\ \hline
\\
\textbf{TOI--238 b} \\
$T_{0}$ -- 2450000 [d] & 10207.0240 $\pm$ 0.0033 \\
$P_{\rm orb}$ [d] & 1.2730991 $\pm$ 0.0000029  \\
$r_p$  [R$\oplus$] & 1.402 $\pm$ 0.086 R$_{\oplus}$ \\
$m_p$  [M$\oplus$] & 3.40 $\pm$ 0.46 M$_{\oplus}$ \\
$a$ [au] &  0.02118 $\pm$ 0.00038 \\
$e$ & \textless~0.18\\
$i$ [º] & $\textgreater$ 84.6  \\
Incident flux [$S_{\oplus}$] & 728 $\pm$ 78  \\
T$_{\rm eq~A = 0.3}$ [K] & 1311 $\pm$ 28  \\ 
$\rho$ [g$\cdot$cm$^{-3}$] & 6.8 $\pm$ 1.6\\ 
R$_{\rm planet}$/R$_{\rm star}$ & 0.01721 $\pm$ 0.00083\\
K [m$\cdot$s$^{-1}$]  & 2.36 $\pm$ 0.32  \\ \\
\textbf{TOI--238 c} \\
$T_{0}$ -- 2450000 [d] & 10204.9675 $\pm$ 0.0050  \\
$P_{\rm orb}$ [d] & 8.465651 $\pm$ 0.000035 \\
$r_p$  [R$\oplus$] &  2.18 $\pm$ 0.18 R$_{\oplus}$\\
$m_p$  [M$\oplus$] &  6.7 $\pm$ 1.1 M$_{\oplus}$ \\
$a$ [au] &   0.0749 $\pm$ 0.0013\\
$e$ & \textless~0.36 \\
$i$ [º] & 87.938 $\pm$ 0.089 \\
Incident flux [$S_{\oplus}$] &  58.3 $\pm$ 6.2 \\
T$_{\rm eq~A = 0.3}$ [K] & 696 $\pm$ 15 \\ 
$\rho$ [g$\cdot$cm$^{-3}$] & 1.83 $\pm$ 0.66 \\ 
R$_{\rm planet}$/R$_{\rm star}$ & 0.0267 $\pm$ 0.0021\\
K [m$\cdot$s$^{-1}$]  & 2.46 $\pm$ 0.41  \\
\hline
\end{tabular}
\end{center}
\end{table}

Figures~\ref{data_tess_phase} and ~\ref{data_rv_phase} show the phase-folded plots of the TESS light curve and the RV time-series, at the periods of both planets b and c, after subtracting the activity model and, in each case, the other signal. 

\begin{figure*}[ht]
    \includegraphics[width=18cm]{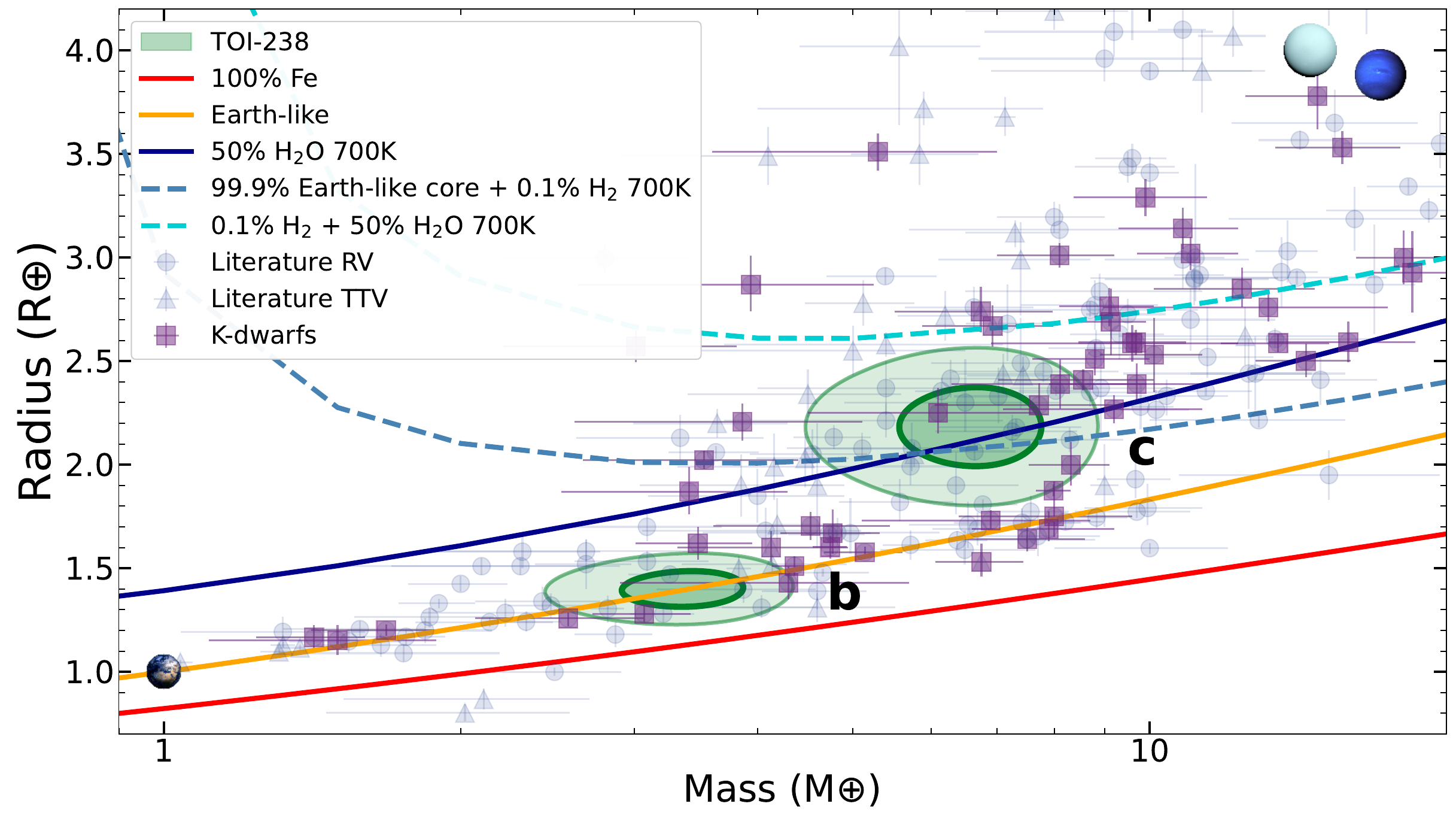}
	\caption{\textbf{Mass-radius diagram} of transiting exoplanets with well established mass and radius from the NASA Exoplanet archive along with some theoretical composition curves \citep{LiZeng2019}. The purple squares highlight the transiting planets orbiting K-dwarfs. The green blobs show the mass and radius posteriors (1- and 2-$\sigma$ contours) of  TOI--238 b and c. The planets of the solar system are included as reference.}
	\label{mass_rad}
\end{figure*}

Figure~\ref{mass_rad} shows the mass-radius diagram of known exoplanets, with mass and radius measured with uncertainties smaller than 33\%, in the range between 1 and 20 M$_{\oplus}$ and 0.7 and 4.2 R$_{\oplus}$. TOI--238 b is a super-Earth whose mass-radius configuration is consistent with an Earth-like composition. TOI--238 c on the other hand is consistent with a water world \citep{LegerSelsis2004} with $\sim$ 50\% water or a rocky core with a very small hydrogen envelope. The planet falls exactly at the intersection of these two tracks, making it impossible to distinguish the correct composition based on the mass and radius alone. The large uncertainty in the radius ($\sim$ 9\%) prevents us from fully rejecting both Earth-like compositions or a water-rich planet with small Hydrogen envelope (both within 3$\sigma$). A better determination of the radius, e.g. using high-precision photometry of missions such as the CHaracterising ExOPlanet Satellite (\textit{CHEOPS}, \citealt{Benz2021}), could constrain better the nature of planet c, however it is unlikely that it would be sufficient to uniquely establish it. Using the parameters described above, we estimate densities for planets b, and c, of 6.8 $\pm$ 1.6 g$\cdot$cm$^{-3}$, and 1.83 $\pm$ 0.66 g$\cdot$cm$^{-3}$, respectively. Planet b has a density larger than Earth, expected for super-Earths with Earth-like composition. Planet c has a density similar to Neptune.

\begin{figure}[ht]
    \includegraphics[width=9cm]{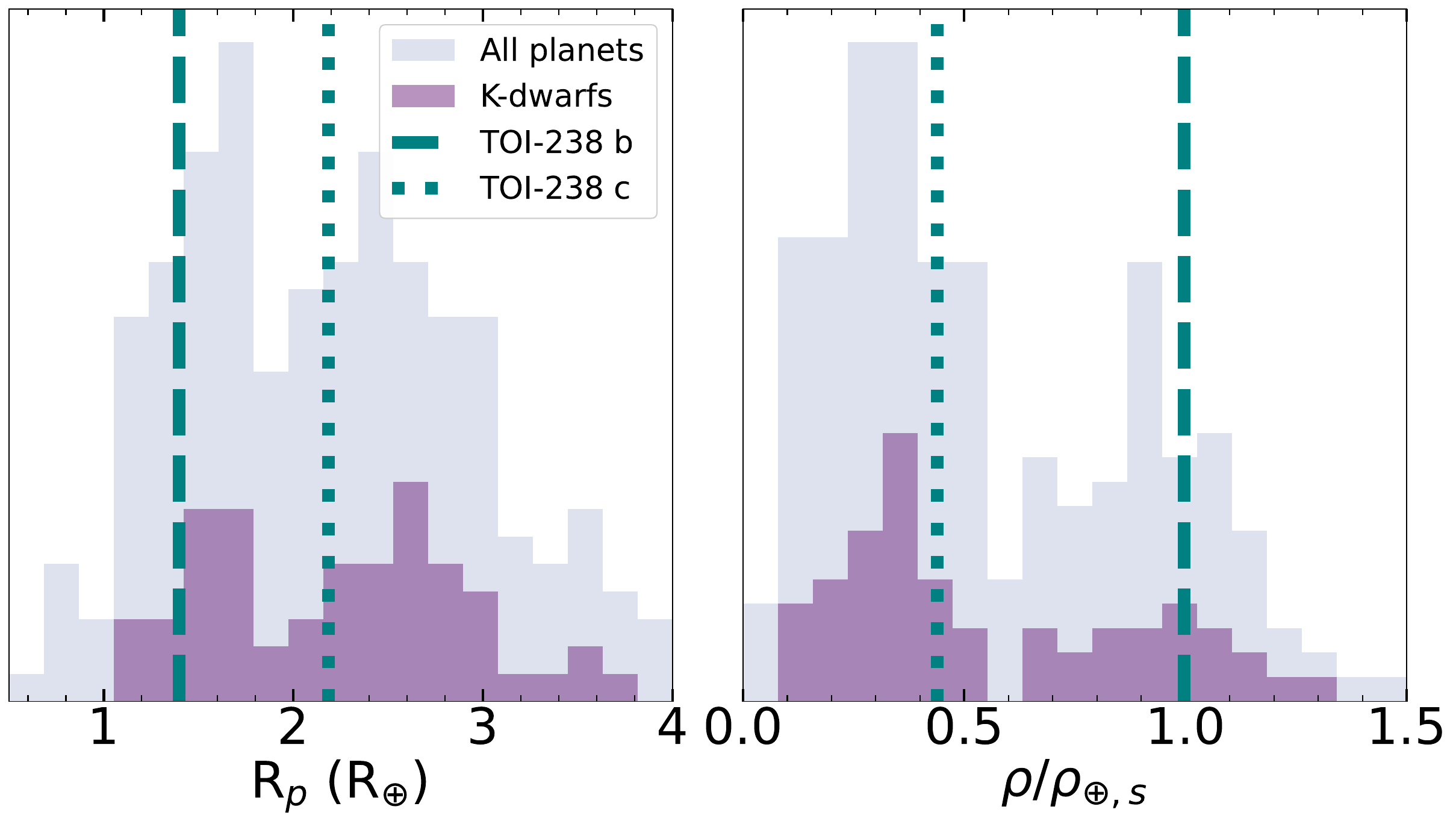}
	\caption{\textbf{Radii and density histograms} of transiting exoplanets with well established mass and radius from the NASA Exoplanet archive. Panel \textbf{a} shows the distribution of radii of all planets, with those orbiting K-dwarfs highlighted in purple. Panel \textbf{b} shows the distribution of densities normalized to the Earth-like density track. The teal vertical lines highlight the position of TOI--238 b and c.}
	\label{hist_rad_den}
\end{figure}

\begin{figure*}[ht]
    \includegraphics[width=18cm]{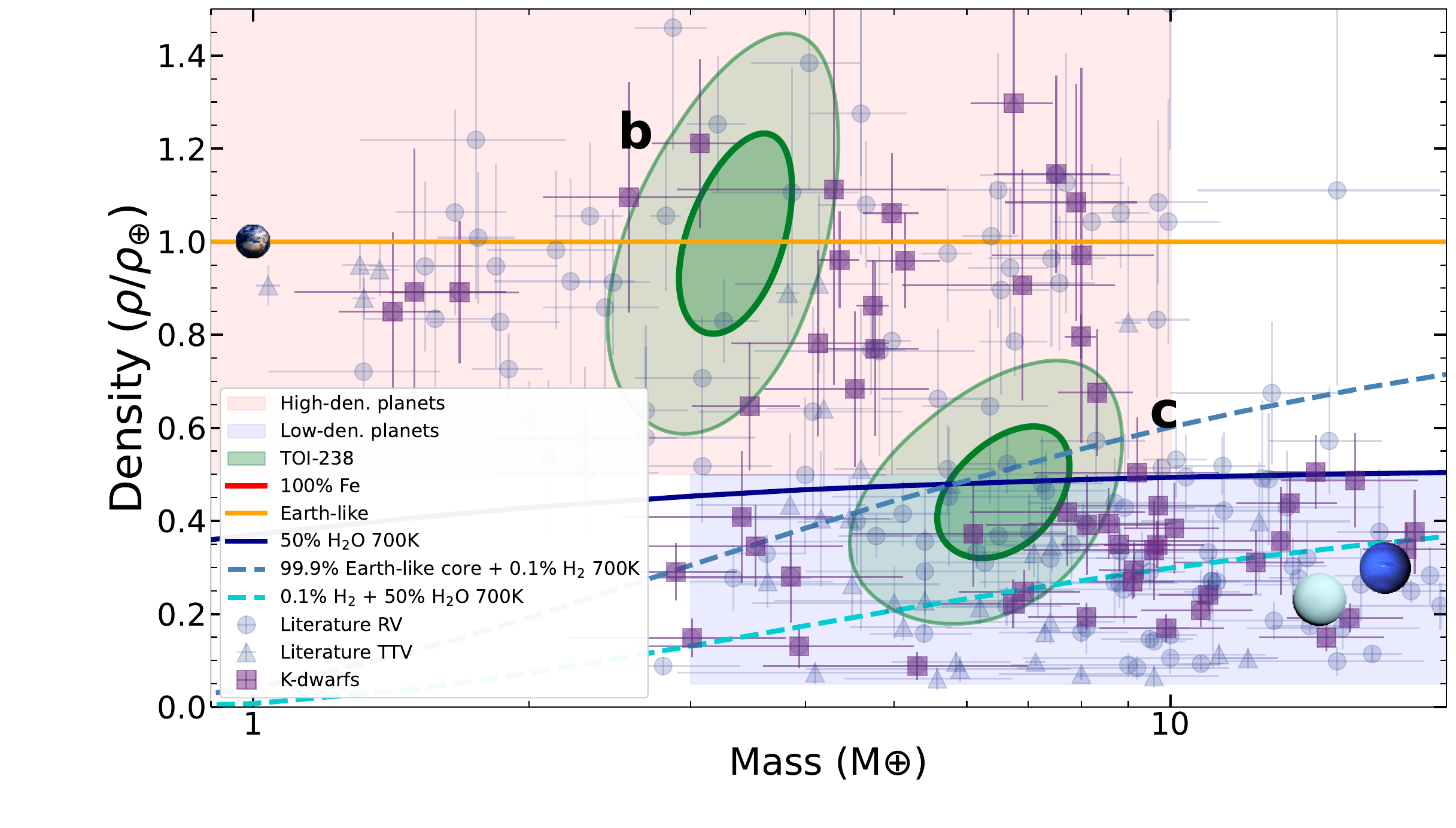}
	\caption{\textbf{Mass-density diagram} of transiting exoplanets with well established mass and radius from the NASA Exoplanet archive along with some theoretical composition curves. All densities are normalized to the Earth-like density track. The purple squares highlight the transiting planets orbiting K-dwarfs. The green blobs show the mass and radius posteriors (1- and 2-$\sigma$ contours) of  TOI--238 b and c. The shaded regions highlight the high-density and low-density groups of exoplanets. The planets of the solar system are included as reference.}
	\label{mass_den}
\end{figure*}

One of the goals of the ESPRESSO-GTO is probing the thresholds leading to significant evaporation of planetary envelopes. For this purpose, the project investigated systems in which small rocky planets and sub-Neptunes coexisted under intense irradiation. TOI--238 was originally not part of this sample, as there was a single identified planet candidate, but the discovery of the second planet grants it a place inside. To study the dependence of the density of these planets with the respective flux received from their host star, we calculated the incident flux from the parent star (S) as a function of the effective temperature, the radius of the star, and the semi-major axis. We estimate a S = 728 $\pm$ 78 S$_{\oplus}$ for TOI--238 b, and S = 58.3 $\pm$ 6.2 S$_{\oplus}$ for TOI--238 c. Both planets fit comfortably within the usual radius vs irradiation regions (e.g. \citealt{Toledo-Padron2020}). 

Figure~\ref{hist_rad_den} shows the histogram of the distribution of radii of well characterized transiting planets. The distribution is bimodal, showing the known radius valley \citep{Fulton2017}. TOI--238 b falls right in the peak of low-radius planets, while TOI--238 c is at the inner edge of the large-radius mode. The figure also shows the histograms of the densities, normalized to the density of the Earth-like composition track, similar to what \citet{LuquePalle2022} showed. Once again we find a bimodal distribution, separating high-density and low-density planets, with a somewhat deeper valley than what is seen in radius. The planets of TOI--238 fall one on each side. If we restrict the histogram to the densities of planets orbiting K-dwarfs, there is a gap at $\rho/\rho_{\oplus,s}$ $\sim$ 0.5, separating rocky planets from puffier planets. However, the amount of data is low enough for it to be a statistical anomaly. In both samples, it is difficult to identify a frontier between different types of low-density planets. Figure~\ref{mass_den} shows density of planets included in Figure~\ref{mass_rad}, normalized to the density of the mass-radius track of Earth-like composition. It is possible to visually identify the two populations hinted in the density histogram (highlighted as shaded regions). 

There are currently several hypotheses behind the origin of these distinct populations of low-mass exoplanets, such as photo-evaporation \citep{OwenWu2017}, core-powered mass loss \citep{Ginzburg2018} or migration after formation beyond the ice-line \citep{Mordasini2009}. Following \citet{OwenCamposEstrada2020} as described in \citet{Cloutier2020} we estimate that the system is consistent with photevaporation as long as planet c is larger than 1.04 $\pm$ 0.11 M$_{\oplus}$ and planet b is smaller than 12.2 $\pm$ 1.7 M$_{\oplus}$. In accordance with \citet{Cloutier2020} we estimate that the system is consistent with core-powered mass loss as long as planet c is heavier than $\sim$ 2.78 M$_{\oplus}$ and planet b is lighter than $\sim$ 8.05 M$_{\oplus}$. With their current mass measurements, the planetary system of TOI--238 is comfortably consistent with both photo-evaporation and core-powered mass loss scenarios. Additionally, the position of both planets within the mass-radius diagram is consistent with the observation of \citet{LuquePalle2022} that in the case of multi-planetary systems including water-worlds, the inner planet is usually a less massive rocky planet, which would be in agreement with the predictions from type I migration models \citep{Mordasini2009}.

We modelled the interior structure of the planets using the machine learning inference model \texttt{ExoMDN} \citep{Baumeister2023}. \texttt{ExoMDN} provides full inference for the interior structure of low-mass exoplanets, based on synthetic data created with the TATOOINE code \citep{Baumeister2020, MacKenzie2023}, using the mass, radius and equilibrium temperature of the planets. The determination of interior structures, based only on those parameters, is a degenerate problem. A potential path to overcome this issue is the use of the elemental abundances of the host star, which may be representative of the bulk abundances of the planet and its atmosphere \citep{Dorn2015}. This approach requires some assumptions about the history of formation and evolution of the system. Another possible approach is through the measurement of the fluid Love numbers. Fluid Love numbers describe the shape of a rotating planet in hydrostatic equilibrium. The second-degree Love number $k2$ depends on the interior density distribution \citep{Kellermann2018}. In a body with $k2$ = 0, the entire mass is concentrated in the center, while $k2$ = 1.5 corresponds to a fully homogeneous body. For a number of exoplanets, the fluid numbers are potentially measurable through second-order effects on the shape of the transit light curve \citep{Akinsanmi2023}. \texttt{ExoMDN} uses the $k2$ number to break some of the degeneracy between interior configurations.

For the case of the planets of TOI--238, it is not possible to measure the $k2$ number directly. Based on the position of the planets in the mass-radius diagram (see Fig.~\ref{mass_rad}), we assume the $k2$ number of Earth for planet b, with an uncertainty of 10\% (0.933 $\pm$ 0.093). This is roughly equivalent to assuming Earth composition. Planet c does not lie on the track of any planet with a known $k2$. We opt to not include this constraint in the analysis, which widens the set of potential interior configurations. To account for the uncertainties in the planetary parameters, we draw normal distributions with a $\sigma$ equal to the uncertainties of each parameter. We run 1000 predictions with 1000 samples each. 

For TOI--238 b we estimate a core-mass fraction of 0.43 $\pm$ 0.24, a mantle-mass fraction of 0.57 $\pm$ 0.24, and negligible mass fractions for water and the atmosphere. We estimate a core-radius fraction of 0.65 $\pm$ 0.18 and a mantle-radius fraction of 0.34 $\pm$ 0.17. We estimate radii fractions of water and volatiles smaller than 0.01. This interior structure is very similar to the interior structure of the Earth. This is not surprising, since the planet falls in the mass-radius track of the Earth and since we have used Earth's $k2$ number as a proxy of its composition. 

\begin{figure*}[ht]
\begin{center}
    \begin{minipage}[b]{0.45\textwidth}
    \includegraphics[width=8cm]{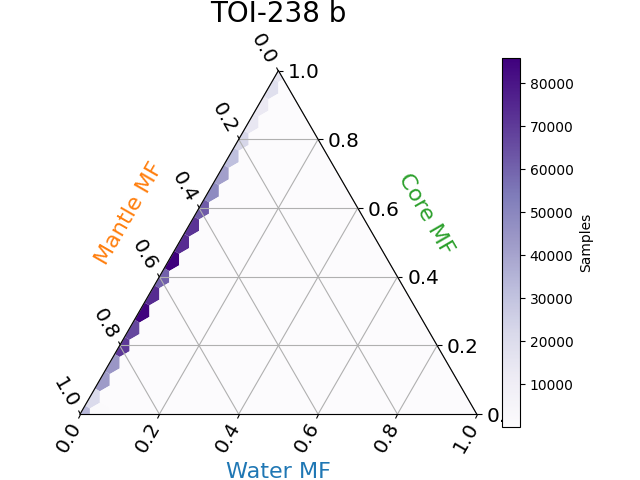}
    \end{minipage}    
    \begin{minipage}[b]{0.45\textwidth}
    \includegraphics[width=8cm]{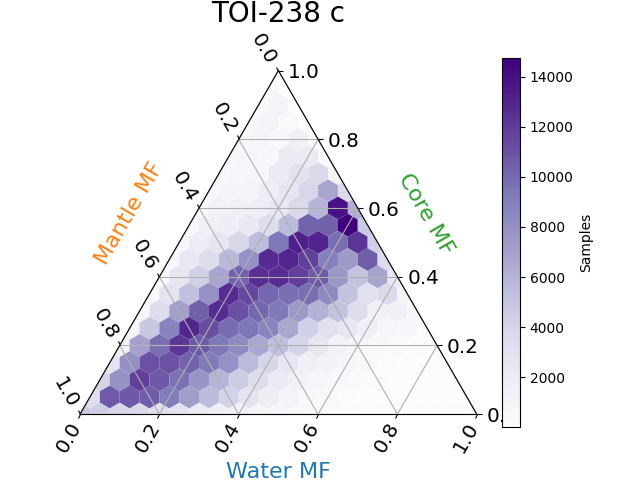}
    \end{minipage}

    \begin{minipage}[b]{0.45\textwidth}
    \includegraphics[width=8cm]{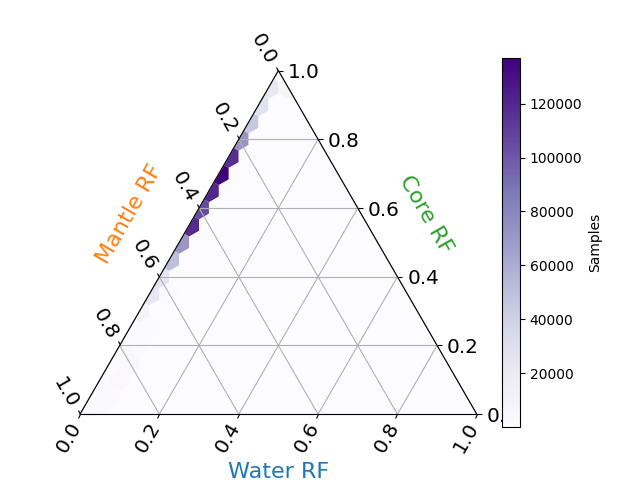}
    \end{minipage}    
    \begin{minipage}[b]{0.45\textwidth}
    \includegraphics[width=8cm]{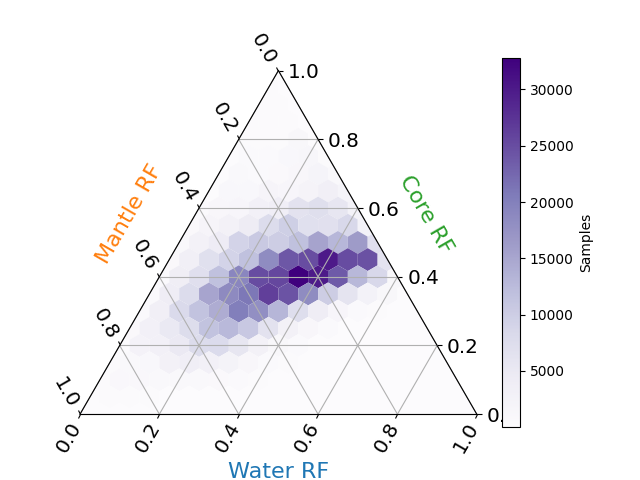}
    \end{minipage}
    
	\caption{\textbf{Ternary plots of the interior structure of the planets of TOI--238 }. The \textbf{top-left} panel shows the mass-fraction structure of planet b. The \textbf{bottom-left} panels shows its radius fraction configuration. The \textbf{top-right} panel shows the mass-fraction structure of planet c. The \textbf{bottom-right} panels shows its radius fraction configuration.}
	\label{interior}
\end{center} 
\end{figure*}

For TOI--238 c we compute a core-mass fraction 0.30 $\pm$ 0.18, a mantle-mass fraction of 0.39 $\pm$ 0.25, a water-mass fraction of 0.27 $\pm$ 0.15 and a mass fraction of volatile elements $\textless$ 0.018 (99.7\%).  We estimate a core-radius fraction 0.38 $\pm$ 0.10, a mantle-radius fraction of 0.25 $\pm$ 0.17, a water-radius fraction of 0.26 $\pm$ 0.12, and atmosphere-radius fraction of 0.10 $\pm$ 0.06. The large uncertainty in the parameter allow for very different configurations. For a more sophisticated analysis, it would be important to obtain a more precise determination of the radius, which would require additional high-precision photometric observations. Figure~\ref{interior} shows the ternary plots of the core-mantle-water mass and radii fraction configurations for both planets. 

\begin{figure}[ht]
    \includegraphics[width=9cm]{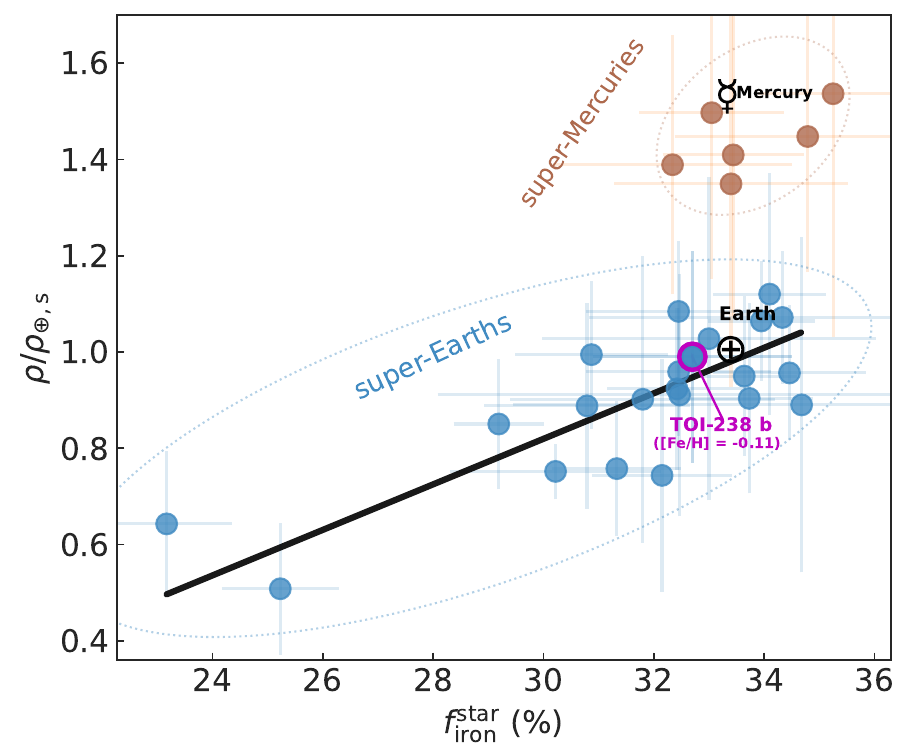}
	\caption{\textbf{Planet density normalized by the expected density of an Earth-like composition} versus the
    estimated iron-to-silicate mass fraction of the protoplanetary disk for the sample studied in \citet{Adibekyan-21} and for TOI-238 b. The Earth and Mercury are indicated with their respective symbols in black. The black line indicates the original correlation for the super-Earths.}
	\label{vardan_plot}
\end{figure}

Recently, \citet{Adibekyan-21} demonstrated that the scaled density of small-sized (R$_{p}$ $\textless$ 2 R$_{\oplus}$) rocky planets correlates with the iron-to-silicate mass fraction ($f^{star}_{iron}$) of the planet building blocks, as estimated from their host star abundances. Using the abundances of Mg, Si, and Fe as proxies for the composition of the protoplanetary disk, we calculated $f^{star}_{iron}$ \citep{Santos-15, Santos-17} to be 32.7 $\pm$ 1.8\%, which is similar to thevalue determined from the composition of the Sun ($f^{sun}_{iron}$ = 33.2 $\pm$ 1.7\%). The scaled density of TOI-238 b, obtained from its radius and mass and normalized to the density of a planet with the same mass but Earth-like composition \citep{Dorn-17}, was calculated to be 0.99 $\pm$ 0.22. With these parameters, TOI--238 b aligns well with the regression line established in \citet{Adibekyan-21} (see Fig.~\ref{vardan_plot}).

The true structure of planets similar to TOI--238 c is very difficult to establish by their mass and radius alone. These planets lie in a region of the parameter space in which several composition tracks are close to each other. The mass-radius configuration TOI--238 c is equally consistent with water-rich world with no extended volatile atmosphere and a rocky core with a small H$_{2}$ layer. To fully disentangle these scenarios requires the characterisation of the atmosphere of the planet. 

Following \citet{Kempton2018}, we compute the transmission spectroscopy metric (TSM) and the emission spetroscopy metric (ESM). The TSM is a metric proportional to the expected transmission spectroscopy S/N ratio and is based on the strength of spectral features and the brightness of the host star. With our results, we estimate a TSM of 5.2 $\pm$ 1.1 for planet b, and 35 $\pm$ 11 for planet c. With these numbers, and considering the cutoffs by \citet{Kempton2018}, we expect the analysis of the atmospheres of either of the planets through transmission spectroscopy to be a challenging effort. The ESM is a metric proportional to the expected S/N ratio of a JWST secondary eclipse at mid-IR wavelengths. We estimate a ESM of 3.03 $\pm$ 0.32 for planet b, and 1.82 $\pm$ 0.30 for planet c. Once again, none of them are truly favourable candidates. 

We compared the results obtained for the TOI--238 system with the values obtained for all planets shown in Fig.~\ref{mass_rad}, using the planetary and stellar parameters from the Planetary Systems Composite Data table from the NASA Exoplanet archive. Figure~\ref{tsm_esm} in Appendix~\ref{append_additional} shows the TSM and ESM of TOI--238 compared to known low-mass transiting planets with radii and masses measured with a precision of 33\%, or better. TOI--238 c ranks average in TSM compared to most planets of its class. Both planets rank average in ESM compared to other similar exoplanets.

\subsection{Stellar activity of TOI--238}

TOI--238 is a K2-dwarf for which we measured a log$_{10}$ (R$_{\rm HK}^{'})$ of -- 4.74 $\pm$ 0.27. This corresponds to an expected rotation period of 28 $\pm$ 16 days. To model the activity-induced RV  variations without leaving structures that could be related to stellar activity, we needed to use a combination of two multi-dimensional GP-terms: one for FWHM+RV and one for BIS+RV. We obtained different timescales in both cases, although not very different. 

In our global analysis, we measured GP periods of 24.42 $\pm$ 0.86 days and 26.64 $\pm$ 0.42 days for the combinations of FWHM+RV, and BIS+RV, respectively. Both these periods are consistent with the expectations. The two measurements are within 2$\sigma$ of each other, which could indicate that the difference is related to the data and/or modelling procedure, and not to a true astrophysical difference. We measured different timescales of evolution ($L$). For the FWHM+RV, we measure a $L = 35.2^{+6.9}_{-5.8}$ days, while for the BIS+RV we measure $L = 
 48.9^{+10.9}_{-8.9}$ days. The signal in the bisector seems more stable than the signal in FWHM, but the differences in evolutionary timescale are within $1\sigma$ of each other. The posterior distributions are fairly similar to the imposed prior (see Fig.~\ref{posterior_2pk}), which suggests that the data does not carry enough information to constrain the evolutionary timescale. We tested the behaviour of modelling with wider priors. The overall results of the model were similar. However, in some cases the GP would heavily overfit the data. We also measured small differences in the length scale of the periodic component ($\omega$). For the FWHM+RV we measure a $\omega$ of 0.71$^{+0.29}_{-0.20}$, while for the BIS+RV we measure a $\omega$ of 0.12$^{+0.21}_{-0.10}$. While the differences are not large, all of them point to the variations measured in the BIS being more stable than those measured in the FWHM, the same as their counterpart signals in the RV data. This picture fits with the data seen in Figs.~\ref{fwhm_model} and ~\ref{bis_model}, in which the FWHM shows larger changes of amplitude and shape over the course of the observing campaigns. This is particularly noticeable in the last year of observations. The origin of this difference is, however, not so clear. One possible explanation is that we are measuring the effect of different phenomena on the surface of the star, with different timescales of evolution. However, given that the data sampling is not always optimal to follow an irregularly-shaped $\sim$ 25 days variations, it might also be that the sampling itself is creating apparent variations by sometimes measuring the data at times of maximum variation and sometimes at times of minimum variation. The signals in FWHM and BIS having a lag with respect to one another could be the culprit behind the observed differences in shape. We estimate the stellar rotation period by averaging the two measurements we obtained in the global model. We obtain a measurement of the P$_{\rm rot}$ of 25.51 $\pm$ 0.49 days. This value is fully consistent with the measured log$_{10}$ (R$_{\rm HK}^{'})$ of -- 4.74 $\pm$ 0.27. 

\begin{figure}[ht]
    \includegraphics[width=9cm]{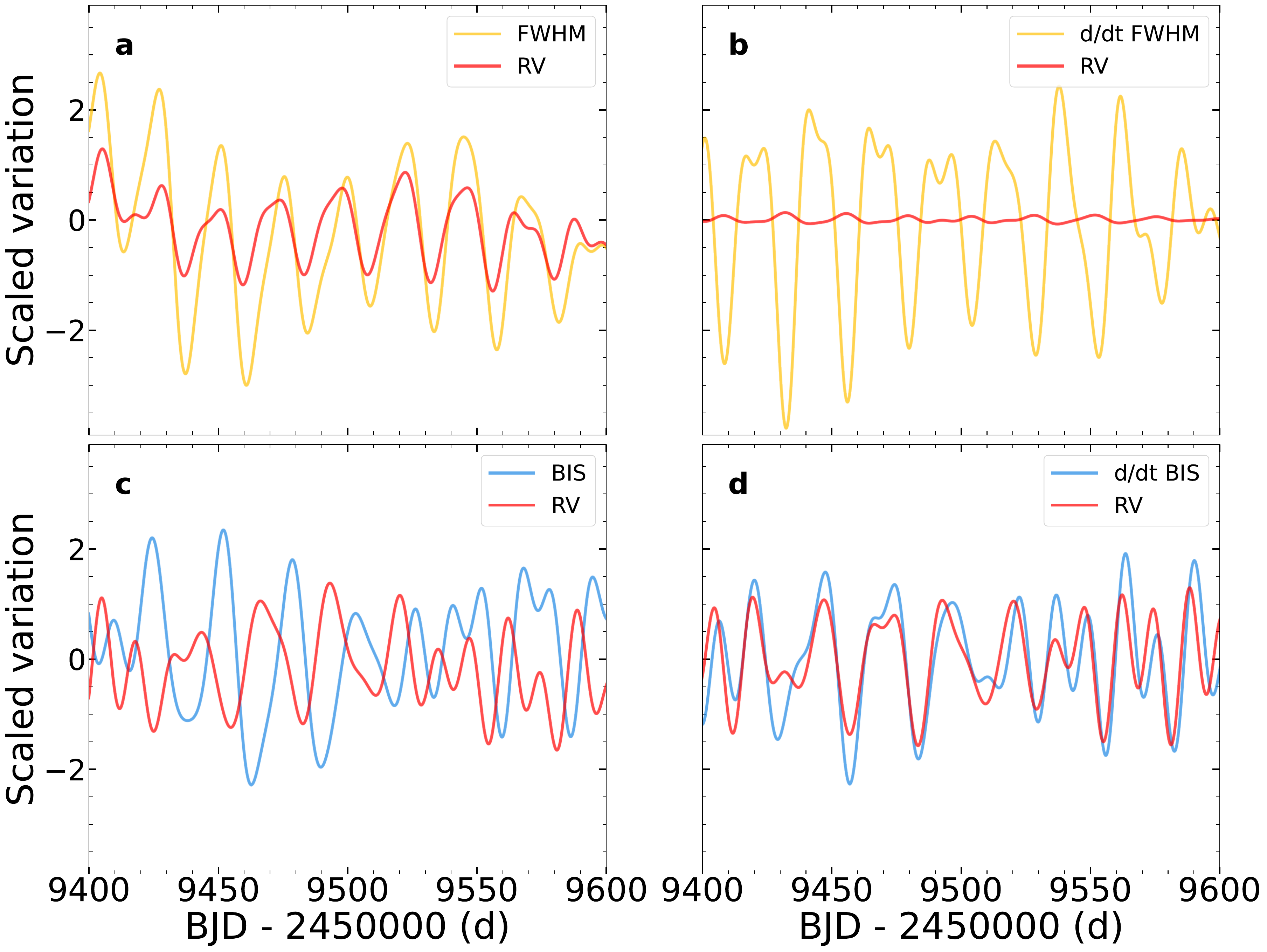}
	\caption{\textbf{Individual components of the activity model.} Panel \textbf{a} shows the scaled variations of the FWHM model and their effect in RV. Panel \textbf{b} show the scaled variations of the gradient of the FWHM and their effect in RV. Panels \textbf{c} and \textbf{d} show the same for the bisector model.}
	\label{activity_decomp}
\end{figure}

We measure a stellar activity standard deviation (stdev) of 22.7$^{+6.9}_{-4.5}$ m$\cdot$s$^{-1}$ in FWHM and a stdev of 5.5$^{+1.3}_{-1.2}$ m$\cdot$s$^{-1}$ in the BIS time-series. In the RV data, activity-induced RV has stdev of 4.5 m$\cdot$s$^{-1}$. The value is consistent with what would be expected for a star with a log$_{10}$ (R$_{\rm HK}^{'})$ of -- 4.74 \citep{SuarezMascareno2017}, however, the large uncertainty in the log$_{10}$ (R$_{\rm HK}^{'})$ and the differences in methodology make the numbers difficult to compare (a standard deviation, as measured with a GP is not strictly the same as an amplitude, measured with a sinusoidal model). We find the activity-induced RV stdev related to the variations correlated to the BIS to be the dominant contribution, with 3.3 m$\cdot$s$^{-1}$, followed by the variations correlated with the gradient of the BIS, with 2.9 m$\cdot$s$^{-1}$. The variations linked to the BIS are anti-correlated, which is expected in the case of spot-dominated variations. The variations correlated with the FWHM account for 2.1 m$\cdot$s$^{-1}$. There are no significant variations correlated with the gradient of the FWHM. Figure~\ref{activity_decomp} shows the scaled variations of the FWHM, BIS, their gradients and the associated RV variations. FWHM, d/dt FWHM, BIS and d/dt BIS are scaled to their variance. All individual RV components are scaled to the variance of the global RV, to show the comparison between them. The different behaviour of the RV variations linked with FWHM and BIS, coupled with the slight difference in timescales, hint at the two indicators tracking different activity regions. The anti-correlation between the BIS and RV, combined with a correlation between its gradient and the RV, are typical signs of spot-dominated variations \citep{Queloz2001, Aigrain2012}. However, the relation between the FWHM and RV variations is more reminescent of plague-dominated variations \citep{Dumusque2014}.

We do not find evidence of the presence of a magnetic cycle. However, the current dataset might not span long enough to allow such a detection. \citet{Sairam2022} showed that the $P_{\rm cyc}$ vs $P_{\rm rot}$ relation showed in \citet{Masca2016} could be used to provide a prediction on the cycle period of the star. Using the slope and zero point identified for FGK stars, we obtain the relationship $log_{10} (P_{\rm cyc}/P_{\rm rot}) = 3.22 + 0.89 \cdot log_{10} (1/P_{\rm rot})$. We obtain a prediction for the stellar cycle period of 2340 $\pm$ 410 days (6.41 $\pm$ 1.12 years). This period is more than three times the baseline of observations, which would explain the lack of a detection. The baseline is, however, long enough to hint towards the lack of a large-amplitude cycle, given that we do not find a significant slope in any of the activity indicators.

\section{Conclusions}

We performed a high-precision radial velocity campaign on the system TOI--238, which hosts the planet candidate TOI--238.01, using the high-resolution spectrographs ESPRESSO and HARPS. We confirmed the presence of the planet, both in TESS photometry and radial velocity. TOI--238 b is a super-Earth with a radius of 1.402$^{+0.084}_{-0.086}$ R$_{\oplus}$ and a mass of 3.40$^{+0.46}_{-0.45}$ M$_{\oplus}$. It orbits at a distance of 0.02118 $\pm$ 0.00038 au of its host star, with an orbital period of 1.2730988 $\pm$ 0.0000029 days. It has an equilibrium temperature of 1311 K $\pm$ 28 and receives a flux 728 $\pm$ 78 times Earth's irradiation. We estimate the planet's internal structure to be consistent with Earth's internal structure. We estimate a core-mass fraction of 0.43 $\pm$ 0.24, a mantle-mass fraction of 0.57 $\pm$ 0.24, and negligible mass fractions for water and the atmosphere.

We discovered a second planet, both in TESS photometry and radial velocity. TOI--238 c is most likely a water-world or a rocky core with a small volatile envelope. It has a radius of 2.18$^{+0.18}_{-0.18}$ R$_{\oplus}$ and a mass of 6.7$^{+1.1}_{-1.1}$ M$_{\oplus}$. It orbits at a distance of 0.0749 $\pm$ 0.0013 au of its host star, with an orbital period of 8.465652 $\pm$ 0.000031 days. It has an equilibrium temperature of 696 $\pm$ 15 K and receives a flux of 58.3 $\pm$ 6.2 S$_{\oplus}$. With the current data available, the interior structure of this planet is difficult to constrain. A more precise radius and, most likely, a study of its atmosphere are needed to improve the characterisation of the planet.  

The RV data gives hints of a potential non-transiting planet. However, the current dataset does not allow us to properly disentangle the signal from the activity-induced RV signal. 

The planets lie at both sides of the radius and density valleys. The current status of both planets seems to be compatible with both photo-evaporation and core-powered mass loss, and with migration of the outer planet after having formed beyond the ice line. 

We estimate the transit spectroscopy metric (TSM) of both planets. While it should be possible to study their atmospheres via transmission spectroscopy with JWST, their comparatively low TSM indicates it would be challenging to obtain any significant detection of the planetary features. 

We study the stellar activity variations of TOI--238. We measured a rotation period of 25.51 $\pm$ 0.49 days, and the activity-induced RV signal to have a standard deviation of 4.5 m$\cdot$s$^{-1}$, which is consistent with the expectations of a star of its spectral type and its mean level of chromospheric activity. We find that the largest components of the activity-induced RV signal are correlated with the bisector span of the CCF and its gradient. We do not detect the presence of a magnetic cycle. However, we estimate that if there is a magnetic cycle it would have a period of $\sim$ 6.4 years. 

\section*{Acknowledgements}
We thank Prof. Didier Queloz for providing us the information on the creation of the original HARPS masks. A.S.M thanks Dr. Emily L. Hunt for the advice on using VS Code, and the suggestions on its configuration. 

A.S.M., V.M.P.,  J.I.G.H., and R.R. acknowledge financial support from the Spanish Ministry of Science and Innovation (MICINN) project PID2020-117493GB-I00 and from the Government of the Canary Islands project ProID2020010129. 

The project that gave rise to these results received the support of a fellowship from the ”la Caixa” Foundation (ID 100010434). The fellowship code is LCF/BQ/DI23/11990071.

A.C.-G. is funded by the Spanish Ministry of Science through MCIN/AEI/10.13039/501100011033 grant PID2019-107061GB-C61. 

F.P.E. and C.L.O. would like to acknowledge the Swiss National Science Foundation (SNSF) for supporting research with ESPRESSO through the SNSF grants nr. 140649, 152721, 166227, 184618 and 215190. The ESPRESSO Instrument Project was partially funded through SNSF's FLARE Programme for large infrastructures. 

Co-funded by the European Union (ERC, FIERCE, 101052347). Views and opinions expressed are however those of the author(s) only and do not necessarily reflect those of the European Union or the European Research Council. Neither the European Union nor the granting authority can be held responsible for them. This work was supported by FCT - Funda\c{c}{\~a}o para a Ci\^{e}ncia e a Tecnologia through national funds and by FEDER through COMPETE2020 - Programa Operacional Competitividade e Internacionaliza\c{C}{\~a}o by these grants: UIDB/04434/2020; UIDP/04434/2020; 2022.06962.PTDC. 

This work was financed by Portuguese funds through FCT - Funda\c c\~ao para a Ci\^encia e a Tecnologia in the framework of the project 2022.04048.PTDC (Phi in the Sky, DOI 10.54499/2022.04048.PTDC). CJM also acknowledges FCT and POCH/FSE (EC) support through Investigador FCT Contract 2021.01214.CEECIND/CP1658/CT0001.

A.M.S acknowledges support from the Fundação para a Ciência e a Tecnologia (FCT) through the Fellowship 2020.05387.BD.

R. A. is a Trottier Postdoctoral Fellow and acknowledges support from the Trottier Family Foundation. This work was supported in part through a grant from the Fonds de Recherche du Qu{\'e}bec - Nature et Technologies (FRQNT). This work was funded by the Institut Trottier de Recherche sur les Exoplan{\'e}tes (iREx). This work has been carried out within the framework of the National Centre of Competence in Research PlanetS supported by the Swiss National Science Foundation. The authors acknowledge the financial support of the SNSF. 

D.J.A. acknowledges funding from the UKRI, (Grants ST/X001121/1, EP/X027562/1).

K.A.C. and C.N.W. acknowledge support from the TESS mission via subaward s3449 from MIT.

This work is based on data obtained via the  HARPS public database at the European Southern Observatory (ESO). We are grateful to all the observers of the following ESO projects, whose data we are using: 072.C-0488, 085.C-0019, 183.C-0972, and 191.C-087. We are grateful to the crews at the ESO observatories of Paranal and La Silla. 

We acknowledge the use of public TESS data from pipelines at the TESS Science Office and at the TESS Science Processing Operations Center. Resources supporting this work were provided by the NASA High-End Computing (HEC) Program through the NASA Advanced Supercomputing (NAS) Division at Ames Research Center for the production of the SPOC data products.

This paper made use of data collected by the TESS mission and are publicly available from the Mikulski Archive for Space Telescopes (MAST) operated by the Space Telescope Science Institute (STScI).

This research has made use of the Exoplanet Follow-up Observation Program (ExoFOP; DOI: 10.26134/ExoFOP5) website, which is operated by the California Institute of Technology, under contract with the National Aeronautics and Space Administration under the Exoplanet Exploration Program. Funding for the TESS mission is provided by NASA's Science Mission Directorate.

Part of the LCOGT telescope time was granted by NOIRLab through the Mid-Scale Innovations Program (MSIP). MSIP is funded by NSF. 

This research has made extensive use of the SIMBAD database, operated at CDS, Strasbourg, France, and NASA's Astrophysics Data System. 

This research has made use of the NASA Exoplanet Archive, which is operated by the California Institute of Technology, under contract with the National Aeronautics and Space Administration under the Exoplanet Exploration Program. This work makes use of observations from the LCOGT network.

The manuscript was written using \texttt{Overleaf}, \texttt{Texmaker} and \texttt{VS Code}. 
Main analysis performed in \texttt{Python3} \citep{Python3} running on \texttt{Ubuntu} \citep{Ubuntu} systems and \texttt{MS. Windows} running the \texttt{Windows subsystem for Linux (WLS)}.
Extensive use of the DACE platform \footnote{\url{https://dace.unige.ch/}}
Extensive usage of \texttt{Numpy} \citep{Numpy}.
Extensive usage of \texttt{Scipy} \citep{Scipy}.
\texttt{AstroimageJ} \citep{Collins:2017}.
\texttt{TAPIR} \citep{Jensen:2013}.
All figures built with \texttt{Matplotlib} \citep{Matplotlib}.

The bulk of the analysis was performed on desktop PC with an AMD Ryzen$^{\rm TM}$ 9 3950X (16 cores, 2 threads per core, 3.5--4.7 GHz) and a server hosting 2x Intel$^{\rm (R)}$ Xeon$^{\rm (R)}$ Gold 5218 (2x16 cores, 2 threads per core, 2.3--3.9 GHz). Additionally, this article used flash storage and GPU computing resources as Indefeasible Computer Rights (ICRs) being commissioned at the ASTRO POC project that Light Bridges will operate in the Island of Tenerife, Canary Islands (Spain). The ICRs used for this research were provided by Light Bridges in cooperation with Hewlett Packard Enterprise (HPE) and VAST DATA.

% WARNING
%-------------------------------------------------------------------
% Please note that we have included the references to the file aa.dem in
% order to compile it, but we ask you to:
%
% - use BibTeX with the regular commands:
%   \bibliographystyle{aa} % style aa.bst
%   \bibliography{Yourfile} % your references Yourfile.bib
%
% - join the .bib files when you upload your source files
%-------------------------------------------------------------------
\bibliography{biblio_toi238}

\begin{appendix}
\section{Comparison of RV extractions} \label{append_rv}

We performed the RV extraction with three different algorithms. We used the CCF~\citep{Fellgett1953,Baranne1996} built-in into the ESPRESSO and HARPS DRS~\citep{Mayor2003,Pepe2021} and two template-matching algorithms~\citep{Bouchy2001,AngladaEscude2012}, SERVAL~\citep{Zechmeister2018} and \texttt{S-BART}~\citep{SilvaAM2022}.

The ESPRESSO CCF velocities were obtained using the default K2 mask available in the ESPRESSO DRS. This mask is based on a high signal-to-noise spectrum of the low-activity K2V star HD 144628 and uses 3470 individual lines to compute the RV. For the HARPS CCF velocities we used the K5 mask available in the HARPS DRS, built using a synthetic spectrum following \citet{Queloz1995IAUS} and \citet{Pepe2002}, which uses 5129 individual lines. The CCF is computed order by order and summed to obtain the final CCF. In both cases the velocities were compute by fitting a Gaussian function to the CCF and computing its centre. We measured an RV RMS of 6.3 m$\cdot$s$^{-1}$ and 5.22 m$\cdot$s$^{-1}$ for HARPS and ESPRESSO data, respectively. We measure RV internal uncertainties of 1.6 m$\cdot$s$^{-1}$ and 0.73 m$\cdot$s$^{-1}$, respectively.

SERVAL builds a high signal-to-noise template by co-adding all the existing observations, and performs a least-squares minimisation of each observed spectrum against the template, yielding a measure of the Doppler shift and its uncertainty. Each order is fitted separately. The final measurement is a weighted mean of the individual RVs of all the orders. SERVAL automatically masks telluric features deeper than 5$\%$ and sky emission lines using a predefined line list. We measured an RV RMS of 6.1 m$\cdot$s$^{-1}$ and 8.09 m$\cdot$s$^{-1}$ for HARPS and ESPRESSO data, respectively. We measure RV internal uncertainties of 1.2 m$\cdot$s$^{-1}$ and 0.66 m$\cdot$s$^{-1}$, respectively.  

\texttt{S-BART}, similar to SERVAL, builds a high signal-to-noise template by co-adding all the existing observations. Then, unlike usual template-matching algorithm, \texttt{S-BART} uses a single RV shift to describe simultaneously the RV differences between all orders of a given spectrum and the template. The algorithm estimates the posterior distribution for the RV shifts after marginalising with respect to a linear model for the continuum levels of the spectra and template, using a Laplace approximation. From the Laplace approximation to the posterior distribution, the mean is used as the estimated RV and the standard deviation as the estimated RV uncertainty for each epoch. Using \texttt{S-BART}, we measured an RV RMS of 6.4 m$\cdot$s$^{-1}$ and 5.65 m$\cdot$s$^{-1}$ for HARPS and ESPRESSO data, respectively. We measure RV internal uncertainties of 1.1 m$\cdot$s$^{-1}$ and 0.58 m$\cdot$s$^{-1}$, respectively.  

\begin{figure}[ht]
	\includegraphics[width=9cm]{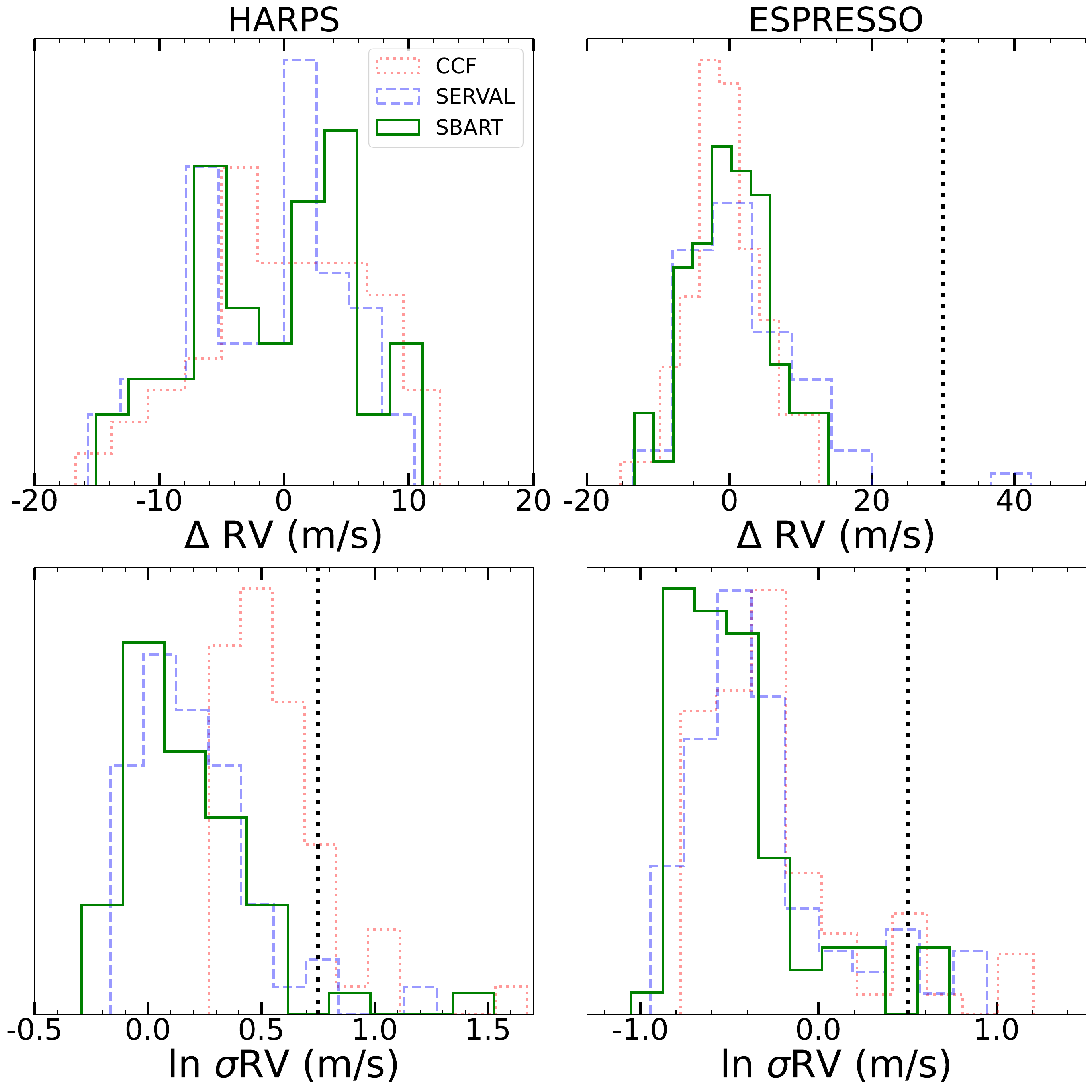}
	\caption{\textbf{Histogram of the RV and error measurements for all RV extractions tested.} \textit{Top panels} show the distribution of RV measurements obtained for HARPS (left) and ESPRESSO (right). \textit{Bottom panels} show the distribution of internal errors for HARPS (left) and ESPRESSO (right), in logarithmic scale. The vertical lines show the thresholds used for the outlier rejection procedure.}
	\label{rv_hist}
\end{figure}

Figure~\ref{rv_hist} shows the distribution of RV measurements obtained with the three algorithms, as well as the distribution of uncertainties measured, in logarithmic scale. We use the distributions shown in ~\ref{rv_hist} to reject some poor quality measurements. First we reject the large outlier in the SERVAL data and then the few spectra where the RV uncertainty deviate from the distribution of RV uncertainties in the \texttt{S-BART} measurements, which shows the most compact distribution of uncertainties. The vertical dotted lines in each panel shows the threshold applied. We reject 2 HARPS spectra and 5 ESPRESSO spectra. After the data rejection we  measure an RV RMS for the HARPS data of 6.3 m$\cdot$s$^{-1}$, 6.1 m$\cdot$s$^{-1}$ and 6.2 m$\cdot$s$^{-1}$, for the CCF, SERVAL and \texttt{S-BART} measurements, respectively. We measure an RV RMS for the  ESPRESSO data of 5.30 m$\cdot$s$^{-1}$, 6.56 m$\cdot$s$^{-1}$ and 5.55 m$\cdot$s$^{-1}$, for the CCF, SERVAL and \texttt{S-BART} measurements, respectively. Table ~\ref{tab:rv_stats} shows the complete set of statistics of the RV data. 

\begin{table}
\begin{center}
\caption{RV dispersion and typical error of the raw RVs and the RVs after the sigma-clip. \label{tab:rv_stats}}
\begin{tabular}[center]{l c c c c}
\hline
Instrument & N. Spec & CCF & SERVAL & \texttt{S-BART}  \\ 
           &         & [m$\cdot$s$^{-1}$] & [m$\cdot$s$^{-1}$] & [m$\cdot$s$^{-1}$] \\ \hline
\textit{Raw RVs} \\
RV RMS HARPS & 50& 6.3 & 6.1& 6.4  \\
$\sigma$ RV HARPS & 50 & 1.6 & 1.2 & 1.1 \\
RV RMS ESPRESSO & 73& 5.22 & 8.09 & 5.66  \\
$\sigma$ RV ESPRESSO & 73 & 0.73 & 0.66 & 0.58 \\ \\
\textit{Sigma-clipped} \\
RV RMS HARPS & 48 & 6.3 & 6.1 & 6.2  \\
$\sigma$ RV HARPS & 48 & 1.6 & 1.1 & 1.1 \\
RV RMS ESPRESSO & 68 & 5.30 & 6.56 & 5.55  \\
$\sigma$ RV ESPRESSO & 68 & 0.72 & 0.66 & 0.58 \\
\hline
\end{tabular}
\end{center}
\end{table}

\begin{figure}[ht]
	\includegraphics[width=9cm]{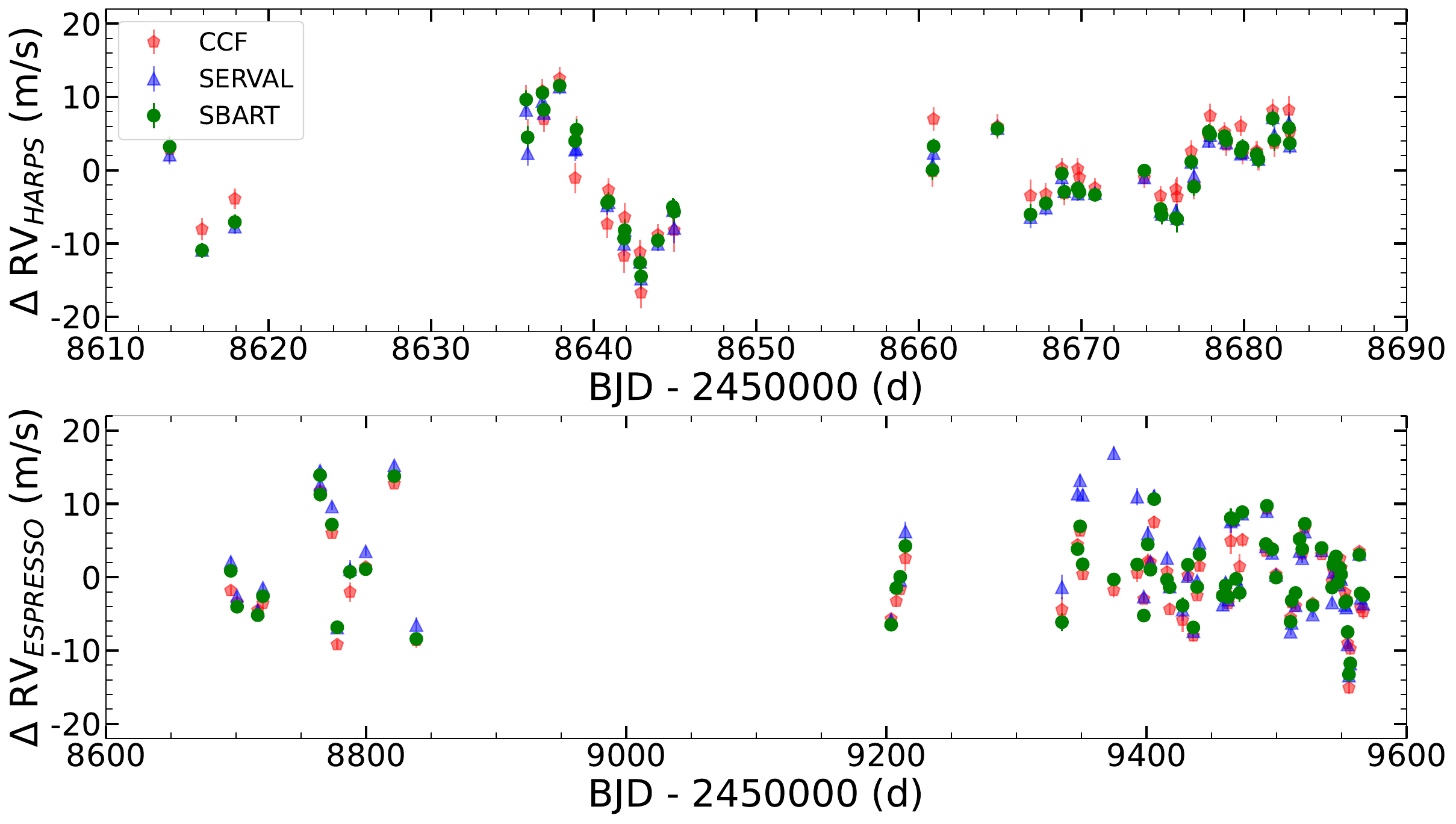}
	\caption{\textbf{Comparison of the velocities coming from the three extraction algorithms as a function of time.} \textit{Top panel} shows the HARPS RV measurements from the CCF method (red), SERVAL (blue) and SBART (green). \textit{Bottom panel} show the same but for the ESPRESSO measurements.}
	\label{rv_comp}
\end{figure}

\begin{figure}[ht]
	\includegraphics[width=9cm]{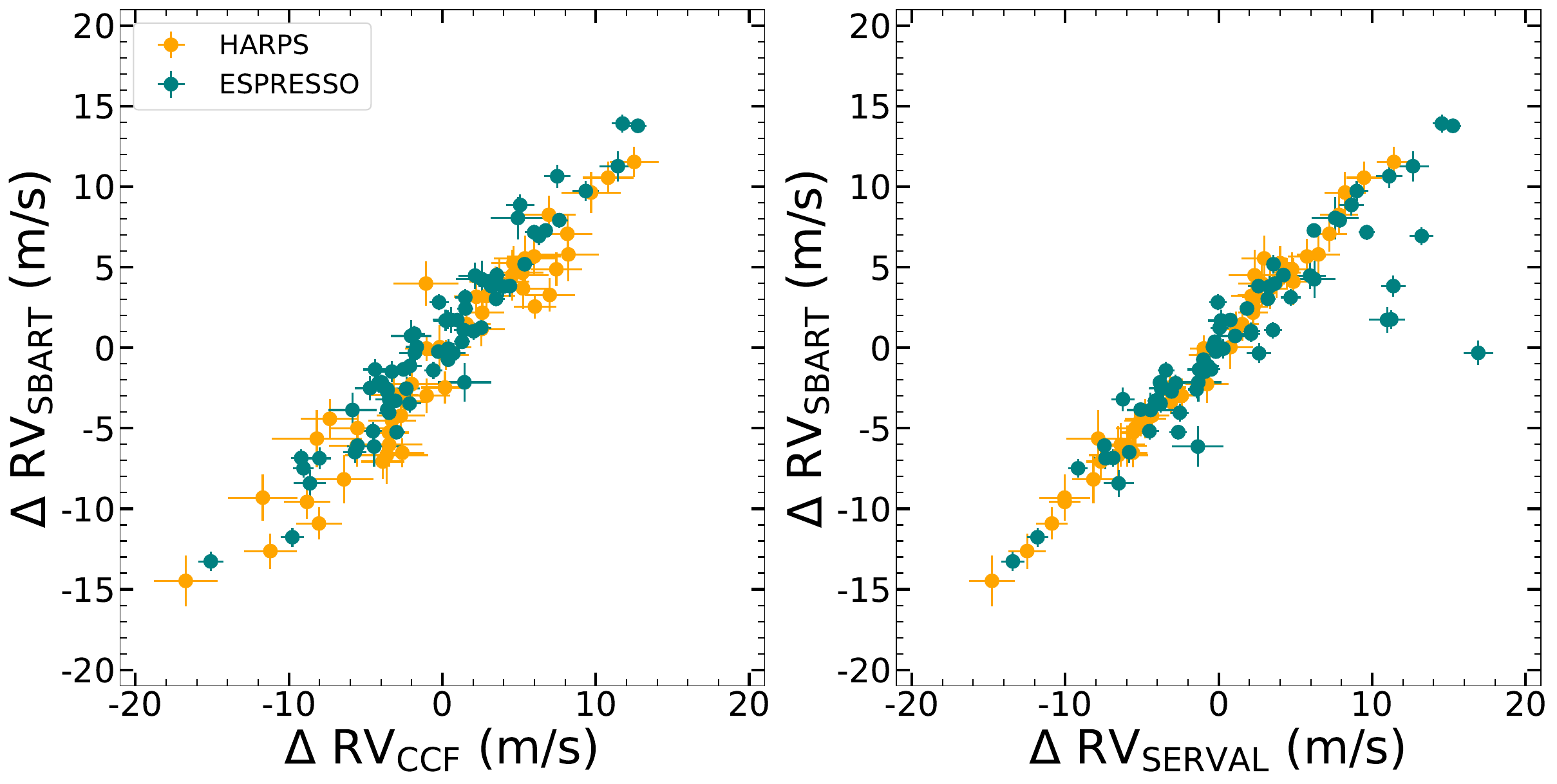}
	\caption{\textbf{Comparison of the velocities coming from SBART with the CCF and SERVAL.} \textit{Left panel} shows the HARPS (orange) and ESPRESSO (teal) RVs computed with SBART compared to the CCF RVs. \textit{Right panel} shows the HARPS (orange) and ESPRESSO (teal) RVs computed with SBART compared to the SERVAL RVs.}
	\label{rv_comp_2}
\end{figure}

Figures~\ref{rv_comp} and ~\ref{rv_comp_2} show the visual comparison of the three RV datasets. Figure~\ref{rv_comp} shows the RV measurements of HARPS and ESPRESSO as a function of time. The three algorithms produce virtually identical measurements except for a few points in which the SERVAL measurements deviate from the rest. This becomes more evident in figure~\ref{rv_comp_2}. The \texttt{S-BART} and CCF velocities show good correspondence, with some noise. The \texttt{S-BART} and SERVAL velocities show a tighter correspondence, except for a few clear outliers. 

The velocities obtained with all three algorithms show similar distributions, with the SERVAL RVs having a few big outlier in the ESPRESSO RVs, which explains the significantly larger RMS. Similarly to the case of \citet{Faria2021}, the \texttt{S-BART} measurements show the smaller uncertainties for both instruments, while having a very similar distribution of RVs. Based on the improved formal precision of the measurements, we opt for the \texttt{S-BART} RVs as our main RV measurements, although we acknowledge that any of the three methods would provide similar results. 

\section{Effect of tellurics in the RV data}
\label{append_tellurics}

We investigated the effect of tellurics in the radial velocity measurements by comparing the measurements we obtained with measurements corrected following \citet{Allart22} and with uncorrected measurements.

\texttt{S-BART} provides RV measurements corrected for the influence of telluric features. To achieve it, \texttt{S-BART} creates a transmittance profile using the night with the highest amount of water content in the atmosphere. From that night, it takes the water content, airmass, 
temperature to generate a transmittance profile with \texttt{TELFIT} \citep{Gullikson2014}. It then flags the regions of the transmittance profile where the flux drops under 1\% of the continuum to create a telluric mask. It then shifts the mask over the complete range of barycentric earth radial velocities for the target and rejects all spectral region that fall within those limits. \texttt{S-BART}'s masking is designed to be conservative. The regions rejected are sometimes quite large, which can impact the radial velocity precision, but should remove most of the effect of the tellurics. To calculate uncorrected velocities we switched off the telluric correction of \texttt{S-BART}.  

 \citet{Allart22} provides a procedure to correct, rather than mask, the telluric features in the spectra. It uses a line-by-line radiative transfer codes assuming a single atmospheric layer. Using the sky conditions and the physical properties of the lines it creates a high-resolution telluric spectrum, which then convolves with the instrumental resolution and samples to the instrumental wavelength grid. It finally uses a subset of selected telluric lines to fit the spectrum and subtract the telluric contribution. This allows to avoid masking regions and losing RV content in the spectrum, thus allowing for potentially better RV precision. A drawback of this algorithm is that, in its current implementation, it is only compatible with ESPRESSO and NIRPS data. We cannot use it to correct the HARPS RVs. We re-compute the RVs using the corrected spectra and with \texttt{S-BART}'s correction switched off. We obtain a slightly larger RMS of the data (6.0~m$\cdot$s$^{-1}$) with smaller error bars (0.5 m$\cdot$s$^{-1}$). 

\begin{figure}[ht]
	\includegraphics[width=9cm]{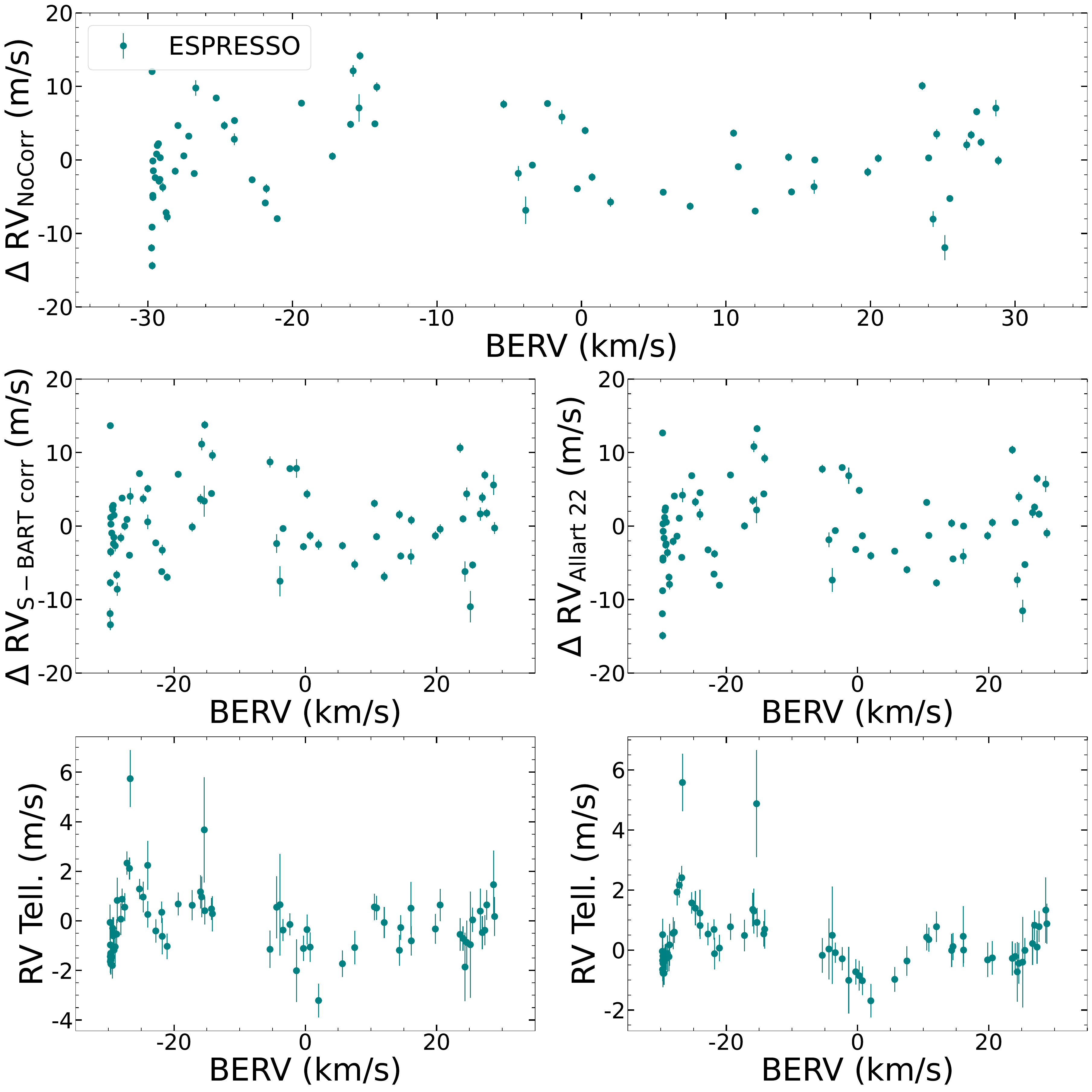}
	\caption{\textbf{ESPRESSO Radial velocity measurements as a function of the Barycentric Earth Radial Velocity of their measurements.} \textit{Top panel} shows the uncorrected RV measurements. \textit{Middle panels} show the telluric-corrected RV measurements using the S-BART default correction (left) and the method of \citet{Allart22} (right). \textit{Bottom panels} show the telluric-induced RV variations estimated by both methods.}
	\label{rv_tell}
\end{figure}

Telluric contamination in the RV measurements often appears as a coherent signal with respect to the barycentric earth radial velocity (BERV) of up to a few m/s \citep{Allart22}. Figure~\ref{rv_tell} shows the uncorrected ESPRESSO RVs as a function of the BERV, the corrected RVs using both methods and the difference, which corresponds to the suspected telluric effect in the RVs. While the difference between the uncorrected and corrected velocities is difficult to see, the residuals after subtracting the corrected RVs show a structure that is similar for both corrections. The biggest difference being that the correction of the algorithm by \cite{Allart22} seems cleaner. Tellurics in our ESPRESSO data induce an RMS of $\sim$ 1.3 m$\cdot$s$^{-1}$, large enough to take it into account. 

\begin{figure}[ht]
	\includegraphics[width=9cm]{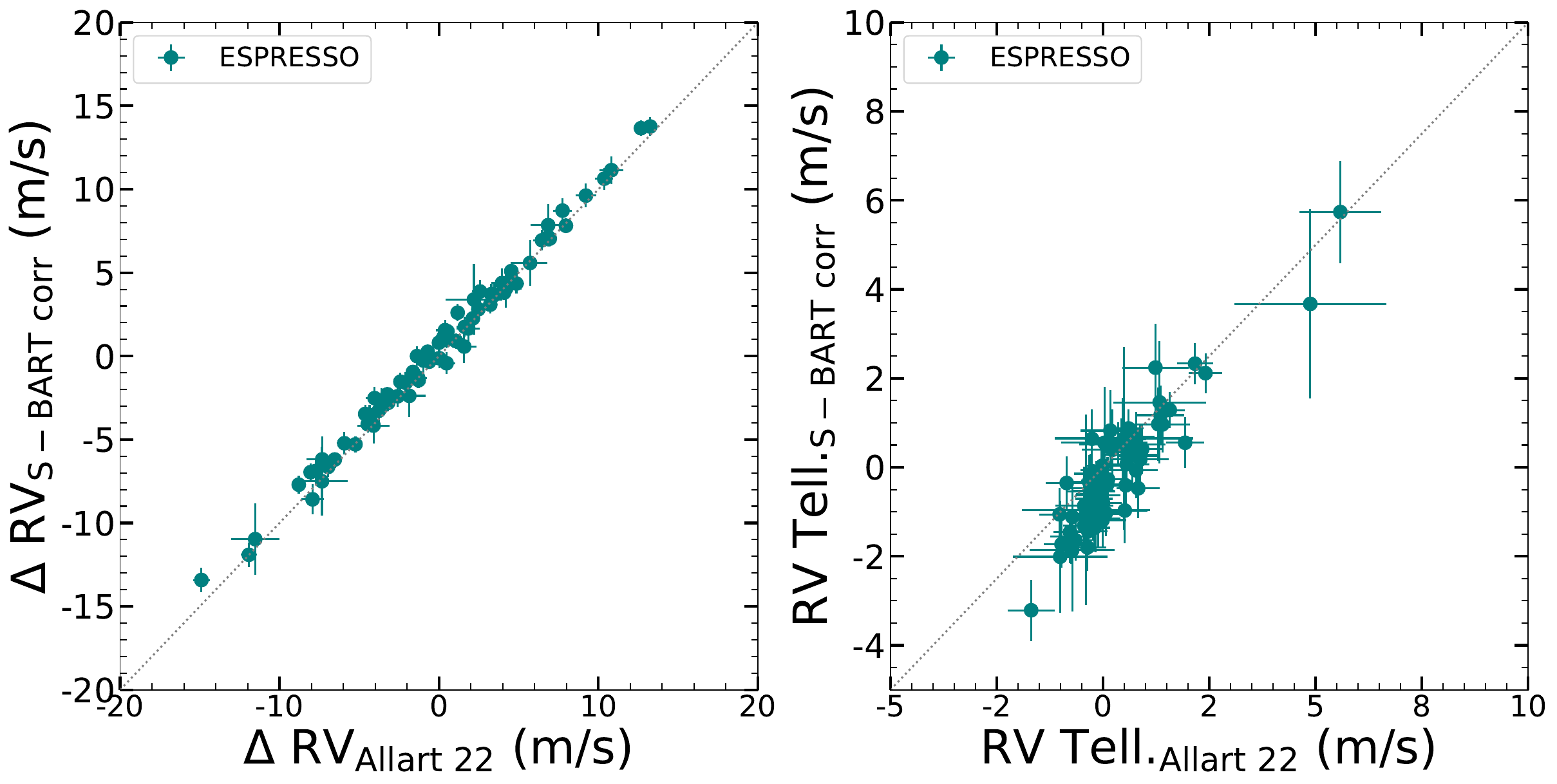}
	\caption{\textbf{Comparison of the RV derived from different telluric correction algorithms.} \textit{Left panels} show comparison between the corrected RVs obtained with the telluric correction of \texttt{S-BART} and with the correction of \citet{Allart22} \textit{Right panels} show the comparison between the telluric RVs. The grey dotted lines show the 1--1 relationship.}
	\label{rv_tell_comp}
\end{figure}

Both algorithms produce very similar corrections and provide very similar corrected RVs. The telluric RVs have an Spearman's correlation coefficient \citep{Spearman1904} of 0.84, while the corrected RVs of 0.99. Figure~\ref{rv_tell_comp} show the direct comparison between both sets of RVs. Both corrected RVs follow a 1--1 relationship almost perfectly. The telluric RVs are also well correlated, however there is a small shift between the computations of both algorithms. Nevertheless, the shift is at the level of the precision of the data. 

\begin{figure}[ht]
	\includegraphics[width=9cm]{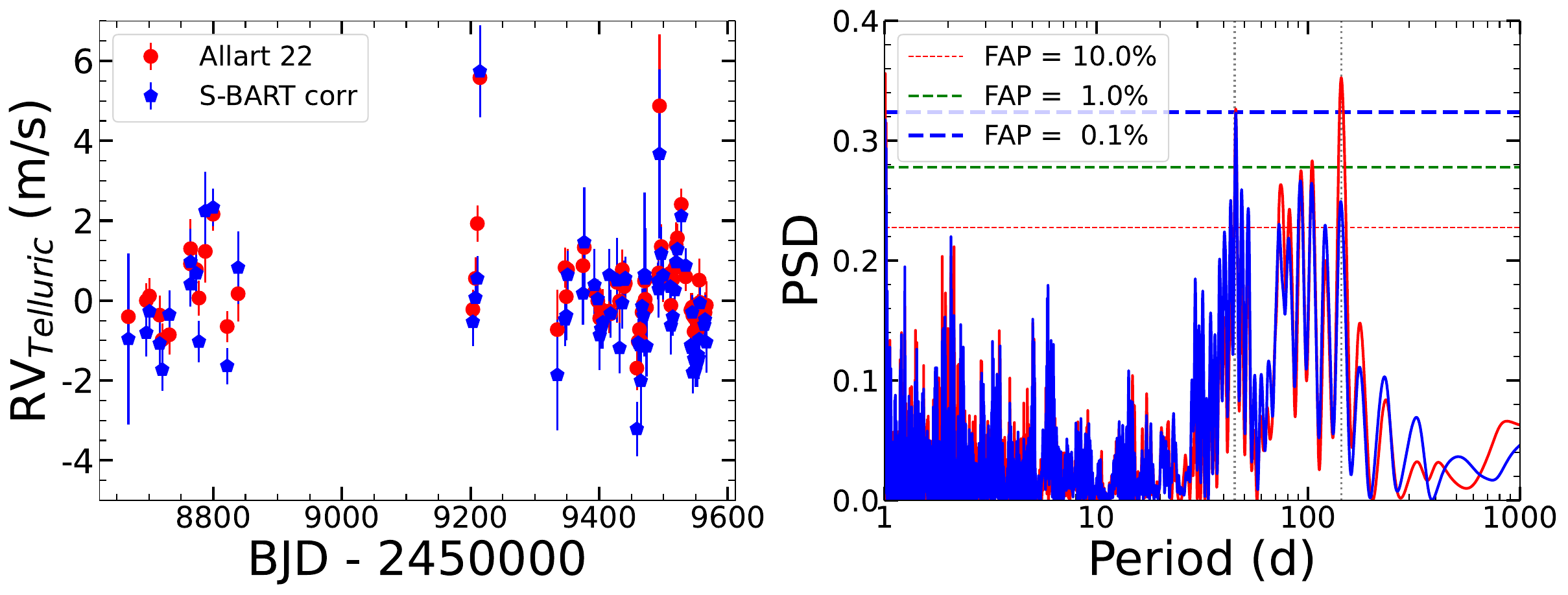}
	\caption{\textbf{Telluric RVs and their periodogram.} \textit{Left} show the telluric RV measurements obtained with the algorithm of \citet{Allart22} and with \texttt{S-BART}. \textit{Right panel} show the GLS periodogram of both datasets. The grey vertical lines highlight the most significant periods, of 45 and 140 days.}
	\label{rv_tell_gls}
\end{figure}

The effect of telluric contamination typically produces periodicities in the data related in some way to Earth's year. Figure~\ref{rv_tell_gls} shows both sets of telluric RVs as a function of time and their GLS periodogram. The periodicities seen in both datasets seem very similar, with the most prominent peaks in the periodogram arising at 140 and 45 days. 

Both telluric correction algorithms tested in the ESPRESSO data provide very similar results. The algorithm by \citet{Allart22} seems to provide a cleaner correction and allows us to avoid masking regions of the spectrum, producing higher precision RV measurements. However, the differences are small in our dataset. For consistency between the ESPRESSO and HARPS RV computations, we opted to rely on \texttt{S-BART}'s default correction with \texttt{TELFIT}.  

\section{Telemetry data of ESPRESSO and HARPS} \label{append_tel}

To prevent the possibility of false-positives or biases due to instrumental effects, we study the telemetry of HARPS and ESPRESSO to search for variations that could be correlated with scientific measurements. \citet{SuarezMascareno2023} identified the temperature of the optical elements as a source of variations that could be mistaken for an astrophysical effect. The temperature of the \'Echelle grating is a typical measurement that can be found in most \'Echelle spectrographs, making it a good proxy to follow the thermal behaviour of the instruments. Another possible source of systematic effects is related to telluric contamination, imperfections on the correction of the Earth's motion or imperfections of the construction of the detectors \citep{Dumusque2015}. These effects manifest as yearly signals and are usually correlated with the Earth's barycentric radial velocity (BERV). 

To determine whether there is an effect in the science measurements, we compute Spearman's correlation coefficient \citep{Spearman1904} between these measurements and the temperature of the \'Echelle grating and the BERV, independently for each instrument. As science measurements we use the RV measurements, FWHM of the CCF, bisector span of the CCF, Ca II H\&K S-index, H$\alpha$ index and NaI index. Same as in \citet{SuarezMascareno2023}, we find no correlation between the temperature changes and the RV measurements, neither for HARPS nor for ESPRESSO. We do find weak but significant correlations with the FWHM of the CCF, bisector span and S-index, both for HARPS and ESPRESSO, and with H$\alpha$ in the case of HARPS. We find no correlation between the BERV and the RV or FWHM measurements. We find a correlation with the bisector span of ESPRESSO, the contrast of HARPS, the S-index of ESPRESSO and the NaI index of HARPS. Table~\ref{tab:tel_corr} shows the measured correlation coefficients and their p-values for all measurements. We used these information in appendix~\ref{append_act}, by including a polynomial model linking the science time-series and the telemetry time-series in the activity model.

\begin{table}
\begin{center}
\caption{Spearman's correlation coefficient between the science measurements and the temperature of the telemetry proxies. Correlations considered significant are highlighted in bold font.\label{tab:tel_corr}}
\begin{tabular}[center]{l l c c}
\hline
 & Instrument & Correlation & P-value\\ \hline
RV vs T$_{\rm Ech.}$ & HARPS & 0.06 & 0.67  \\
RV vs T$_{\rm Ech.}$ & ESPRESSO & -0.04 & 0.70  \\
FWHM vs T$_{\rm Ech.}$ & HARPS & 0.29 & \textbf{0.05}  \\
FWHM vs T$_{\rm Ech.}$ & ESPRESSO & 0.22 & \textbf{0.07} \\
BIS vs T$_{\rm Ech.}$ & HARPS & 0.28 & \textbf{0.05}  \\
BIS vs T$_{\rm Ech.}$ & ESPRESSO & 0.22 & \textbf{0.07} \\ 
Contrast vs T$_{\rm Ech.}$ & HARPS & -0.18 & 0.23 \\ 
Contrast vs T$_{\rm Ech.}$ & ESPRESSO & -0.11 & 0.36 \\ 
S-index vs T$_{\rm Ech.}$ & HARPS & 0.25 & \textbf{0.08} \\ 
S-index vs T$_{\rm Ech.}$ & ESPRESSO & 0.23 & \textbf{0.05} \\ 
H$\alpha$ vs T$_{\rm Ech.}$ & HARPS & 0.42 & \textbf{\textless 0.01} \\ 
H$\alpha$ vs T$_{\rm Ech.}$ & ESPRESSO & 0.06 & 0.62 \\
Na I vs T$_{\rm Ech.}$ & HARPS & 0.09 & 0.56 \\ 
Na I vs T$_{\rm Ech.}$ & ESPRESSO & -0.05 & 0.66 \\ \\

RV vs BERV & HARPS & -0.17 & 0.25  \\
RV vs BERV & ESPRESSO & 0.14 & 0.25  \\
FWHM vs BERV & HARPS & -0.02 & 0.87  \\
FWHM vs BERV & ESPRESSO & 0.08 & 0.49  \\
BIS vs BERV & HARPS & 0.42 & \textbf{\textless 0.01}  \\
BIS vs BERV & ESPRESSO & 0.04 & 0.74 \\
Contrast vs BERV & HARPS & 0.27 & \textbf{0.07}  \\ 
Contrast vs BERV & ESPRESSO & 0.06 & 0.64 \\ 
S-index vs BERV & HARPS & -0.09 & 0.53 \\ 
S-index vs BERV & ESPRESSO & -0.26 & \textbf{0.03} \\ 
H$\alpha$ vs BERV & HARPS & 0.10 & 0.48\\ 
H$\alpha$ vs BERV & ESPRESSO & -0.14 & 0.27\\
Na I vs BERV & HARPS & 0.01 & 0.93\\ 
Na I vs BERV & ESPRESSO & 0.28 & \textbf{0.02} \\
\hline
\end{tabular}
\end{center}
\end{table}

\section{Extended analysis of stellar activity} \label{append_act}

Here we include the analysis of stellar activity indicators not explicitly described in section~\ref{sect_tel_stel} to search for periodicities related to stellar activity which could create false-positive detections in RV or bias the amplitude measurements in RV. 

\subsubsection{Contrast of the CCF}

\begin{figure*}[ht]
	\includegraphics[width=18cm]{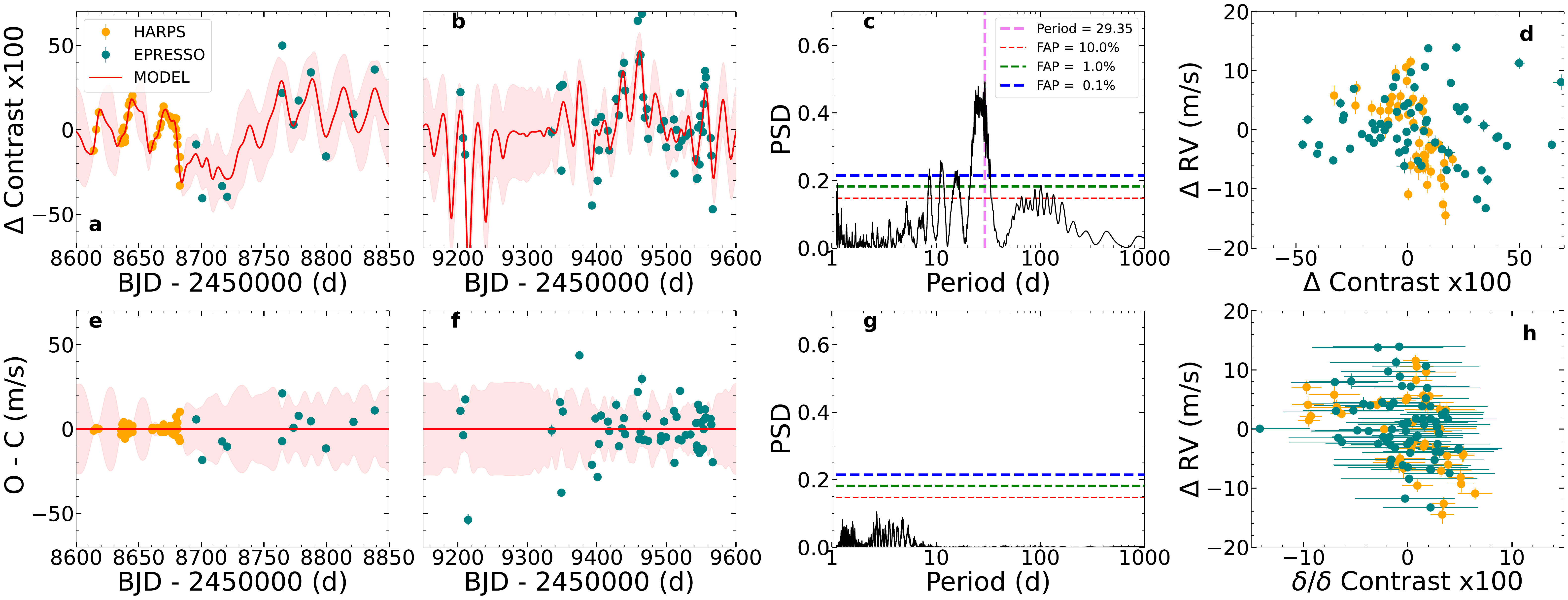}
	\caption{\textbf{Analysis of the contrast of the CCF.} Same as Figure~\ref{fwhm_model}, only with the contrast of the CCF instead.}
	\label{cont_model}
\end{figure*}

Figure~\ref{cont_model} shows the data, model, periodogram and correlation plots of the variations contrast of the CCF. The contrast is a dimensionless measure of the difference between the continuum and the minimum in the normalized CCF. To avoid numerical issues and ease visibility, we chose to work with $\Delta$ contrast $\times$ 100. We measure an RMS of the data of  24.9, and a median uncertainty of 1.1. We get a period of 25.1$^{+2.0}_{-2.8}$ days, with an evolutionary timescale of 29$^{+12}_{-10}$ days. We obtain a fit that is more stochastic than that for the other parameters of the CCF. After subtracting the model, we measure an RMS of the residuals of 11.3 and find no significant signals in their periodogram. We do not find a significant relationship between the contrast, or its gradient, and the RV data.

\subsubsection{S$_{MW}$ Index (Ca II H\&K)}

\begin{figure*}[ht]
	\includegraphics[width=18cm]{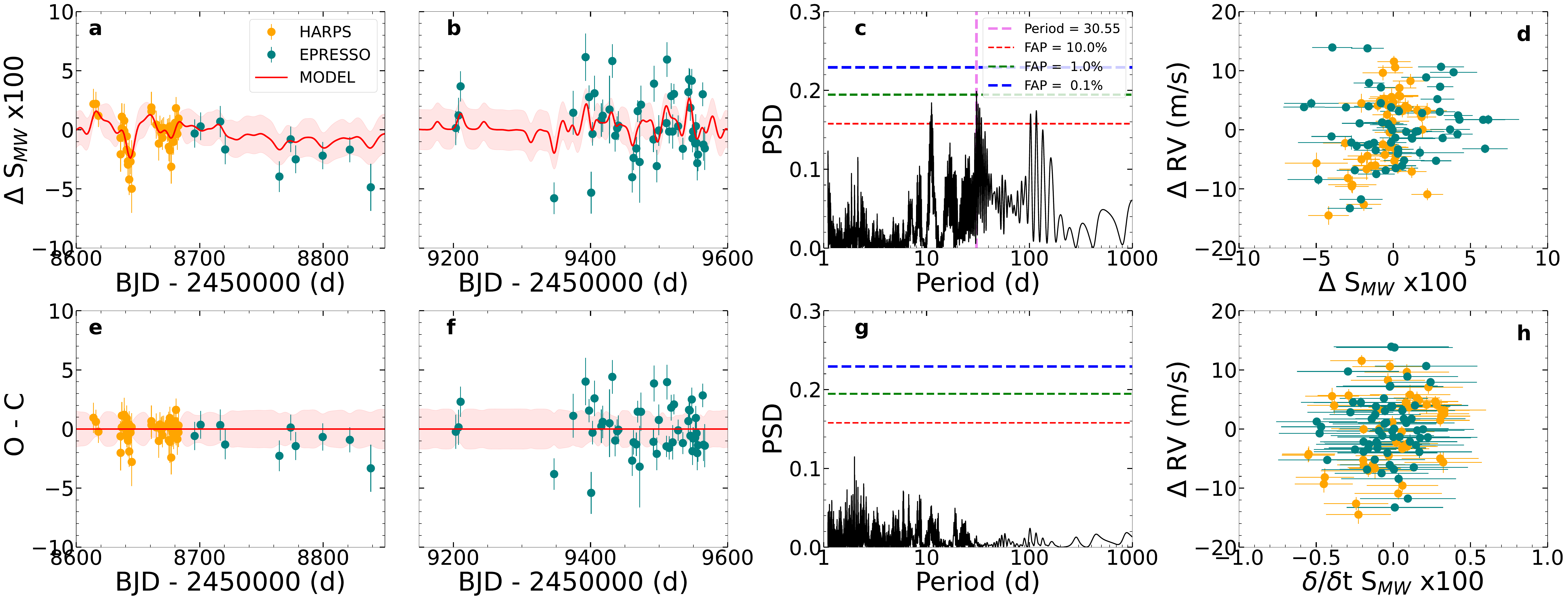}
	\caption{\textbf{Analysis of the S$_{MW}$ Ca II H\&K index.} Same as Figure~\ref{fwhm_model}, only with the S$_{MW}$ Ca II H\&K index instead.}
	\label{smw_model}
\end{figure*}

Figure~\ref{smw_model} shows the data, model, periodogram and correlation plots of the variations of the Mount Wilson S-index. To avoid numerical issues and aid the reader in visualisation, we opted to work with $\Delta$ S$_{MW}$ $\times$ 100.  We measure an RMS of the data of  0.064. The GLS periodogram of the data shows significant structure at periods longer than the rotation period. We attempted to include different strategies to account for this, but none of them gave good results. Among the tested models are low-order polynomials, sinusoidal models with several different priors for the periods and low-order polynomials against the measurements of the telemetry of the instruments. With the GP model, we obtain a period of 29$^{15}_{-10}$ days, with an evolutionary timescale of 28$^{+22}_{-10}$ days. We obtain a fit that is mostly stochastic. After subtracting the model, we find no significant signals in their periodogram. We do not find a significant relationship between the S$_{MW}$, or its gradient, and the RV data.

\subsubsection{H$\alpha$ Index}

\begin{figure*}[ht]
	\includegraphics[width=18cm]{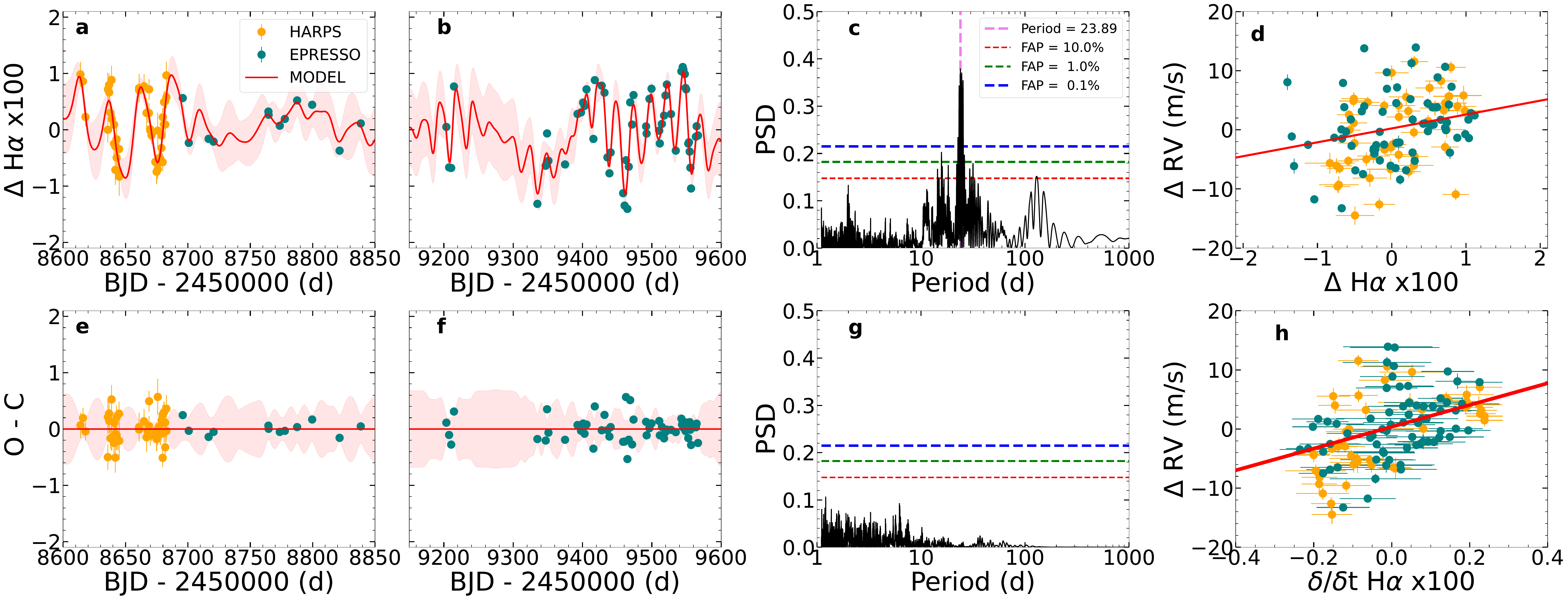}
	\caption{\textbf{Analysis of the H$\alpha$ index.} Same as Figure~\ref{fwhm_model}, only with the S$_{MW}$ H$\alpha$ index instead.}
	\label{ha_model}
\end{figure*}

Figure~\ref{ha_model} shows the data, model, periodogram and correlation plots of the variations of the H$\alpha$ index. To avoid numerical issues and ease visibility, we opted to work with $\Delta$ H$\alpha$ $\times$ 100. The GLS periodogram of the data shows significant peak at the same period detected in FWHM and bisector span. With the GP model, we obtain a period of 24.8$^{+1.6}_{-1.4}$ days, with an evolutionary timescale of 27.6$^{+7.7}_{-6.0}$ days. We obtain a fairly stable fit. After subtracting the model. We find significant positive relationships between the variations in H$\alpha$ and their gradient, and the RV data. 

\subsubsection{Na I Index}

\begin{figure*}[ht]
	\includegraphics[width=18cm]{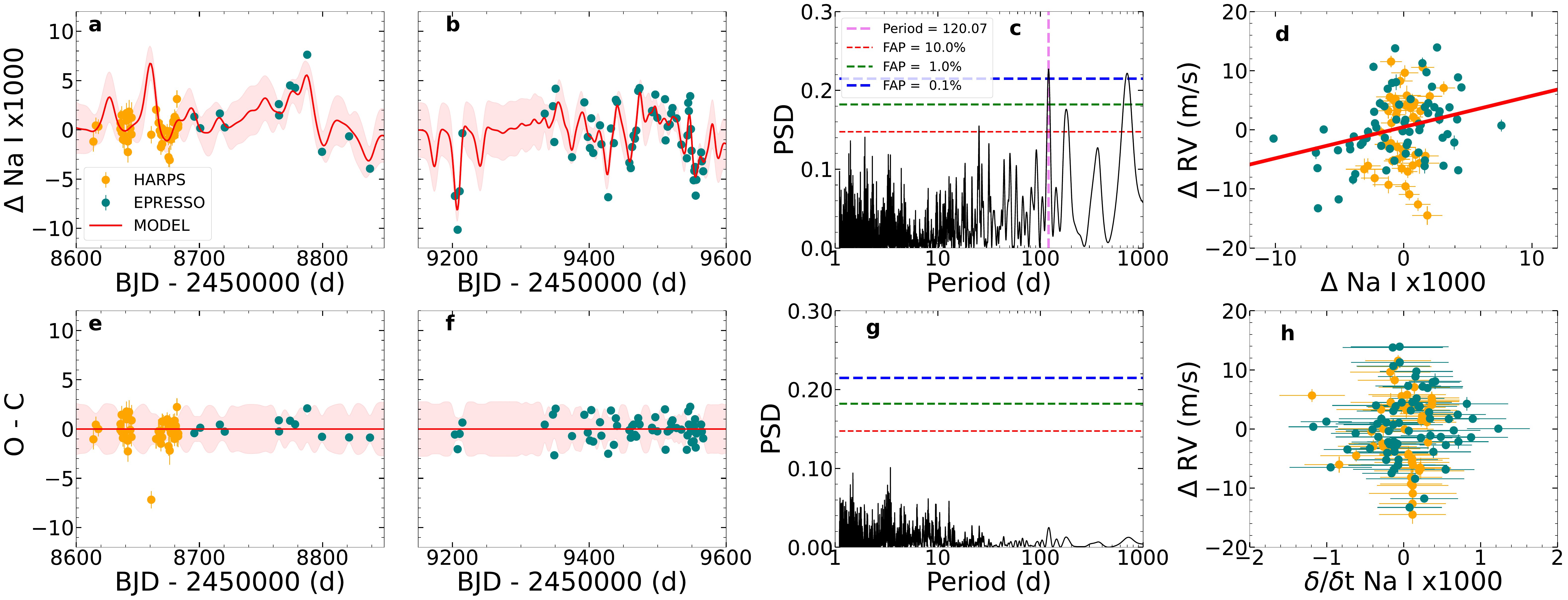}
	\caption{\textbf{Analysis of the Na I index.} Same as Figure~\ref{fwhm_model}, only with the Na I index instead.}
	\label{nai_model}
\end{figure*}

Figure~\ref{nai_model} shows the data, model, periodogram and correlation plots of the variations of the Na I index. To avoid numerical issues and ease visibility, we opted to work with $\Delta$ Na I $\times$ 1000. The GLS periodogram of the data shows significant structure at periods longer than the rotation period. We attempted to include different strategies to account for this, but none of them gave good results. Among the tested models are low-order polynomials, sinusoidal models with several different priors for the periods and low-order polynomials against the measurements of the telemetry of the instruments. With the GP model, we obtain a period of 34.0$^{+4.2}_{3.8}$ days, with an evolutionary timescale of 27.3$^{+10.9}_{-7.4}$ days. After subtracting the model, we find no significant signals in the periodogram of the residuals. We find a significant relationship between the NaI data and the RVs, but not with its gradient. 

\subsubsection{GP timescales} \label{gp_timescales}

\begin{figure}[ht]
	\includegraphics[width=9cm]{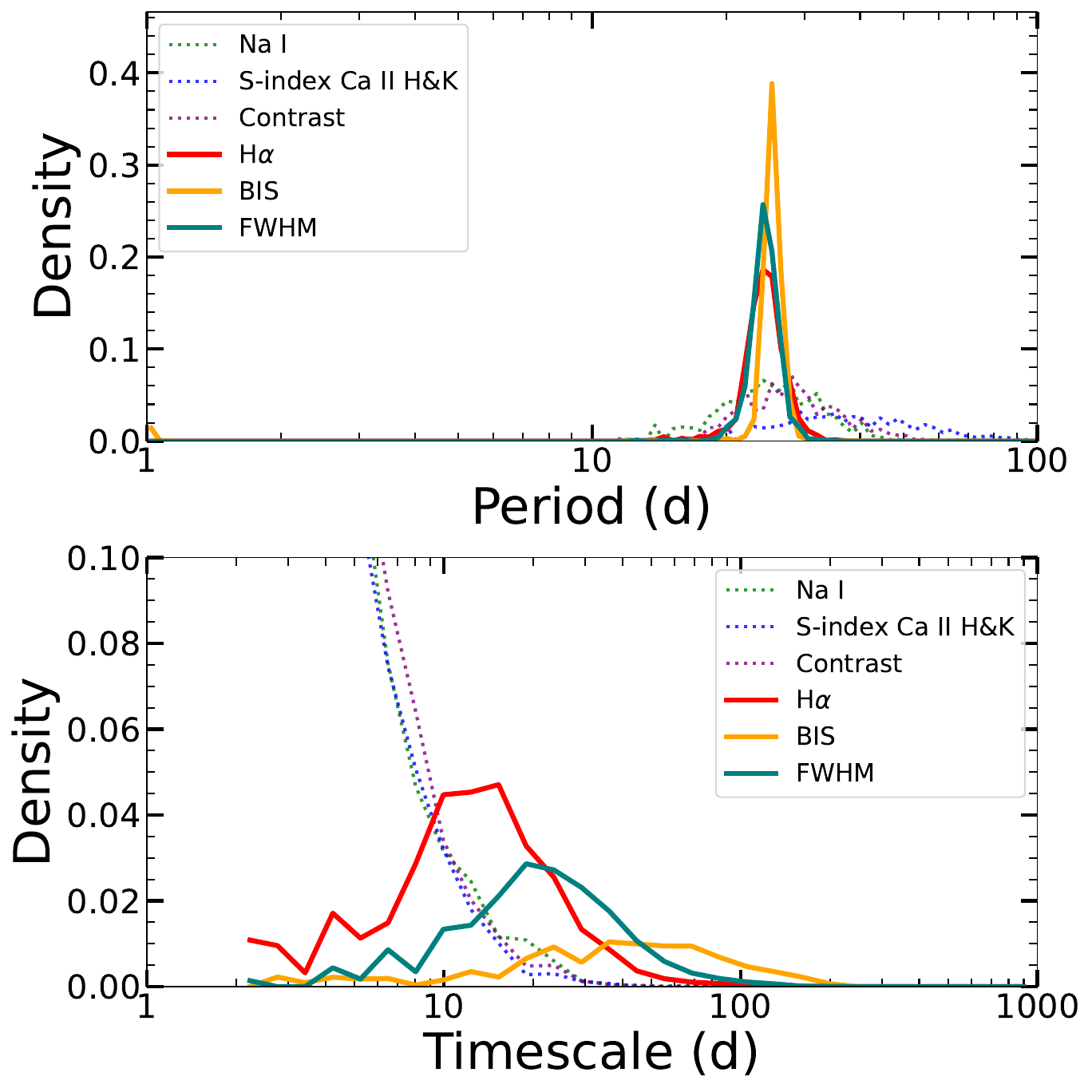}
	\caption{\textbf{GP timescales measured in the activity indicators.} \textbf{Top panel}: Posterior distribution of the periodicity measured for each activity indicator. \textbf{Bottom panel}: Posterior distribution of the damping timescale measured for each activity indicator. The y-axis is cut at 0.1 to ease visibility of the distributions with a peak different from 2 (edge of the prior).}
	\label{posterior_gp}
\end{figure}

The previous models provided a variety of measurements of the timescales of stellar activity of the system. Figure~\ref{posterior_gp} shows the posterior distributions for the GP period and damping timescale. All activity indicators roughly agree with the periodicity, with the FWHM, bisector span and H$\alpha$ index providing the tightest determinations of the period. In those three cases, the period determination is 1$\sigma$ consistent. In these three indicators we also  measured damping timescales significantly different from zero. The specific timescale varies, but the posteriors are very broad. For the case of the contrast, S-index and Na I index, we obtain a timescale that sits at the edge of the prior (2 days), which indicates a mostly stochastic behaviour of the models. 

\clearpage

\onecolumn

\section{Parameters of the models}\label{append_tables}

%\fontsize{8.6}{8.6}\selectfont

\renewcommand{\arraystretch}{1.36}
\begin{longtable}{llccc} 
\caption{Priors and measured parameters of the models described in sections~\ref{sect_gp}, ~\ref{sect_gp_1p} and ~\ref{sect_gp_2p}. All Upper limits are 99.7\% limits.} \label{table_measured} \\
\endfirsthead
\multicolumn{5}{c}{\tablename\ \thetable\ -- \textit{Continued from previous page}} \\
\hline
\endhead
\hline \multicolumn{5}{r}{\textit{Continued on next page}} \\
\endfoot
\hline
\endlastfoot
\hline
Parameter  & Priors & 0-Planets & 1-Planet (Kep) &  2-Planets (Kep)\\ \hline

\textbf{Zero points} \\
V0 HARPS$_{FWHM}$ [m~s$^{-1}$] &  
$\mathcal{N}$ (0 , 17) &   
4.2$^{+8.9}_{-9.3}$ &   
5.8$^{+7.6}_{-7.8}$ & 
4.6$^{+8.3}_{-9.0}$\\

V0 ESP$_{FWHM}$ [m~s$^{-1}$] &  
$\mathcal{N}$ (0 , 17) &   
-0.9$^{+6.5}_{-5.8}$ & 
-0.6$^{+5.2}_{-5.4}$ &  
-0.3$^{+6.3}_{-5.8}$ \\

V0 HARPS$_{BIS}$ [m~s$^{-1}$] &  
$\mathcal{N}$ (0 , 6.2) &  
0.1$^{+2.0}_{-2.0}$ &  
0.8$^{+2.3}_{-2.0}$ &  
0.4$^{+2.0}_{-1.9}$ \\
 
V0 ESP$_{BIS}$ [m~s$^{-1}$] &  
$\mathcal{N}$ (0 , 6.2) &   
-0.1$^{+1.3}_{-1.3}$ &  
-0.3$^{+1.2}_{-1.3}$ & 
0.2$^{+1.4}_{-1.3}$   \\ 

V0 HARPS$_{RV}$ [m~s$^{-1}$] & 
$\mathcal{N}$ (0 , 5.9) &   
-0.6$^{+2.3}_{-2.4}$ & 
-1.9$^{+2.9}_{-3.2}$ & 
-1.0$^{+2.0}_{-2.3}$\\

V0 ESP$_{RV}$ [m~s$^{-1}$] &  
$\mathcal{N}$ (0 , 5.9) &   
0.1$^{+1.6}_{-1.7}$ &  
0.7$^{+2.0}_{-2.1}$ & 
-0.2$^{+1.5}_{-1.5}$\\

V0 TESS [ppm] &  
$\mathcal{N}$ (0 , 3070) &   
700$^{+2600}_{-2700}$ &  
500$^{+2600}_{-2500}$ & 
400$^{+2700}_{-2600}$\\

\\
\textbf{GP Parameters} \\
A$_{FWHM}$ [m~s$^{-1}$] & 
$\mathcal{U}$ (0 , 85) &   
23.7$^{+9.5}_{-5.1}$ &  
21.2$^{+7.2}_{-4.0}$ & 
22.7$^{+6.9}_{-4.5}$\\

A11$_{RV}$ [m~s$^{-1}$] & 
$\mathcal{U}$ (-29.3 , 26.3) &   
4.7$^{+2.1}_{-1.5}$ &  
6.5$^{+2.0}_{-1.6}$ &   
4.1$^{+1.7}_{-1.3}$ \\

A12$_{RV}$ [m~s$^{-1}$] & $\mathcal{U}$ (-58.6 , 58.6) &   
-1.6$^{+4.5}_{-5.2}$ &  
-3.9$^{+5.0}_{-6.3}$ & 
-1.4$^{+4.2}_{-4.7}$\\

$P_{\rm rot~FWHM}$ [d] & $\mathcal{N}$ (25.0 , 2.5) &
24.50$^{+0.86}_{-0.84}$ &  
24.28$^{+0.67}_{-0.62}$ & 
24.42$^{+0.85}_{-0.82}$\\

ln $L_{\rm FWHM}$ [d] & $\mathcal{N}$ (4.0 , 0.4) &   
3.57$^{+0.17}_{-0.17}$ &  
3.62$^{+0.18}_{-0.18}$ & 
3.56$^{+0.18}_{-0.18}$ \\

ln $\omega_{\rm FWHM}$ & $\mathcal{U}$ (-5 , 5) &   
-0.27$^{+0.39}_{-0.41}$ &  
-0.50$^{+0.36}_{-0.37}$ & 
-0.34$^{+0.34}_{-0.37}$ \\

A$_{BIS}$ [m~s$^{-1}$] & 
$\mathcal{U}$ (0 , 31) &   
5.8$^{+1.2}_{-1.1}$ & 
6.1$^{+1.3}_{-1.1}$ & 
5.5$^{+1.3}_{-1.2}$\\

A21$_{RV}$ [m~s$^{-1}$] & $\mathcal{U}$ (-29.3 , 29.3) & 
-5.1$^{+1.1}_{-1.3}$ & 
-7.4$^{+1.4}_{-1.5}$ & 
-5.0$^{+1.2}_{-1.4}$\\

A22$_{RV}$ [m~s$^{-1}$] & $\mathcal{U}$ (-58.6 , 58.6) &
8.7$^{+3.6}_{-3.7}$ &  
-2.0$^{+8.7}_{-7.8}$ & 
10.6$^{+4.0}_{-3.5}$\\

$P_{\rm rot~BIS}$ [d] & $\mathcal{N}$ (25.0 , 2.5) &
26.53$^{+0.45}_{-0.46}$ &  
26.06$^{+0.79}_{-0.73}$ & 
26.64$^{+0.41}_{-0.43}$\\

ln $L_{\rm BIS}$ & $\mathcal{N}$ (4.0 , 0.4) &   
3.87$^{+0.19}_{-0.22}$ &  
3.36$^{+0.27}_{-0.20}$ & 
3.89$^{+0.20}_{-0.23}$ \\

ln $\omega_{\rm BIS}$ & $\mathcal{U}$ (-5 , 5) &   
-2.4$^{+1.3}_{-1.7}$ &  
-1.56$^{+0.74}_{-2.08}$ & 
-2.1$^{+1.0}_{-1.8}$ \\

ln A$_{TESS}$ [ppm] & $\mathcal{U}$ (0 , 20) &
10.60$^{+0.20}_{-0.21}$ &  
10.55$^{+0.20}_{-0.19}$ & 
10.62$^{+0.20}_{-0.20}$\\

$P_{\rm rot~TESS}$ [d] & $\mathcal{N}$ (25 , 2.5) &
23.8$^{+2.5}_{-2.5}$ &  
24.2$^{+2.5}_{-2.6}$ & 
26.7$^{+2.8}_{-2.8}$\\

ln $L_{\rm TESS}$ [d] & $\mathcal{N}$ (4.0 , 0.4) &   
3.58$^{+0.30}_{-0.30}$ &  
3.60$^{+0.31}_{-0.32}$ & 
3.57$^{+0.29}_{-0.30}$ \\

\\
\textbf{Planet b} \\
$Phase$ & 
$\mathcal{U}$ (0 , 1) &  
& 
0.1212$^{+0.0025}_{-0.0026}$ & 
0.1213$^{+0.0025}_{-0.0027}$\\

$P_{\rm orb}$ [d] & 
$\mathcal{N}$ (1.2731 , 0.1) & 
&
1.2730988$^{+0.0000027}_{-0.0000028}$ & 
1.2730989$^{+0.0000027}_{-0.0000028}$\\

ln R$_{p}$ [R$_{\oplus}$] &  
$\mathcal{U}$ (-5 , 5) &
& 
0.327$^{+0.057}_{-0.065}$ & 
0.337$^{+0.059}_{-0.061}$\\

ln M$_{p}$ [M$_{\oplus}$] &  
$\mathcal{U}$ (-5 , 5) &
& 
1.37$^{+0.12}_{-0.14}$ & 
1.23$^{+0.13}_{-0.15}$\\

b &  
$\mathcal{U}$ (0 , 1) &
& 
$\textless$ 0.61 & 
$\textless$ 0.57\\

$\sqrt{e} ~cos(\omega)$ & 
$\mathcal{N}$ (0 , 0.3) &
&
-0.06$^{+0.16}_{-0.14}$ &
0.02$^{+0.17}_{-0.16}$ \\

$\sqrt{e} ~sin(\omega)$ & 
$\mathcal{N}$ (0 , 0.3) &
& 
-0.02$^{+0.16}_{-0.16}$ &
-0.02$^{+0.15}_{-0.15}$ \\

\\
\textbf{Planet c} \\
$Phase$ & 
$\mathcal{U}$ (0 , 1) &  
& 
& 
0.624938$^{+0.00063}_{-0.00050}$\\

$P_{\rm orb}$ [d] & 
$\mathcal{N}$ (8.5 , 0.5) & 
&
&
8.465651$^{+0.000035}_{-0.000029}$\\

ln R$_{p}$ [R$_{\oplus}$] &  
$\mathcal{U}$ (-5 , 5) &
& 
& 
0.776$^{+0.084}_{-0.092}$\\

ln M$_{p}$ [M$_{\oplus}$] &  
$\mathcal{U}$ (-5 , 5) &
& 
& 
1.89$^{+0.16}_{-0.19}$\\

b&  
$\mathcal{U}$ (0 , 1) &
& 
& 
0.774$^{+0.052}_{-0.057}$\\

$\sqrt{e} ~cos(\omega)$ & 
$\mathcal{N}$ (0 , 0.3) &
& 
& 
0.22$^{+0.13}_{-0.17}$ \\

$\sqrt{e} ~sin(\omega)$ & 
$\mathcal{N}$ (0 , 0.3)  &
& 
& 
0.00$^{+0.19}_{-0.19}$ \\

\\
\textbf{Limb darkening} \\
q$_{1}$& 
$\mathcal{N}$ (0.4676 , 0.0067) &   
&   
0.4677$^{+0.0021}_{-0.0021}$ & 
0.4679$^{+0.0021}_{-0.0021}$\\

q$_{2}$& 
$\mathcal{N}$ (0.1442 , 0.011) &   
&   
0.1443$^{+0.0034}_{-0.0033}$ & 
0.1441$^{+0.0033}_{-0.0034}$\\

\\
\textbf{Stellar parameters} \\
M$_{*}$ [M$_{\odot}$]& 
$\mathcal{N}$ (0.790 , 0.043) &   
&   
0.783$^{+0.041}_{-0.042}$ & 
0.788$^{+0.040}_{-0.041}$\\

R$_{*}$ [R$_{\odot}$]& 
$\mathcal{N}$ (0.733 , 0.034) &   
&   
0.746$^{+0.029}_{-0.028}$ & 
0.746$^{+0.027}_{-0.028}$\\

\\
\textbf{Jitter} \\
ln Jit HARPS$_{FWHM}$ [m~s$^{-1}$] & 
$\mathcal{U}$ (-5 , 5) &   
1.36$^{+0.19}_{-0.21}$ &   
1.44$^{+0.22}_{-0.22}$ & 
1.36$^{+0.21}_{-0.22}$\\

ln Jit ESP$_{FWHM}$ [m~s$^{-1}$] & 
$\mathcal{N}$ (-5 , 5) &   
2.27$^{+0.13}_{-0.13}$ & 
2.27$^{+0.12}_{-0.12}$ &  
2.28$^{+0.13}_{-0.14}$ \\

ln Jit HARPS$_{BIS}$ [m~s$^{-1}$] & 
$\mathcal{N}$ (-5 , 5) &
1.11$^{+0.19}_{-0.21}$ &  
1.23$^{+0.17}_{-0.19}$ &  
1.19$^{+0.17}_{-0.18}$ \\

ln Jit ESP$_{BIS}$ [m~s$^{-1}$] & 
$\mathcal{N}$ (-5 , 5) &   
1.45$^{+0.15}_{-0.15}$ &  
1.40$^{+0.14}_{-0.14}$ & 
1.58$^{+0.14}_{-0.15}$   \\ 

ln Jit HARPS$_{RV}$ [m~s$^{-1}$] & 
$\mathcal{N}$ (-5 , 5) &   
1.12$^{+0.16}_{-0.15}$ & 
0.74$^{+0.22}_{-0.21}$ & 
0.45$^{+0.20}_{-0.22}$\\

ln Jit ESP$_{RV}$ [m~s$^{-1}$] & 
$\mathcal{N}$ (-5 , 5) &   
1.25$^{+0.13}_{-0.13}$ &  
0.37$^{+0.32}_{-0.29}$ & 
0.69$^{+0.17}_{-0.17}$\\

\\
\textbf{Model statistics} \\
lnZ & & 
-82110.6 & 
-81614.9 & 
-81582.5  \\

$\Delta$ lnZ vs 0p & &
& 
497.7 &  
528.1 \\ \\

RMS O - C RV [m~s$^{-1}$] & & 2.96 & 1.53 & 1.26 \\
RMS O - C RV$_{ESP}$ [m~s$^{-1}$] & & 2.91 & 0.95 & 1.05  \\
RMS O - C RV$_{HARPS}$ [m~s$^{-1}$] & & 3.02 & 2.11 & 1.51 \\
\hline
\end{longtable}

\begin{figure*}[ht]
	\includegraphics[width=18cm]{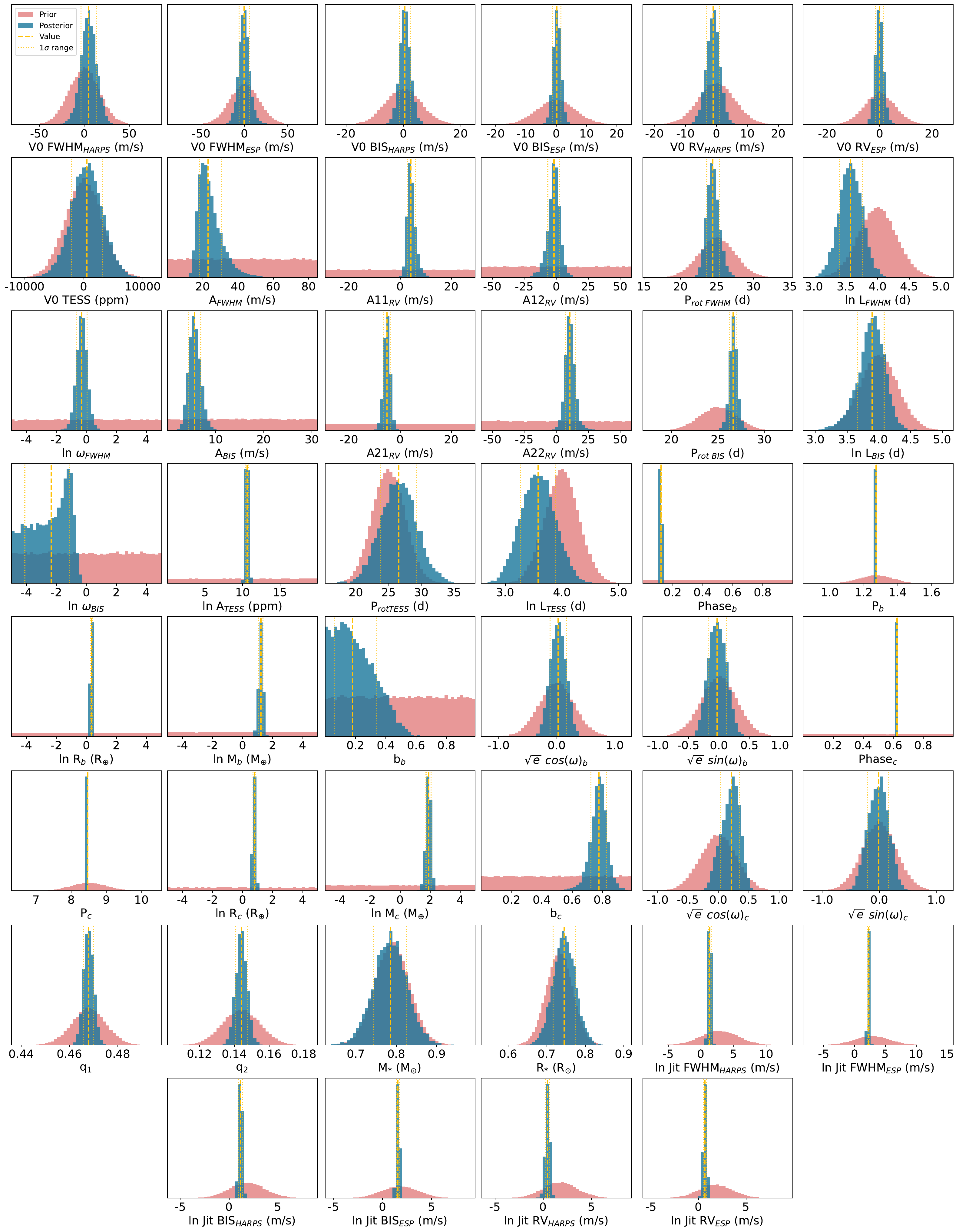}
 	\caption{\textbf{Posterior distribution} (blue distributions) of every parameter sampled in the definitive model. The red distributions show the prior used in the Nested Sampling. The yellow dashed and dotted lines show the median value and 1$\sigma$ range, respectively.}
	\label{posterior_2pk}
\end{figure*}

\clearpage
\section{Additional figures} \label{append_additional}

\begin{figure*}[ht]
	\includegraphics[width=18cm]{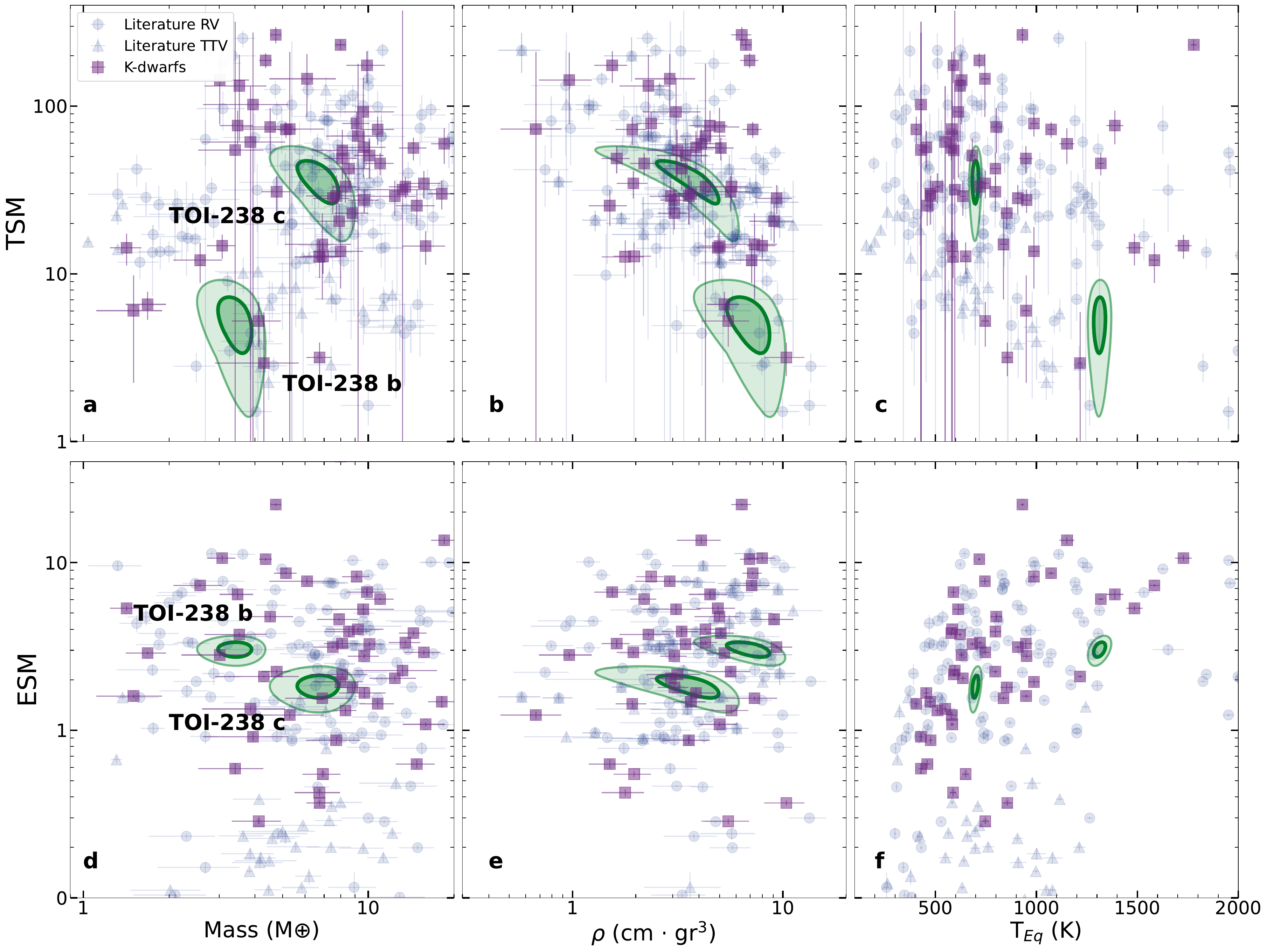}
 	\caption{\textbf{TSM and ESM of TOI--238} compared with the rest of transiting exoplanets with precise measurements of mass and radius. Panels \textbf{a}, \textbf{b} and \textbf{c} show the TSM of all planets against their mass, density and equilibrium temperature. Panels \textbf{d}, \textbf{e} and \textbf{f} show the equivalent plots for the ESM.}
	\label{tsm_esm}
\end{figure*}

\end{appendix}
\label{lastpage}

\end{document}